\documentclass[twocolumn,amsmath,amssymb]{revtex4-1}
\usepackage[T1]{fontenc}
\usepackage{etoolbox}

\pdfoutput=1

\usepackage{upgreek}
\usepackage{graphicx}
\usepackage{soul}
\usepackage{color}
\usepackage{pdfpages}

\begin{document}

\title{Bulk crystalline optomechanics}

\author{W. H. Renninger,$^\dag$}
\email{william.renninger@yale.edu}

\author{P. Kharel}
\altaffiliation{These authors contributed equally to this work.}

\author{R. O. Behunin}

\author{P. T. Rakich}
\email{peter.rakich@yale.edu}

\affiliation{Department of Applied Physics, Yale University, New Haven, Connecticut 06520}

%\titlepage

%\twocolumngrid

%Brillouin useful for lasers, spectroscopy, rf filtering
%Optomechanics as the basis for quantum information processing, fundamental quantum mechanics and sensitive metrology.
%\begin{itemize}
%\item surface physics like defects..etc
%\item Noninvasive- intrinsic material limits
%\item Any transparent material
%\item Wavelength agnostic
%\item any size
%\item ugram mass
%\item high f, low occupation number
%\item long-lived phonon limit of Brillouin
%\item radical Brillouin enhancement
%\item record FQ, Q
%\item high phonon intensities
%\item new type of symmetry breaking
%\item nonlocal phonons
%\end{itemize}
\begin{abstract}%\fontsize{14}{18.6}\selectfont
Brillouin processes couple light and sound through optomechanical three-wave interactions. Within bulk solids, this coupling is mediated by the intrinsic photo-elastic material response yielding coherent emission of high frequency (GHz) acoustic phonons. This same interaction produces strong optical nonlinearities that overtake both Raman or Kerr nonlinearities in practically all solids. In this paper, we show that the strength and character of Brillouin interactions are radically altered at low temperatures when the phonon coherence length surpasses the system size. In this limit, the solid becomes a coherent optomechanical system with macroscopic (cm-scale) phonon modes possessing large ($60\ \upmu \rm{g}$) motional masses.  These phonon modes, which are formed by shaping the surfaces of the crystal into a confocal phononic resonator, yield appreciable optomechanical coupling rates (${\sim}100$ Hz), providing access to ultra-high $Q$-factor ($4.2{\times}10^7$) phonon modes at high ($12$ GHz) carrier frequencies.  The single-pass nonlinear optical susceptibility is enhanced from its room temperature value by more than four orders of magnitude.  Through use of bulk properties, rather than nano-structural control, this comparatively simple approach is enticing for the ability to engineer optomechanical coupling at high frequencies and with high power handling.  In contrast to cavity optomechanics, we show that this system yields a unique form of dispersive symmetry breaking that enables selective phonon heating or cooling without an optical cavity (i.e., cavity-less optomechanics).  Extending these results, practically any transparent crystalline material can be shaped into an optomechanical system as the basis for materials spectroscopy, new regimes of laser physics, precision metrology, quantum information processing, and for studies of macroscopic quantum coherence.
\end{abstract}

\maketitle
%\newpage
%\fontsize{20}{28.6}\selectfont
%\fontsize{16}{20.6}\selectfont
%\fontsize{14}{18.6}\selectfont

\section{Introduction}
In recent years, high-coherence phonons have become an attractive resource for applications like precision measurements and high-fidelity information processing. Such applications, in conjunction with new strategies for the control of phonons using hybrid opto-mechanical \cite{Aspelmeyer2014,Meystre2013,Marquardt2009,Favero2009a,DeLorenzo2014,Lee2012a,Safavi-Naeini2012,Safavi-Naeini2013,Andrews2014,Wilson2014,Chan2011,Poggio2007,Jayich2008a,Bahl2011,Cohen2014},  electro-mechanical \cite{Ekinci2005,Regal2008,Bochmann2013,Mahboob2013,Macquarrie2013,MacQuarrie2014,Weinstein2014,Woolley2016,Bagci2014}, and superconducting circuit QED systems \cite{Chu2017a,Schuetz2015,Cleland2004,Teufel2011a,Fink2015} have sparked a surge of interest in phononic device physics and technologies. % (OM; Vahala, Kippenberg, Painter, etc), (superconducting; Conrad Leonard, Cleland, Painter), (qubits; Surface wave, and Yiwen Chu), (phononic crystal folks; Ihab Elkady, Georgia tech, etc)
In the pursuit of new mesoscopic quantum phenomena, it has become commonplace to operate these systems at cryogenic temperatures to avoid thermal noise.
Low temperatures have another advantage; internal sources of phonon dissipation plummet within crystalline solids permitting acoustic phonons to live for an astounding number ($10^7{-}10^{10}$) of cycles \cite{Thaxter1966,Blair1966}. %can also cite low frequency acoustic measurements that demonstrate 10 Billion.
When harnessed to create high quality-factor ($Q$) modes, such long-lived phonons become a tremendous resource for new hybrid technologies \cite{Chu2017a,Aspelmeyer2014,Bochmann2013}.
Within a variety of micro- and nanoscale systems, high $Q$-factor phonon modes have enabled new platforms for technologies like squeezing \cite{Safavi-Naeini2013}, single phonon detection \cite{Cohen2014}, and microwave to optical conversion \cite{Bochmann2013,Pitanti2015,Balram2015}. % (OM, cite Painter, etc), (superconducting, Conrad, Painter), (Superconducting cite Surface wave, and Yiwen Chu)

%Such extended phonon lifetimes drastically improve the performance of micro-and nanoscale optomechanical and superconducting systems, enabling everything from squeezing to single phonon detection to microwave to optical conversion at cryogenic temperatures. % (OM, cite Painter, etc), (superconducting, Conrad, Painter), (Superconducting cite Surface wave, and Yiwen Chu)
Recently, the highest $Q$-factor (${\sim}10^9$) microwave frequency (${\sim}200\,$MHz) phonon modes have been created using elegant high quality bulk-crystalline resonator structures with minimal material interface loss \cite{Goryachev2013b,Goryachev2012,Galliou2013}.
These ultrahigh $Q$-factor bulk-acoustic modes have enabled new precision metrology studies that examine anomalously strong nonlinearities \cite{Goryachev2014b}, gravitational wave detection \cite{Goryachev2014}, and Lorentz symmetry \cite{Lo2016}.
To date, these high-$Q$ bulk acoustic modes have only been accessible through electro-mechanical (piezoelectric) interactions within crystalline quartz structures. % to allow linear coupling between microwave photons and phonons.
Alternatively, a range of new optomechanical interactions may be possible in an expanded array of materials if these phonon modes were accessible with light.  %(more punchy motivation here...)
%Alternatively, if these phonon modes could be accessed using light, many new regimes of optomechanical interaction may be possible using a variety crystalline media that support distinct dyanmics. (need to work on simple punchy motivation here...) %e.g. microwave to optical converters, new regimes of parametric interaction
\begin{figure*}[]
\centerline{
\includegraphics[width=12.0cm]{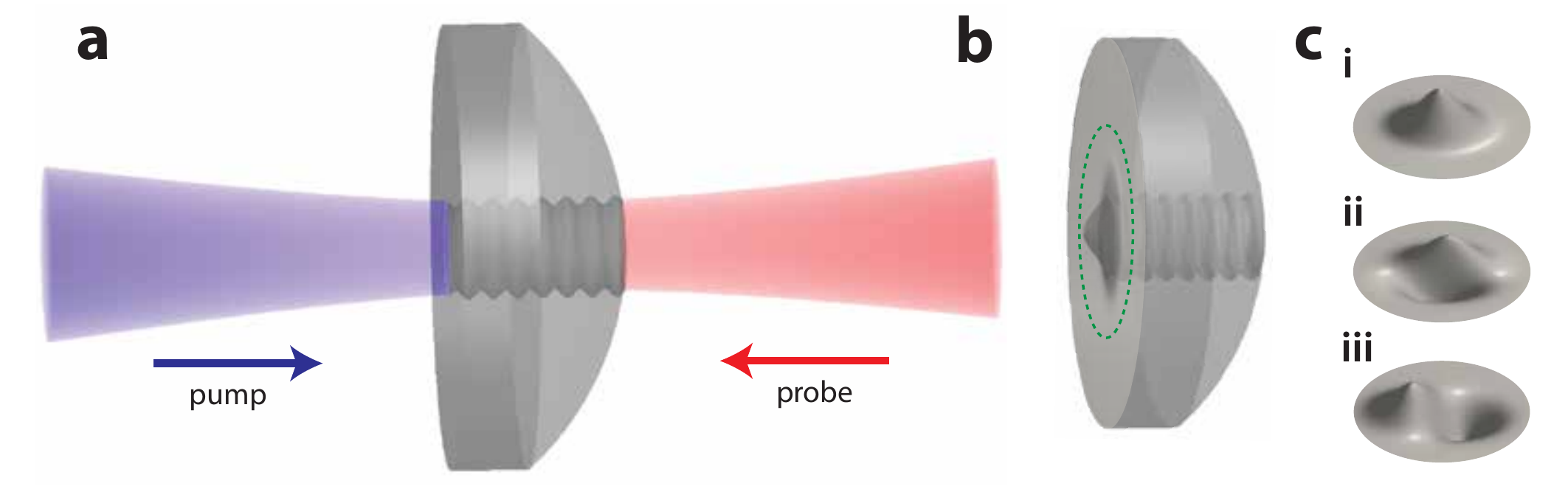}}
\caption{\textbf{Bulk crystalline optomechanical system:} \textbf{a} Traveling optical pump- and probe-beams impinge on the bulk crystalline resonator; interaction between these optical waves is mediated by a phonon-cavity mode that is confined to crystalline confocal resonator. \textbf{b} Schematic of acoustic resonator showing deformation of crystal associated with motion of fundamental longitudinal acoustic phonon-mode; the displacement, beam-waist, and spatial period are exaggerated for conceptual clarity.  \textbf{c} Surface plots showing the displacement amplitude profile for the first three spatial longitudinal acoustic modes.  The (\textbf{i}) fundamental and (\textbf{ii-iii}) first two higher spatial order longitudinal acoustic modes are shown. }
\label{3dfig}
\end{figure*}

\begin{figure*}[]
\centerline{
\includegraphics[width=16.0cm]{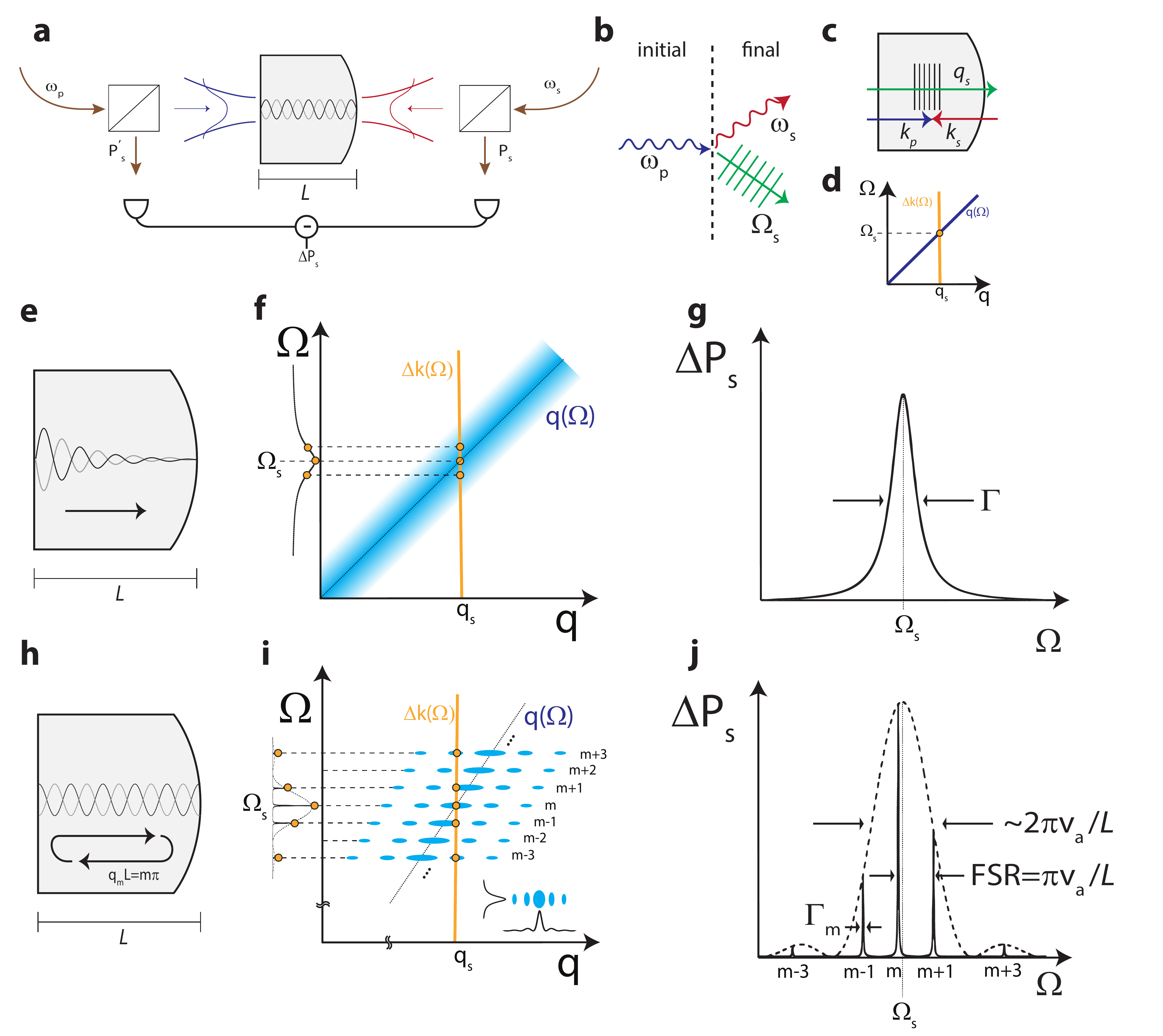}}
\caption{\textbf{Anatomy of Stokes signatures produced by bulk crystalline system:} \textbf{a} Simplified schematic showing the basic experimental configuration used to perform stimulated Stokes scattering measurements; counter-propagating pump- and probe-beams traverse the crystal while the change in the probe (Stokes) power ($\Delta P_s (\Omega)$) is monitored using balanced detection.   \textbf{b} Nonlinear scattering diagram identifying the initial and final states of the energy conserving three-wave Stokes scattering interaction. \textbf{c} Phase-matching diagram that describes the process summarized in \textbf{b}. \textbf{d} Diagram showing the phonon dispersion relation, $q(\Omega)$ (blue),  plotted atop the optical wavevector mismatch, $\Delta k(\Omega)$ (orange),  in an infinite lossless system; energy conservation and phase-matching occurs at the point of intersection. Panels \textbf{e}, \textbf{f}, and \textbf{g} describe the behavior of this system in the Brillouin-limit ($l_{coh}\ll L$). \textbf{e} Sketch comparing the phonon coherence (or attenuation) length to the cavity length ($L$). \textbf{f} 2D intensity map (blue) showing the phononic response as a function of wavevector ($q$) and frequency ($\Omega$) to an applied forcing function. The optical wavevector mismatch, $\Delta k(\Omega)$ (orange), is plotted on top of the intensity map and the points of intersection indicate the frequencies over which coupling occurs, which  yields a Lorenzian frequency response (projected left). Since each emitted phonon coincides with a stimulated Stokes photon, $\Delta P_s (\Omega)$ exhibits the same Lorenzian response, as displayed in panel \textbf{g}.   Panels \textbf{h}, \textbf{i}, and \textbf{j} describe the behavior of this system in the coherent-phonon limit ($l_{coh}\gg L$). \textbf{h} Sketch of discrete acoustic cavity-modes in the coherent-phonon limit.  The acoustic modes satisfy the resonance condition, $q_mL = m\pi$, which gives a frequency spacing or free-spectral range (FSR) of FSR$=\pi v_a/L$. \textbf{i} 2D intensity map (blue) showing the phononic response of acoustic cavity modes (indicated with index $m$) as a function of wavevector ($q$) and frequency ($\Omega$) of an applied forcing function.  Each acoustic mode response (inset) is broadened in frequency because of dissipation and exhibits a sinc$^2$ broadening in wavevector because of the finite spatial extent of the crystal.  The optical wavevector mismatch, $\Delta k(\Omega)$ (orange), is plotted on top of the intensity map and the points of intersection indicate the frequencies over which coupling occurs which yields the discretely sampled sinc$^2$ response (projected left). Since each emitted phonon coincides with a stimulated Stokes photon, $\Delta P_s (\Omega)$ exhibits the same sinc$^2$ modulated multi-peaked response, as displayed in panel \textbf{j}.}
\label{Teach}
\end{figure*}

In this paper, we demonstrate a new form of optomechanical coupling that grants access to ultra long-lived bulk-acoustic modes at cryogenic temperatures.
Our system consists of a pristine crystalline solid whose surfaces are shaped to create a stable phonon resonator geometry that traps ultra-high $Q$-factor ($4.2{\times}10^7$) phonon modes.
By engineering the optomechanical coupling between these high $Q$-factor phonon modes and an incident laser-field, we create Brillouin-like optomechanical interactions which are greatly enhanced compared to traditional Brillouin interactions.
Using only the intrinsic photoelastic material response, we engineer efficient optomechanical coupling to massive ($60\ \upmu$g) high frequency (${>}12$ GHz) bulk acoustic phonon modes.
%Since this system produces effective nonlinear optical susceptibility is orders of magnitude larger than traditional stimulated Brillouin spectroscopy methods.
This bulk crystalline system supports high optical powers (${>}$kW) and can be designed with practically any transparent crystalline medium.   These unique properties open the door to new forms of cryogenic phonon spectroscopy, optomechanical parametric oscillators, and precision metrology techniques.

In prior work, Ohno \emph{et al.} demonstrated that extended phonon coherence alters the character of the Brillouin interactions within crystalline substrates at cryogenic temperatures \cite{Ohno2006,Ohno2015}.  In this paper, we bridge the gap between cryogenic Brillouin physics and mesoscopic optomechanical interactions by engineering bulk phonons and their interaction with light. Using new optomechanical design principles, simulation techniques, and precision spectroscopy methods, we create ultra high $Q$-factor macroscopic (cm-scale) phonon modes that can be addressed individually using a laser field, expanding the range of possible optomechanical interactions in bulk crystals. We also show that, in contrast to cavity optomechanics, this system exhibits a new form of symmetry breaking that enables phonon heating or cooling \textit{without} an optical cavity (or cavity-less optomechanics).

%\begin{figure*}[tbh!]
%\centerline{
%\includegraphics[width=18.0cm]{Fig2.eps}}
%\caption{\textbf{Bulk crystalline optomechanics:} Optical pump and probe beams interact with a phonon beam confined in a crystalline confocal resonator.}
%\label{Teach}
%\end{figure*}

%\newpage
\section{Optomechanical coupling in crystals}

\subsection{Physical System and Approach}
In what follows, we examine the physics of light-sound coupling in a bulk crystalline system (Fig. \ref{3dfig}a).
The surfaces of this crystalline solid are fashioned into a plano-convex geometry that produces stable Hermite-Gaussian-\textit{like} acoustic phonon modes (Fig. \ref{3dfig}b-c) with ultra-high  $Q$-factors at cryogenic temperatures.
These phononic cavity-modes are designed to permit strong optomechanical coupling with an incident laser beam.

At room temperatures, intrinsic forms of phonon dissipation limit the phonon mean-free path $(l_{ph})$ to ${\sim}100$ microns \cite{Braginsky1985,Cleland2003}.
Since this mean free path ($l_{ph}$) is much smaller than the crystal length ($L = 5\,$mm), the phonons decay before a round trip oscillation can be completed.
However, at cryogenic temperatures, intrinsic phonon dissipation is dramatically reduced \cite{Braginsky1985,Cleland2003,Thaxter1966,Blair1966}, permitting the phonons to traverse multiple cavity round-trips, forming ultra-high $Q$-factor phonon modes \cite{Goryachev2012,Galliou2013}.
In the limit as the phonon coherence is much greater than the system length ($l_{ph}\gg L$), the dynamics of photon-phonon coupling becomes quite different from what has historically been termed Brillouin interactions. To differentiate these two regimes of interaction we term them the Brillouin-limit  ($l_{ph}\ll L$) and the coherent-phonon limit  ($l_{ph}\gg L$) respectively.
In the coherent-phonon limit we will show that the nature of this coupling is altered, giving rise to greatly enhanced nonlinear susceptibilities as well as a form of dispersive symmetry breaking that enable new approaches for phonon self-oscillation or phonon cooling within this bulk crystalline system.

%  \textit{Note that while the phonons are confined in the resonator, the light passes through the crystal; there is no optical cavity. In contrast to cavity optomechanical systems, light makes a single pass }

%\begin{figure*}[htb]
%\centerline{
%\includegraphics[width=16.0cm]{modecompare.eps}}
%\caption{\textbf{Shaped resonator:} \textbf{a} Theory.  \textbf{b}  Experiment for misaligned z-cut quartz.  The transverse acoustic modes are identified by red for L00, green for L10, and blue for L20.  The free-spectral range and mode spacing is indicated.  Peak normalized 2D plots of the three modes are displayed.  The red dashed line indicates a theoretical plot of the geometrically imposed phase-matching uncertainty bandwidth.}
%\label{mdcompare}
%\end{figure*}

We experimentally explore the strength and character of the optomechanical coupling using continuous-wave pump-probe spectroscopy.  We quantify the phonon-mediated energy transfer between counter-propagating pump-wave, with frequency $\omega_p$, and probe-wave, with frequency $\omega_s$, using the basic experimental configuration illustrated in Fig \ref{Teach}a.
Both the pump- and probe-beams pass once through the phononic cavity; the light-field is not resonantly enhanced within the crystal.
In this experiment, stimulated emission of a phonon is observed as energy transfer between the two incident optical beams via the three-wave interaction of Fig. \ref{Teach}b-c.
Optomechanical coupling  is mediated by photoelastic interactions within the bulk of the crystal; however, energy transfer only occurs when the phase matching and energy conservation conditions are met, as described in the following section.

%%% This should be evident from our discussion below... too redundant%%%%
%The functional form of this energy transfer is derived from the effective nonlinear susceptibility of the system, which is distinct in the two regimes of operation.  Pump-probe measurements are used to quantify of the optomechanical ($g_0$) coupling or Brillouin gain ($G_B)$ of the system from the susceptibility, depending on which description is appropriate. \textit{In the coherent phonon limit, the susceptibility is related to the coupling rate ($g_o$) between the light field a macroscopic phonon mode, whereas in the Brillouin limit, the cyrstal can be subdivided in to volume elements that each have an identical brillouin susceptibility characterized by a local material parameter $g_B$.}

\subsection{Optomechanical Coupling}
\label{Ocsec}
In general, acoustic phonons can scatter an incident pump photon ($\omega_p$) through either a Stokes process (creating a phonon) or an anti-Stokes process (annihilating a phonon). Of these interactions, only the Stokes process produces stimulated energy transfer between counter-propagating waves as the basis for the driven pump-probe studies employed here.
Through a Stokes process, a right-moving pump photon, with frequency and wavevector ($\omega_p$, $k_p$), scatters to a red-shifted left-moving Stokes photon ($\omega_s$, $-k_s$) and a right-moving Stokes phonon ($\Omega_s$, $q_s$).
Here, the optical wavevector ($k_j$) and frequency ($\omega_j$) are related by the optical dispersion relation of the solid, $k(\omega)$, as $k_j=k(\omega_j)$; the phonon wavevector ($q_s$) and frequency ($\Omega_s$) are related by the acoustic dispersion relation, $q(\Omega_s)$, as $q_s=q(\Omega_s)$.
For this scattering process to occur, both energy conservation ($\Omega_s = \omega_p - \omega_s$) and phase-matching ($q_s = k_p+k_s$) must be satisfied (Fig. \ref{Teach}b-c).
Combined, these conditions require that the Stokes phonon have a frequency and wavevector component ($\Omega_s$, $q_s$) that satisfy the equation, $q(\Omega_s)= \Delta k(\Omega_s)$, where $\Delta k(\Omega)=k(\omega_p)+k(\omega_p-\Omega)$ is the optical wavevector mismatch.
We use this requirement as the basis for a succinct diagrammatic representation that includes both phase-matching and energy conservation.
%Energy conservation and phase matching are satisfied the phonon dispersion relation, $q(\Omega)$, (blue) intersects  the optical phase-mismatch, $\Delta k_s(\Omega_s)$, (orange) as seen in Fig. \ref{Teach}d.

For example, within an infinite (lossless) system, the optical waves and phonon-modes can be viewed as plane-waves with perfectly well-defined frequencies and wavevectors;
Fig. \ref{Teach}d shows the phonon dispersion relation, $q(\Omega)$ (blue), plotted atop the optical wavevector mismatch, $\Delta k(\Omega)$ (orange). Energy conservation and phase-matching are satisfied by the set of states where these curves intersect.
In this idealized case, $q(\Omega_s)= \Delta k(\Omega_s)$ is satisfied at a single point, $(\Omega_s,q_s)$, meaning that Brillouin coupling occurs at only a single frequency. However, in practice, we will see that the system geometry and phonon dissipation shape the coupling bandwidth.

In bulk media, the Brillouin frequency and coupling strength are determined by the material properties.
Making the approximation $\Delta k(\Omega_s)\cong 2k(\omega_p)$, the Brillouin frequency is estimated from our phase-matching conditions as $\Omega_s \cong 2\, \omega_p (v_a /v_o)$,  where $v_o$ ($v_a$) is the phase velocity of light (sound) within the crystal.
In typical materials (and with near-infrared light), the Brillouin frequency ($\Omega_s/2\pi$) is between 10 and 50 GHz, and the photoelastic tensor predominantly couples to longitudinal acoustic phonons.
Through stimulated Stokes scattering, phonon emission can be viewed as the result of time modulated optical forces that are produced by electrostriction (or photoelastic response) within the crystal.
Through phase-matched coupling, pump- and probe optical fields interfere to a produce a time-modulated force distribution (of beat-frequency $\Omega$) with a wavevector $\Delta k(\Omega)$. When phase-matching is satisfied, these forces resonantly drive phonons with wavevector $q(\Omega_s)$.

Next, we consider optomechanical coupling within the finite crystalline system of Fig. \ref{Teach}e. At room temperatures high-frequency ($ 10$ GHz) Brillouin-active elastic waves decay rapidly (${\sim}1000$ cycles), meaning that $l_{ph} \ll L$ and the system is in the Brillouin regime of interaction (Fig. \ref{Teach}e).
Rapid phonon decay ($l_{ph}\sim100$ microns) permits us to neglect propagation of the envelope of the phonon field in the theoretical description of the gain dynamics; the dynamics of photon-phonon coupling is faithfully represented using the local nonlinear optical susceptibility following conventional Brillouin treatments  \cite{boyd}. In other words, each volume element within the crystal can be viewed as an independent system, which supplies distributed Brillouin gain as the pump and probe waves interact within the crystal.

%\newpage
In the Brillouin-limit, because elastic waves are heavily damped in comparison to the system size ($l_{ph}\ll L$), the density of phonon modes (or density of states) can be viewed as continuous.
Damping also effectively broadens the phonon dispersion curve.
This is because the damped system has complex eigen-values that correspond to exponential temporal (spatial) decay.
A Fourier transform reveals that these exponentially damped elastic-wave solutions take on a Lorentzian frequency bandwidth given by the phonon decay rate, $\Gamma$.  In other words, the linear phononic dispersion curve of Fig. \ref{Teach}d effectively broadens as seen in Fig. \ref{Teach}f.
This 2D intensity map now represents the magnitude of the phonon response as a function of the wavevector and frequency of an applied forcing function (i.e. dissipation permits coupling to phonons with a finite phase-mismatch).
In this picture, a \textit{band} of frequencies over which coupling occurs is identified by the intersection of the loss-broadened dispersion curve (blue intensity map) with the optical wavevector mismatch (orange), as seen in  Fig. \ref{Teach}f.
Projecting these points of intersection to frequency axis of Fig. \ref{Teach}f, we find a Lorentzian line-shape $\propto [(\Omega-\Omega_s)^2+(\Gamma/2)^2]^{-1}$; since phonons and Stokes photons are generated synchronously, the phase-matched optical response (Fig. \ref{Teach}f-g) is identical in form.  This well-known Lorentzian response is characteristic of Brillouin interactions \cite{boyd}. See Supplement Section 1.1 for more information.

At cryogenic temperatures, new behaviors emerge as we enter the coherent-phonon limit ($l_{ph} \gg L$).
%At cryogenic temperatures the phonon dissipation rate drops to the point where the phonon coherence far exceeds that of light.
In the absence of phonon dissipation, standing-wave phonon modes are formed between the front and back faces of the crystal (Fig. \ref{Teach}h) creating a discrete set of cavity modes  $\{\Omega_m\}$ with characteristic wavevectors $\{q_m\}$.
In this case, the continuous dispersion curve (Fig. \ref{Teach}f) becomes discretized by the formation of cavity modes, with mode spacing $\Delta \Omega/2\pi  \cong v_a (2L)^{-1} $, where $v_a$ is the acoustic velocity (see Fig. \ref{Teach}i).
As before, each cavity mode response will also have Lorentzian frequency broadening due to exponential temporal decay.
The $m^{th}$ cavity mode, with decay rate $\Gamma_m$, will have have a Lorentzian width ($\Gamma_m$) in the frequency domain.
However, in contrast with the Brillouin limit, because the spatial extent of the standing-wave is finite (i.e., abruptly terminates at the crystal faces), the phonon wavevector occupies a band of spatial frequencies and takes on a $\textrm{sinc}^2$ form. This can be seen from the spatial Fourier transform of a truncated sinusoid.
Therefore, each acoustic cavity mode has a frequency/wavevector response depicted by the lower right inset of Fig. \ref{Teach}i.
Following the same procedure as described above, the optical response of this acousto-optical interaction is given by the intersection of the discretized phonon response curve (blue) with the optical wavevector mismatch (orange). The formation of cavity modes within this finite system then yields a response of the form $ \sim\sum_m [(\Omega-\Omega_m)^2+(\Gamma_m/2)^2]^{-1}\textrm{sinc}^2 [(q(\Omega_m) -\Delta k_s (\Omega))L/2]$.  Hence, the finite extent of the crystal shapes the response into a series discrete modes which are modulated by a $\textrm{sinc}^2$ distribution (Fig. \ref{Teach}j).

The coherent-phonon limit produces a new characteristic nonlinear susceptibility (e.g. compare Figs. \ref{Teach}g with \ref{Teach}j).  When phonons are trapped to produce high $Q$-factor phonon modes, we will show that in comparison to the Brillouin limit, the nonlinear optical coupling strength can be enhanced by several orders of magnitude, and the dynamics of photon-phonon coupling become nontrivial and depends on phonon propagation.  Moreover, this interaction opens the door to new regimes of dynamics including optomechanical self-oscillation and mode-cooling processes, akin to cavity optomechanical systems.  A key requirement for reaching these new regimes of dynamics is low-loss phonon modes that exhibit appreciable coupling to the light-field. Therefore, as detailed in the following section, the geometry of the crystal and the overlap between the phonon mode and the light-field become crucial considerations.

%Through our development above,  we have considered the simplified case of 1D wave propagation within a finite crystal and varying degrees of acoustic loss.
%In the Brillouin limit, $l_{ph} \ll L$ and in the coherent phonon limit we assumed that  $l_{ph} \gg L$.
%However, spatial evolution of our phonon modes within the experimental (3D) presents an important consideration that cannot be neglected.

\begin{figure*}[]
\centerline{
\includegraphics[width=15.0cm]{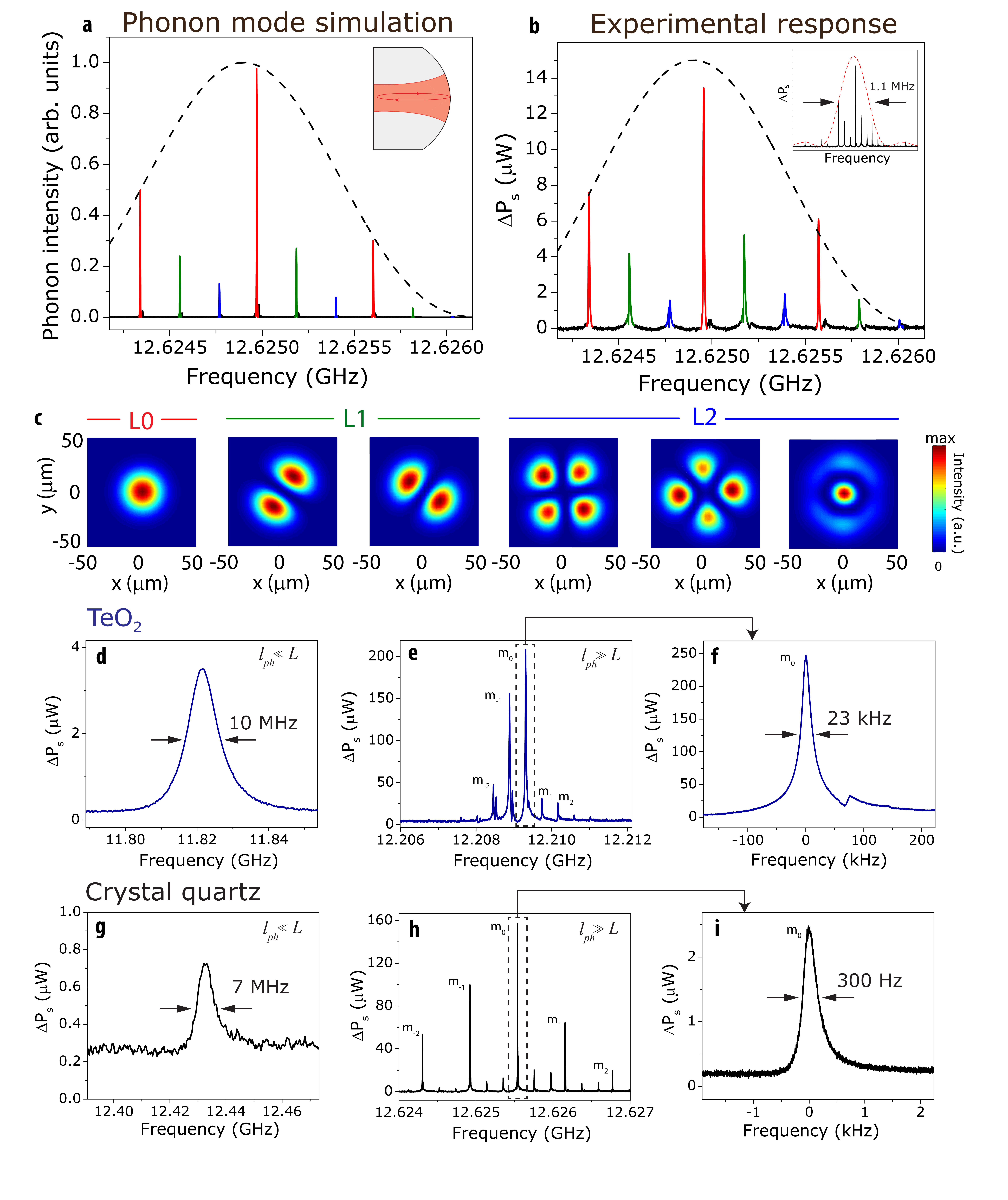}}
\caption{\textbf{Experimental results:} \textbf{a} The phonon mode spectrum of z-cut quartz is calculated with an acoustic beam propagation algorithm accounting for the anisotropic elastic properties of the crystal and the curvature of the cavity endfaces.  The acoustic spatial mode families are identified by color with red corresponding to the fundamental mode (L0), green to the near-degenerate 1st higher-order modes (L1) and blue to the near-degenerate 2nd higher-order modes (L2).  The simulation is seeded with a beam shifted 10 $\upmu$m horizontally and 15 $\upmu$m vertically to couple to higher-order spatial modes for comparison with experiment.  The simulated plano-convex cavity geometry of the acoustic resonator is displayed in the inset.  \textbf{b}  The experimental phonon mode spectrum for z-cut quartz as measured through a stimulated Stokes scattering measurement.  The optical beams are misaligned from the phonon cavity to enhance the response from the higher-order spatial modes.  The wider frequency span of the measurement is displayed in the inset illustrating the sinc$^2$ spectral response.  The dashed lines indicate the geometrically imposed phase-matching bandwidth. \textbf{c} 2D intensity plots of the spatial profiles of the acoustic modes from the simulation results represented in \textbf{a}, normalized by the peak intensity.  The modes are grouped in spatial mode families (L0-L2) by their common resonant frequencies.  Panels \textbf{d}, \textbf{e}, and \textbf{f} represent experimental stimulated Stokes scattering measurements from a z-cut TeO$_2$ crystal.  \textbf{d} At room temperature (${\sim}300$ K) TeO$_2$ exhibits a Lorentzian optical response with a 10-MHz linewidth.  \textbf{e} At cryogenic temperatures (${\sim}10$ K), in the coherent-phonon limit, the optical response exhibits discrete acoustic modes, indicated with index m, with narrow features.  The optical beams are aligned to the center of the crystal, which minimizes the response of the higher-order spatial modes. \textbf{f} The m$_0$ mode exhibits a linewidth of 23 kHz for TeO$_2$.  Panels \textbf{g}, \textbf{h}, and \textbf{i} represent experimental stimulated Stokes scattering measurements from a z-cut quartz crystal.  \textbf{g} At room temperature (${\sim}300$ K) quartz exhibits a Lorentzian optical response with a 7-MHz linewidth.  The signal-to-noise is reduced from the TeO$_2$ case because of the relatively lower photoelastic response of quartz.  \textbf{h} At cryogenic temperatures (${\sim10}$ K), in the coherent-phonon limit, the optical response of quartz exhibits discrete acoustic modes, with modes that are well-resolved owing to the lower acoustic dissipation in quartz. \textbf{i} The m$_0$ mode for quartz exhibits a narrow linewidth of 300 Hz.}
\label{exptex}
\end{figure*}

\section{Experimental Results}

Above,  we have considered the simplified case of 1D wave propagation in a finite crystal with varying degrees of acoustic loss.  In practice, spatial dispersion and the geometry of our 3D crystal strongly impact the lifetime and character of the resonantly excited phonon modes.
%For instance, a flat-flat crystal geometry produces limited enhancement of the susceptibility because it forms an unstable resonator.
For instance, enhancement of the susceptibility is severely limited within a flat-flat crystal because this planar geometry forms an unstable resonator.
In this case, phonon modes excited by a finite laser beam suffer appreciable diffractive losses, limiting the phonon $Q$-factor and nonlinear enhancement at cryogenic temperatures (Supplement Section 5).

%For instance, a flat-flat crystal geometry can only support limited enhancement of the susceptibility; this system is known to be an unstable resonator because it does not support stationary modes that are finite in lateral extent. As a result, the phonon modes excited by a finite laser beam tend to be a collection leaky modes that suffer appreciable diffractive losses, limiting the nonlinear coupling strength. (See supplement Section XX.)
To greatly enhance this coupling, we shape the surfaces of the crystal to form ultra-high $Q$-factor phonon modes that are tightly confined, permitting efficient coupling to a laser beam with a small (${\sim}30\ \upmu$m) spot size at ${\sim}10$ GHz frequencies.
A plano-convex resonator geometry is known to support stable Hermite-Gaussian modes for optical waves propagating in an isotropic medium \cite{siegmanbook}.
%Using Gaussian-beam optics, one can readily design a set of ultra-long lived Hermite-Gaussian resonator modes.
However, for the complex elastic wave dynamics in anisotropic crystals, it is not clear that a comparable stable cavity exists.  Therefore, we develop an acoustic beam propagation algorithm which accounts for the anisotropic elastic properties of the crystal and the curvature of the cavity end-faces.  We demonstrate stable acoustic resonators in a plano-convex geometry, and by adapting optical design principles to the case of anisotropic elastic wave propagation, we create stable resonator geometries (Fig. \ref{3dfig}) that support  Hermite-Gaussian-\textit{like} modes within z-cut quartz and TeO$_2$ crystals.
Simulated spatial profiles for the fundamental (L0), and higher order transverse (L1, L2) modes are shown for quartz in Fig. \ref{exptex}c.

These phonon modes have many unusual properties.  Absent device imperfections and material dissipation, acoustic mode simulations reveal that the $Q$-factor of the fundamental (L0) mode exceeds $10^9$ for each crystal type. These tightly confined phonon modes have negligible anchoring losses because the transverse mode size is more than a hundred times smaller than the crystal diameter.  The modal mass \cite{Pinard1999}, given by the mode size and the mass density is 60 $\upmu$g (Supplement Section 2.4).  This large modal mass represents a unique and valuable aspect of this system compared to less massive cavity optimechanical systems for its value for studies of quantum decoherence (see further information in the discussion). For device parameters and simulation methods, see Supplement Section 5.

To explore optomechanical coupling in both the Brillouin and coherent-phonon limits of operation, we perform phonon-mediated energy transfer measurements in both crystal species. Stimulated energy transfer between counter-propagating pump and probe beams is measured with the basic experimental configuration seen in Fig. \ref{Teach}a both at room temperature and at cryogenic temperatures (see Supplement Sections 3-4, 6).
%Through these measurements, the pump beam is modulated at a frequency of 20 MHz, and the nonlinearly induced modulation of the probe beam is detected using a lock-in amplifier and a balanced detector (20-MHz frequency).
%Additionally, a feedback loop is used to eliminate phase instabilities produced by vibrations of the crystal and the cryostat.
%This technique (described in the SI) permits the measurement of phonon modes with sub-Hz linewidth.
As discussed previously, stimulated Stokes energy transfer between pump ($\omega_p)$ and probe ($\omega_s$) beams provides a direct measurement of the effective nonlinear susceptibility of the system.

The stimulated Stokes energy transfer spectrum is obtained by fixing the pump frequency ($\omega_p$) while the pump-probe frequency detuning ($\Omega = \omega_p-\omega_s$) is swept through the Brillouin resonance ($\Omega_s$). Stimulated Stokes scattering is observed as a fractional increase in the probe power ($\Delta P_s/P_s$) while the probe frequency is tuned.  Since each Stokes-scattering event corresponds to emission of a phonon, the Stokes energy transfer ($\Delta P_s$) also provides a direct measure of the phonon emission rate.
Estimation of the Brillouin frequency from phase-matching considerations ($\Omega_s \cong 2\, \omega_p (v_a /v_o)$) leads us to expect resonant coupling to longitudinal phonons at $\sim$12 GHz frequencies within both quartz and TeO$_2$ when using a pump wavelength of $1549$ nm.

We begin by examining the stimulated Stokes spectrum at room temperature (i.e., in the Brillouin-limit). The Stokes energy transfer measurements seen in Fig. \ref{exptex}g(d) reveal broad Lorentzian line-shapes at 12.43 GHz (11.82 GHz), corresponding to a peak fractional change in the probe intensity of 13 ppm (80 ppm) for z-cut quartz (TeO$_2$); these results agree well with the predicted resonance frequencies. Note that the fractional change in probe power (or gain) produced by TeO$_2$ is larger than that of quartz because TeO$_2$ possesses a larger photo-elastic response. The measured $7$-MHz ($10$-MHz) line-widths for quartz (TeO$_2$) agree with known dissipation rates and correspond to a spatial decay length of ${\sim}140$ (${\sim}70$) microns \cite{Sonehara2007}.

At cryogenic temperatures (5-15 Kelvin) the system enters the coherent-phonon limit ($l_{ph}\gg L$). In this regime, photon-phonon coupling is mediated by a series of standing-wave cavity modes. Fig. \ref{exptex}b shows the stimulated Stokes spectrum generated by quartz at 9 Kelvin temperatures.  The wide spectral scan (inset) reveals a series of regularly spaced resonances modulated by a sinc$^2$ profile. The calculated phase-matching bandwidth (dashed line) is shown atop the data, demonstrating good agreement of the predicted geometric bandwidth with the observed spectral trend. Note that the central frequency of the quartz (TeO$_2$) spectrum shifts to 12.63 GHz (12.21 GHz) due to changes in elastic and optical properties at cryogenic temperatures (e.g. see Ref. \cite{Ohmachi1970a,McSkimin1965}).

Inspection of Fig. \ref{exptex}b reveals three families of resonances (color-coded) that repeat at the acoustic free spectral range ($v_a/(2L) \cong 630$ kHz). Elastic mode simulations of Fig. \ref{exptex}a,c show that this mode structure is consistent with coupling to the fundamental (L0) and higher order (L1, L2) modes supported by the plano-convex geometry. The experimental results of Fig. \ref{exptex}b show the Stokes spectrum obtained when the laser-field is intentionally misaligned from the cavity-axis; this enhances coupling with higher-order spatial modes, increasing their visibility.  The simulated phonon spectrum (Fig. \ref{exptex}a) reproduces the experimental spectrum by seeding the cavity with an acoustic beam with a lateral center offset of $10\ \upmu$m horizontally and $15\ \upmu$m vertically.  As anticipated, improved axial alignment of the laser fields with the plano-convex geometry enhances coupling to the fundamental (L0) mode and suppresses higher order modes (L1,L2), as seen in Fig. \ref{exptex}h.  A similar comparison of  mode structure and spectrum of TeO$_2$ shows good agreement with elastic wave simulations. For further details see Supplement Section 5.

High resolution spectral analysis of individual L0 cavity modes (Fig. \ref{exptex}f,i) reveals a spectral width, $\Gamma_m/2\pi$, of 300 Hz (23 kHz) in the quartz (TeO$_2$) system, corresponding to phonon $Q$-factors of 42 million (0.5 million).
Since the peak susceptibility scales inversely with dissipation rate (or proportional to $Q$-factor), these high-$Q$ cavity modes enhance the peak stimulated Stokes scattering rates by orders of magnitude.
Within quartz (TeO$_2$), Stokes scattering rates are increased by factors of more than $ 10^4$ ($10^2$) over their room temperature values.
In quartz, these ultra-long lived phonon modes are consistent with a coherence lengths ($l_{coh}$) of ${\sim}3$ meters.
%This phonon decay rate is consistent with simulated scattering loss rates obtained by feeding high resolution atomic force microscopy surface profiles into simulations, suggesting that roughness-induced scattering at surfaces is a dominant source of loss in quartz. We speculate that material quality limits the damping rate in  TeO$_2$, as high defect densities were identified through materials characterization.
%Within quartz, these ultralong-lived phonon modes enhance the stimulated Stokes scattering rates by more than 4 orders of magnitude from the room temperature rate.
Moreover, the demonstrated quality-factors in quartz are among the highest measured for phonon frequencies above $10$ GHz \cite{Blair1966}, corresponding to a $(\textrm{frequency})\times (Q\textrm{-factor})$ product of $4.2{\times}10^{17}$ Hz, which is comparable with world-class electromechanical devices demonstrated in Refs. \cite{Goryachev2012,Galliou2013}.

This drastically modified susceptibility (of Fig. \ref{exptex}) highlights the fact that the nature and dynamics of coupling in this bulk optomechanical system are radically altered from that of conventional Brillouin scattering. We see that the laser-field can now be used to selectively couple to an array of macroscopic phonon modes. As we explore the physics of this system further, we will show that this bulk crystalline system admits new regimes of dynamics--including phonon self-oscillation and mode cooling--which ordinarily require an optical cavity in the context of cavity optomechanics.

\section{Bulk Crystal As Optomechanical System}

 %\textit{(However, we will see that important distinctions arise from the continuous and extended nature of the system.)}
To analyze this new system and to leverage the extensive body of work in the context of cavity optomechanics, it is useful to develop a quantum description of optomechanical coupling.  Since the laser field couples to a macroscopic phonon mode, we can define a coupling rate between the optical fields and the quantized phonon-modes, as is typical in cavity optomechanics \cite{Aspelmeyer2014,Meystre2013,Marquardt2009,Favero2009a}.
%However, because there is no optical cavity, coupling occurs between a single phonon in the phonon mode to a single average photon present in the time of flight through the crystal (i.e. the photon flux times the optical travel time is one for a single average photon).    Using this new coupling rate ($g_0$) and the dissipation rate ($\Gamma$), we can examine different regimes of interaction within the framework of cavity optomechanics.

It is instructive to draw an analogy between our bulk crystalline interaction and a  multimode cavity optomechanical interaction, wherein a single phonon mode ($b$) mediates coupling between two distinct optical modes ($a_p$) and ($a_s$). This process is captured by an interaction Hamiltonian of the form $\hbar (g_{12} a_p a_s^\dagger b^\dagger+ g^*_{12} a_p^\dagger a_s b)$ \cite{Aspelmeyer2014,Grudinin2010,Braginsky2001, Law1995}, where $b$, $a_p$, and $a_s$ are phonon, pump-photon, and Stokes-photon annihilation-operators, respectively. This interaction has been used to describe optomechanical coupling in the case when a phonon mode is frequency-matched to the free spectral range (FSR) of  an optical Fabry-P$\acute{\text{e}}$rot resonator; in this case, the phonon mediates coupling between two different longitudinal optical modes.

Coupling between counter-propagating laser beams in our bulk crystalline system (Fig. \ref{3dfig}a) is reminiscent of this multimode cavity optomechanical interaction because a single phonon couples two distinct optical modes--each with different wavevectors. However, since light travels directly (or ballistically) through our bulk crystalline device, no optical cavity modes are formed.  As a consequence, the $m^{th}$ phonon mode admits coupling to a \textit{continuum} of optical waves. %As a result, the coupling rate that we develop here has distinct definition.

The Hamiltonian of the phonon field,  $H^{\text{ph}} = \sum_m \hbar \Omega_m b_m^{\dagger} b_m$, is a sum over the set of phonon-modes, whereas the Hamiltonian in the optical field is given by an integral over all possible wavevectors for both the Stokes- and the pump-waves.  Assuming that the traveling optical waves are sharply peaked about their carrier wavevectors, the Hamiltonian for the optical fields is $H^{\text{opt}} \!\!=\!\!  \int dk \hbar \omega_p(k) a_{p,k}^{\dagger} a_{p,k} + \int dk' \hbar \omega_s(k')a_{s,k'}^{\dagger}a_{s,k'} $.  In this framework, the optomechanical coupling to the $m^{th}$ phonon-mode is given by the interaction Hamiltonian
\begin{equation}
%\begin{aligned}
H_m^{\text{int}} \!=\!\frac{\hbar}{2\pi}\!\int_k\int_{k'} \!\!\left[{g}_{m}(k,k') {a}_{p,k}^{\dagger}{a}_{s,k'} b_m +\textrm{H.c.}\right]dk\, dk',
%G^{\parallel}_i=\frac{\pi Q \xi_i n_g^2(n^2-1)^2}{\lambda X_{1i}^2 v^2 n^4 c \rho R^2}\quad \text{and} \quad G^{\perp}_i=0,
%\end{aligned}
\label{Hint}
\end{equation}

\noindent where, ${g}_{m}(k,k') =g^{m}_0L\textrm{sinc} [(k'-k+q_m)L/2]e^{i\phi} $ is termed the geometric coupling rate, $\phi=((k^{\prime}-k+q_m) L/2)+\pi/2$, and $g^{m}_o$ is a bare coupling rate produced by the zero-point motion of the $m^{th}$ phonon mode. The total interaction, $H^{int}=\sum_mH^{int}_m $,  includes contributions from all of the optomechanically-coupled phonon modes supported by the crystalline resonator. Note that this formulation (Supplement Section 2-2.4) assumes that the optical beam profiles are uniform, and the propagating optical fields do not grow rapidly in space (For a more general treatment, not subject to these constraints, see Supplement Section 2.6). This newly defined coupling rate, $g^{m}_0$, can be related directly to the multimode cavity optomechanical coupling rate, $g_{12}$, under certain limits in which a the cavity mode is equivalent to the traveling wave as detailed in the Section 2.5 of the Supplement.

Notice that phase-matching is implicit in our geometric coupling rate,  ${g}_{m}(k,k')$.
In the case when the incident fields are monochromatic, and the coupling is weak, ${a}_{p,k}$ and ${a}_{s,k'}$ can be replaced with sharply peaked distributions consistent with the frequency and wavevector of each laser-field; in this limit one can readily show that the nonlinear optical susceptibility takes on the form $ \sum_m |g^m_0|^2\,[(\Omega-\Omega_m)^2+(\Gamma_m/2)^2]^{-1}\textrm{sinc}^2 [(q(\Omega_m) -\Delta k_s (\Omega))L/2]$ (see Supplement Section 2.2).  This form of the susceptibility is identical to that obtained through phase-matching considerations (Section \ref{Ocsec}), demonstrating the equivalence between the two pictures.  See Supplement Section 2 for more information.

Using this form of the susceptibility to analyze our stimulated Stokes spectra for both quartz and TeO$_2$ systems, we find zero-point coupling rates ($g^m_0/2\pi$) of $31$ Hz and $82$ Hz, respectively. These values agree with theoretical values of $36$ Hz and $66$ Hz obtained using known material properties and device geometry (within experimental error).  See Supplement Section 6.

The Hamiltonian framework can now be used to derive and interpret the optomechanical cooperativity, an important metric for optomechanical interactions  (Supplement Section 2.3).
The cooperativity, which compares the photon-phonon coupling rate to the photon and phonon dissipation rates, is defined as $\mathcal{C}^{\text{om}} = 4 n_p|g_{12}|^2/(\Gamma_m \kappa)$ in the context of cavity-optomechanics \cite{Aspelmeyer2014,Meystre2013,Marquardt2009,Safavi-Naeini2013a}; here $n_p$ is the number of pump photons in the optical cavity and $\kappa$ is decay rate for the photons.
However, since our system does not have cavity modes, it is more natural to define a free-space cooperativity ($\mathcal{C}^{\text{fs}}$) in terms of the incident pump-power ($P_p$) as $\mathcal{C}^{\text{fs}}=P_p|g^m_0|^2 L^2 /(\Gamma_m \hbar \omega_p v_{gs} v_{gp})$. Here, $v_{gs}$ ($v_{gp}$) is the group velocity of the Stokes (pump) wave. The connection between $\mathcal{C}^{\text{om}}$ and $\mathcal{C}^{\text{fs}}$ are discussed in Supplement Section 2.5. When $\mathcal{C}^{\text{fs}}$ reaches unity, phonon self-oscillation and appreciable mode-cooling are possible.
For phonon self-oscillation, within our quartz crystal, a dissipation rate of 300 Hz translates to a threshold power of 7.2 Watts.  Interestingly, the combination of high power handling and ultra-high $Q$-factor phonon modes within this bulk crystalline system could lead to extraordinary spectral-narrowing as the basis for new oscillator technologies \cite{Vahala2008}.

\section{symmetry breaking}
\label{sbr}
\begin{figure*}[]
\centerline{
\includegraphics[width=15.0cm]{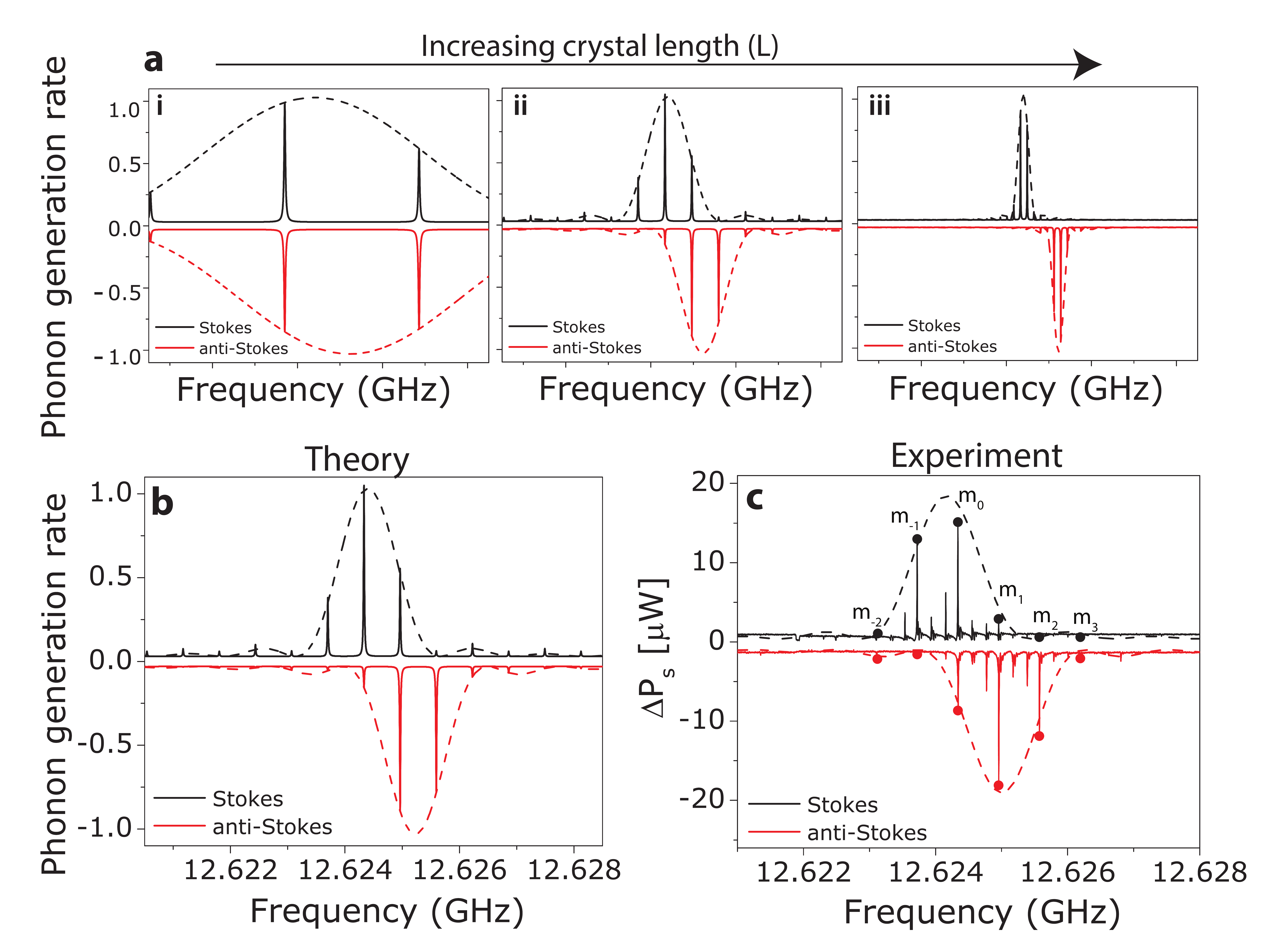}}
\caption{\textbf{Dispersive symmetry breaking in bulk crystalline optomechanics:} The predicted Stokes (black) and anti-Stokes (red) scattering rates are plotted in terms of the corresponding generation rate of the phonons.  The Stokes process corresponds to the generation of a phonon (positive generation rate) and the anti-Stokes process corresponds to the annihilation of a phonon (negative generation rate).  The two processes in the bulk crystalline system have distinct spectral dependencies.  The phase matching bandwidth is depicted with a dashed line.  \textbf{a} Graphic depiction of the length dependence of the relative shift in the frequency to the phase-matching bandwidth for lengths of \textbf{i} 1 mm, \textbf{ii} 5 mm, and \textbf{iii} 20 mm.  Longer crystal lengths enable generation independent of annihilation, or dispersive symmetry breaking.  \textbf{b} Predicted generation rates for both processes for the experimentally accessible 5-mm z-cut quartz crystal. \textbf{c} Experimental stimulated scattering experiments representing the Stokes and the anti-Stokes processes for 5-mm z-cut quartz.  As detailed in Supplement Section 7, measurements representing the anti-Stokes trace are taken with a stimulated Stokes measurement using an appropriately frequency-shifted pump beam.  The phonon mode numbers are indicated with index m.   Measurements reveal that phonon generation is 70 times more likely for $m_{-1}$ and phonon annihilation is 20 times more likely for $m_2$ mode, which illustrates the magnitude of dispersive symmetry breaking without an optical cavity. }
\label{stat}
\end{figure*}

Through optomechanical interactions, a phonon mode can typically mediate Stokes and anti-Stokes processes simultaneously; to engineer useful optomechanical operations, we typically seek to control which process will occur. Hence, as we contemplate the range of possible interactions, it is crucial to identify the relative probability of Stokes and anti-Stokes scattering by these ultra high $Q$-factor phonon modes.   For instance, in order to selectively heat or cool a phonon mode it is necessary to bias the system, giving preference to either Stokes and anti-Stokes processes.  %Stokes and anti-Stokes symmetry breaking is implemented in cavity optomechanics by using the optical cavity to enhance or suppress one process or the other.
In the context of cavity optomechanics, Stokes and anti-Stokes symmetry breaking is achieved by detuning the frequency of incident pump-wave from the optical cavity resonance. However, our bulk crystalline system does not possess an optical cavity, and we will see that a cavity is unnecessary to bias the system.

In stark contrast to cavity optomechanical systems, this bulk crystalline optomechanical system produces a form of \textit{dispersive} symmetry breaking, arising from distinct phase-matching requirements imposed by Stokes and anti-Stokes scattering.  In other words, this system exhibits a new form of symmetry breaking that enables phonon heating or cooling \textit{without} an optical cavity (which we refer to as cavity-less optomechanics).

%Phase matching becomes important because the  phonon mode interacts with light-field over an extended distance ($L \sim 20,000$ optical wavelengths).

Energy conservation ($\omega_{as} = \omega_p+\Omega_{as}$) and phase-matching ($k_{as} = k_p + q_{as}$) conditions for the anti-Stokes process combine to require $q(\Omega_{as}) = k(\omega_p) + k(\omega_p+\Omega_{as})$, whereas Stokes scattering requires $q(\Omega_s) = k(\omega_p) + k(\omega_p-\Omega_s)$.  The sign-difference in these two relations results in a difference between the peak-frequency of the Stokes ($\Omega_s$) and anti-Stokes ($\Omega_{as}$) processes. The Stokes scattering is a maximum for phonon-modes of frequency $\Omega_s = \Omega_0 (1+\frac{v_a}{v_o})^{-1}$ whereas anti-Stokes scattering is maximum for $\Omega_{as} = \Omega_0 (1-\frac{v_a}{v_o})^{-1}$, where $\Omega_0 \equiv 2\, \omega_p (v_a /v_o) $.  For typical parameters (e.g. 5-mm long z-cut quartz), this gives a dispersive frequency difference, $(\Omega_{as}-\Omega_s)/2\pi$ of ${\sim}1$ MHz, which becomes significant in the coherent-phonon limit.

%Since phase matching is implicit in our definition of the geometric coupling rate, ${g}_{m}(k,k')$, it naturally follows that the Stokes coupling rate, ${g}_{m}(k_p,k_s)$, is maximum for phonons of frequency $\Omega_s$ whereas the anti-Stokes rate ${g}_{m}(k_p,k_{as})$ is maximum for phonons of frequency $\Omega_{as}$.

Since phase matching is implicit in our definition of the geometric coupling rate, ${g}_{m}(k,k')$, the effect of the relative Stokes/anti-Stokes frequency shift can be seen by comparing the Stokes (${g}_{m}(k_p,k_s)$) and anti-Stokes (${g}_{m}(k_p,k_{as})$) coupling rates.
Fig. \ref{stat}a(i-iii) compare the Stokes/anti-Stokes coupling rates as a function of phonon frequency (mode number) for crystal lengths ($L$) of 1, 5, 20 mm.
Since the coupling rate for these processes are peaked at distinct frequencies, entire families of modes can be addressed, each with differing degrees of heating and cooling.
In the limiting case of a large $L$, we see that the Stokes and anti-Stokes process are mediated by distinct sets of phonon-modes (Fig. \ref{stat}a(iii)).

Since dispersive symmetry breaking results from the frequency dependence of our geometric coupling, we can quantify the frequency dependence of ${g}_{m}(k_p,k_s)$ by varying the \textit{pump} frequency as we perform stimulated Stokes scattering measurements. By tuning the pump frequency from $\omega_p$ to $\omega'_p \rightarrow \omega_p +\Omega_{as}$, we can quantify the effective geometric couping rate associated with the anti-Stokes processes.
The measured geometric couplings associated with the Stokes and anti-Stokes processes are shown in Fig. \ref{stat}c for z-cut quartz. Comparison of these two measurements reveals a dispersive frequency shift of 830 kHz, which agrees well with calculations.

In conjunction with theory, these measurements reveal that dispersive symmetry breaking produces a very strong differential gain, as the basis for selective mode heating/cooling in this system.  For instance, the data of Fig. \ref{stat}c reveal that phonon mode $m_{-1}$ is $ 70$ times more likely to experience Stokes scattering, whereas mode $m_2$ is $20$ times more likely to experience anti-Stokes scattering. Hence, these measurements illustrate  how dispersive symmetry breaking permits mode selective heating or cooling within this bulk crystalline system. See Supplement Section 7 for more information.

\section{Discussion and Conclusions}

In the preceding section, we have shown that mode-cooling and mode-heating can be readily performed using this cavity-less optomechanical system.  As we examine the range of possible uses for this system, it is also intriguing to note that this crystalline quartz system can support immense optical powers (kW to MW) without adverse effects \cite{Chiao1964a}. Since the dispersive symmetry breaking (Section \ref{sbr}) is compatible with the use of relatively short pulses ($<1$ns), we can use pulses to reach very large cooperativities ($10^3-10^6$) within this cavity-less system.  The idea of using pulses to perform quantum control has been recognized as a fertile direction in numerous prior works \cite{Vanner2011,Vanner2013,Meenehan2015,Bennett2016}, and phase-matched optomechanical coupling within this bulk crystalline system may offer a different way of approaching these ideas. Building on these concepts, one can now consider using this system to perform quantum control or entanglement operations using these ultra-high $Q$-factor phonon modes.

Pulses are not necessary to reach new regimes of nonlinear dynamics, however. Even under continuous-wave (CW) operation, mode cooling and phonon self-oscillation are within reach if cooperativities are increased from their current values ($\mathcal{C}^{\text{fs}}\approx 0.03$) to $\mathcal{C}^{\text{fs}}>1$. The cooperativity, $\mathcal{C}^{\text{fs}}=P_p|g^m_0|^2 L^2 /(\Gamma_m \hbar \omega_p v_{gs} v_{gp})$, is increased by reducing the phonon dissipation rate ($\Gamma_m$) while increasing the coupling rate ($g^m_0$) and the incident pump-power ($P_p$). With fixed material properties (e.g. with quartz) cooperativities of unity can be attained by improving the acousto-optic overlap and boosting the pump power to 2 Watts. It is also interesting to note that greatly reduced phonon dissipation rates ($\Gamma_m$) may be possible with further device refinement.  For example, with improved surface quality and crystal purity, significant reductions in phonon dissipation rates are possible. Since the roughness-induced diffraction losses scale with the root-mean-squared roughness to the 2$^{\text{nd}}$ power, a modest (2-fold) reduction in the surface roughness could translate to a 4-fold reduction in dissipation rate (assuming that surface roughness plays a dominate role in dissipation), potentially reducing the phonon self-oscillation optical threshold ($\mathcal{C}^{\text{fs}} = 1$) to $\sim 0.5$ Watts.

Cooperativities can also be drastically enhanced by constructing our optomechanical system using media with a larger photo-elastic response. For a fixed optical wavelength and mode-matched optical and acoustic waves, the cooperativity scales as $\mathcal{C}^{\text{fs}} \propto P_p n^{7}p_{13}^2/(\Gamma_m  \rho v_a)$. Hence, the coupling rate is enhanced by reducing the acoustic dissipation ($\Gamma_m$) or improving the material photo-elastic response.  For fixed crystal geometry ($L = 5$ mm), enhanced photo-elastic response produced by PbMoO4, Ge, and GaAs produce bare coupling rates ($g^m_0$) of $\sim$144, $\sim$266, and $\sim$292 Hz at center frequencies of 11, 26, and 25 GHz, respectively. Hence, comparable phonon dissipation rates ($\Gamma_m = 300 \times 2 \pi$ Hz) within these media translate to self-oscillation threshold ($\mathcal{C}^{\text{fs}} = 1)$ powers of $\sim$175, $\sim$40 and $\sim$33 mW, respectively.

This bulk crystalline optomechanical system also provides optical access to a large range of phonon-mode sizes and masses.  Phonon resonators can be formed in cavities as short as tens of microns to as long as tens of centimeters with corresponding phonon motional masses spanning from nanograms to grams. Counterintuitively, one finds that cooperativity ($\mathcal{C}^{\text{fs}} \propto P_p n^{7}p_{13}^2/(\Gamma_m  \rho v_a)$) is invariant with volume of the phonon mode (or modal mass).  This is of considerable interest because the combination of coherent quantum states and large masses are required to test fundamental theories of quantum decoherence and quantum gravity (see discussion in Ref. \cite{Aspelmeyer2014} for more information).

It is instructive to contemplate the construction of a cavity-optomechanical system which harnesses the presented bulk crystalline phonon modes by, for example, placing the bulk crystalline system of Fig. 1a within an optical cavity. A cavity-optomechanical system of this type could offer intriguing opportunities.  For instance, the extraordinarily high power handling of this crystalline system permits intracavity power enhancement to reach high cooperativies ($\mathcal{C}^{\text{om}}\gg 1$) with CW operation.  Assuming that the crystal is placed within an optical cavity whose length matches the crystal length ($L$) and maintains the modal overlap, the coupling rate of this cavity-system will be equivalent to the coupling rates ($g^{m}_o$) measured in this paper (See Supplement Section 2.5). In this configuration, the intracavity power will be resonantly enhanced by a factor of the optical finesse ($\mathcal{F}$), permitting extraordinarily high cooperativities ($\mathcal{C}^{\text{om}} =  1000$) within high finesse ($\mathcal{F} = 10^4$) cavities using moderate incident powers (100 mW).

Brillouin-based optomechanical coupling in an optical cavity have been demonstrated in microcavities \cite{Bahl2011, Rokhsari2005,Bahl2012,Tomes2011} at room temperatures as well as in an ultra-cold (15 mK) superfluid \cite{Kashkanova2016}.
The bulk crystalline system, owing to its simple configuration with minimal contribution from surfaces, allows for material-limited phononic dissipation at desirable high frequencies (${>}10$ GHz).  Moreover, since practically any transparent crystalline solid can be shaped into a low loss phononic resonator, a wide range of materials provides a flexible design space for specific applications.  Each material offers a different phonon frequency, wavelength transparency, and may permit coupling to a range of additional excitations.  For example, by utilizing a piezoelectric material, the cavity optomechanical system could combine with the electromechanical coupling techniques of Ref. \cite{Goryachev2012,Galliou2013} for a high frequency RF-to-optical conversion platform  \cite{Bochmann2013,Pitanti2015,Balram2015}.  Phonon modes of the type from this bulk crystalline system can also be readily coupled with superconducting qubits \cite{Chu2017a}.

The task of adapting this bulk crystalline system to a cavity-optomechanical system presents some challenges; one must exercise great care in the design of phonon modes and optical modes depending on the optical and elastic properties of the crystal.  For instance, the cavities required to produce optimal phonon confinement and optical confinement are likely to be very different; this is because nontrivial dispersion surfaces of the elastic medium drastically alter the conditions for stable phonon cavity-mode formation (for further discussion see Supplement Section 5).  Moreover, while it is tempting to consider creating an optical cavity by depositing multi-layer stacks directly on the crystal, one must also be cognizant of the fact that (perhaps large) excess phonon dissipation will be accompanied by deposition of layered media on the crystal surface \cite{Galliou2016b,Galliou2016c}.  Whether the bulk crystalline system has an optical cavity or not, the techniques developed in the present work are essential for optimizing phonon dissipation, material type, material quality, and optomechanical coupling.  In other words, here we have taken the crucial steps required before constructing more sophisticated cavity optomechanical systems.

Looking ahead, it is clear that the field of quantum information and quantum measurement have growing need for ultra-low dissipation modes within optical, acoustic, and electromagnetic systems.  As we evaluate the best approach for obtaining low-loss phonon modes in bulk or micro-scale systems, it remains unclear which materials are most ideal as the basis for high performance optomechanical \cite{Aspelmeyer2014}, electro-optomechanical \cite{Bochmann2013}, or quantum-phononic systems \cite{Chu2017a}; improved understanding of cryogenic phonon dissipation could provide crucial upper bound for device performance.  To date, we have a fragmented understanding of the limits of phonon dissipation; aside from crystalline quartz, very few crystalline media have been exhaustively studied at cryogenic temperatures. % Prior crystalline quartz resonator technologies utilize piezoelectric response to access 100-200 MHz phonons with microwaves, have recently demonstrated record Q \cite{Goryachev2012,Galliou2013}.
With our laser-based approach to analyzing phonon dissipation, we have demonstrated  $(\textrm{frequency})\times (Q\textrm{-factor})$ products of $4.2{\times}10^{17}$ Hz, which are comparable to the world-class quartz electromechanical systems of Ref. \cite{Goryachev2012,Galliou2013}. In addition, since \textit{all} media posses photoelastic response, we can use these non-invasive laser-based optomechanical techniques to study practically any bulk crystalline medium. Hence, beyond new device concepts, bulk crystalline optomechanics has the potential to greatly expand our knowledge of cryogenic phonon physics.

In summary, we have demonstrated bulk crystalline optomechanics with macroscopic pristine single-crystal shaped acoustic resonators.  Through a Brillouin-like coupling, high frequency (${>}10$ GHz) ultra-high $Q$-factor phonon-modes are stimulated and detected with light.  Photon-phonon coupling in this bulk crystalline system presents an array of opportunities for both spectroscopy and optomechanical device design. In the context of nonlinear optics, this process is an extreme limit of Brillouin interactions wherein excited phonons become highly non-local and coherent. In the context of optomechanical phenomena, coupling to such mesoscopic phonon modes can be used as the basis for new quantum-optomechanical interactions. With high frequency (${>}10$ GHz), ultra-high $Q$-factor ($4.2{\times}10^7$) phonon modes, with variable mass ($10^{-9}{-}1$g), and high power handling ($>$kW), it presents an intriguing avenue for optomechanical device design as the basis for quantum information processing, tests of quantum coherence and sensitive metrology.  From either perspective, dramatic ($10^4$-fold) enhancement of the nonlinear coupling is achieved by trapping ultra high $Q$-factor phonon modes in the focus of an incident laser beam. This interaction can be engineered in practically any transparent crystal and can be viewed as a new type of ultra-sensitive Brillouin-like materials spectroscopy. Bulk crystalline optomechanics provides a broadly tunable platform for the study of new parametric processes as well as basic material properties.\\

%\\

\newpage

\section*{Acknowledgments}
Primary support for this work was provided by NSF MRSEC DMR-1119826. This work was supported in part by the Packard Fellowship for Science and Engineering and Yale University. The authors thank Paul Fleury, Yiwen Chu, Eric Kittlaus, Nils Otterstrom, Jack Harris, Kale Johnson, Alexey Shkarin, Anna Kashkanova, and Glen Harris for valuable feedback and discussions.

\section*{Author contributions}
W.H.R. and P.T.R. conceived the device and spectroscopy approach. W.H.R. conducted experiments to produce the initial results.  W.H.R. and P.K. jointly advanced these techniques to produce the final results under the guidance of P.T.R.  W.H.R. and P.K. developed simulation methods with input from R.O.B. and P.T.R.  P.K. and R.O.B. developed the analytical theory with guidance from W.H.R. and P.T.R.  All authors participated in the writing of this manuscript.
%Or more specifically (maybe too much), the manuscript was prepared by W.H.R. and P.T.R. and the supplementary information was prepared by P.K. with guidance from P.T.R., W.H.R. and R.O.B.

\section*{Additional information}
The authors declare no competing financial interests.

%\bibliographystyle{naturemag}
%\bibliography{library}
%\bibliographystyle{unsrt}

%\newpage

\clearpage
%\newpage


\section*{Supplementary material}

# Supplementary Information: Bulk crystalline optomechanics

# Contents

# 1 'Brillouin limit' vs. 'coherent-phonon limit'

In this section, we contrast the non-linear response and the dynamics of the acousto-optic interaction in the 'Brilloun limit' where phonons decay rapidly to that in the 'coherent-phonon limit' where phonons are long-lived.

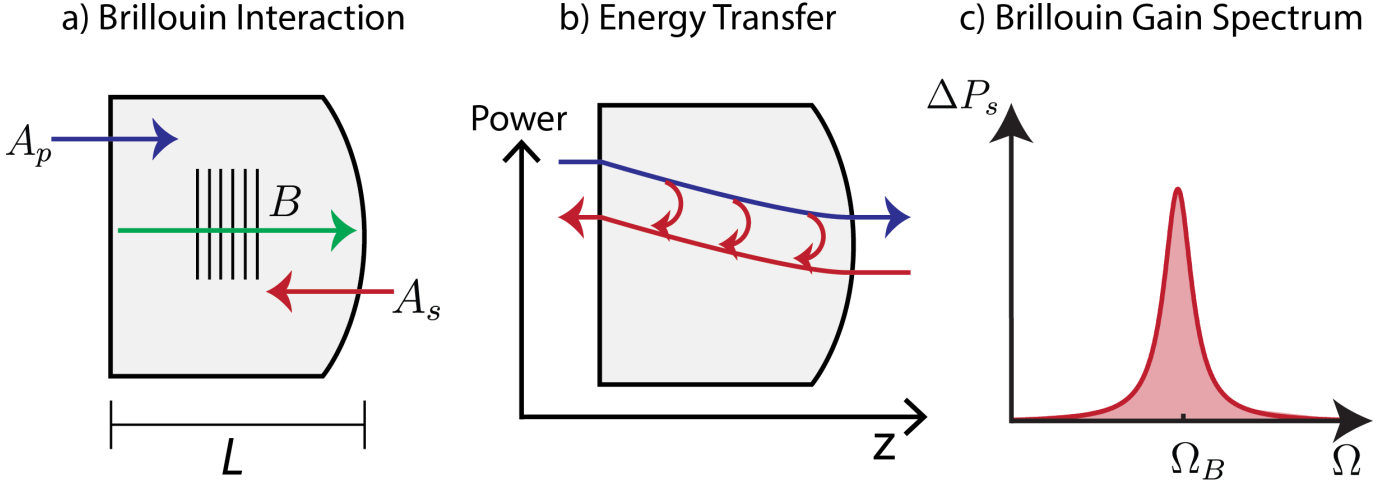

Figure 1: Stimulated Brillouin Scattering in the 'Brillouin limit' for phonons with the phonon coherence length that is much shorter than the crystal size. a) Acousto-optic interaction between counter-propagating pump ($A_p$) and Stokes ($A_s$) light with traveling acoustic wave ($B$). b) Electrostriction and photoelasticity mediates energy transfer between the pump and the Stokes light c) Resonant transfer of energy between the pump and the Stokes light occurs at the Brillouin frequency ($\Omega_B$) set by energy conservation and phase-matching requirements.

## 1.1 Brillouin Scattering: 'Brillouin limit'

Stimulated Brillouin scattering is a nonlinear scattering process involving optical and acoustic waves. This process occurs when two-counter propagating optical waves in a medium generate an optical beat pattern which then produces periodic density variation and refractive index modulation through electrostriction and photoelasticity, respectively (see Fig. 1). Frequency detuning of the optical waves generates a moving density wave which acting like a moving Bragg grating, Doppler shifts and back-scatters the pump light to Stokes (or probe) light. Resonant enhancement of both the acoustic wave and the back-scattered Stokes light occur when the frequency detuning between the pump light and Stokes light is such that the velocity of the beat pattern moves at the velocity of acoustic waves. In the backward scattering geometry, a forward moving pump photon with frequency and wavevector ($\omega_p, k(\omega_p)$) is scattered to a backward moving Stokes photon ($\omega_s, -k(\omega_s)$) and a forward moving acoustic phonon ($\Omega, q(\Omega)$) (see Fig. 2). Here, $k(\omega) = \omega/v_o$ ($q(\Omega) = \Omega/v_a$) is the frequency-dependent optical (acoustic) wavevector and $v_o(v_a)$ is the phase velocity of light (sound). For resonant energy transfer, the optical and acoustic waves involved in Brillouin scattering satisfy stringent energy conservation and phase matching requirements: $\omega_p = \omega_s + \Omega$ and $k(\omega_p) = -k(\omega_s) + q(\Omega)$. These requirements along with the dispersion relations (assumed linear for a bulk system) set the frequency for the resonant Brillouin energy transfer:

$$\Omega_B = \frac{2\omega_p v_a/v_o}{1 + v_a/v_o} \approx 2\omega_p v_a/v_o. \tag{1}$$

The coupled-mode equations governing the three-wave interaction between the forward moving pump wave with electric displacement field $\mathbf{D}_p(\mathbf{r},t) = \mathbf{D}_p(\mathbf{r})\bar{A}_p(z,t)e^{i(k_p z - \omega_p t)} + c.c.$, backward moving Stokes wave with electric displacement field $\mathbf{D}_s(\mathbf{r},t) = \mathbf{D}_s(\mathbf{r})\bar{A}_s(z,t)e^{i(-k_s z - \omega_s t)} + c.c.$ and the acoustic wave with elastic displacement field $\mathbf{u}(\mathbf{r},t) = \mathbf{u}(\mathbf{r})\bar{B}(z,t)e^{i(qz - \Omega t)} + c.c.$ involved in the classical Brillouin

3

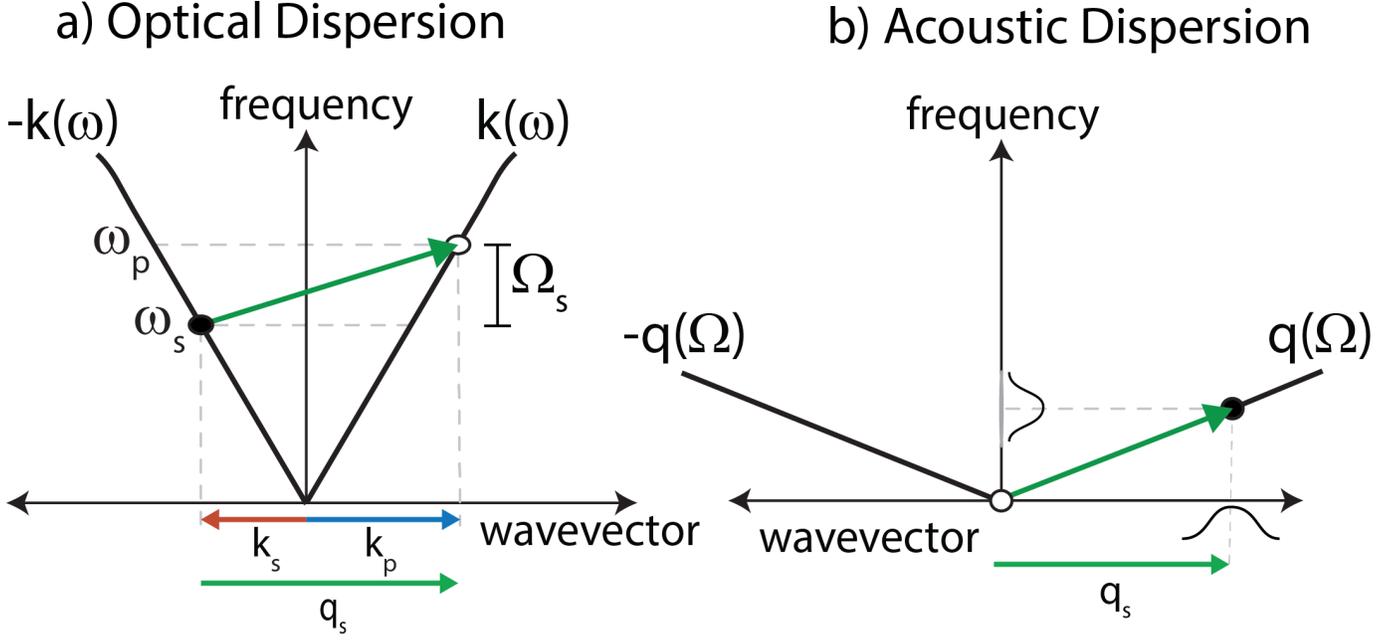

Figure 2: Phase matching and energy conservation requirements for traveling wave acousto-optic interactions.

scattering process is described by these coupled mode equations [1, 2]

$$\frac{\partial \bar{B}}{\partial t} + v_a \frac{\partial \bar{B}}{\partial z} = -i(\Omega_s - \Omega)\bar{B} - ig_B^* \bar{A}_s^* \bar{A}_p - \frac{\Gamma_B}{2}\bar{B}, \tag{2}$$

$$\frac{\partial \bar{A}_p}{\partial t} + v_o \frac{\partial \bar{A}_p}{\partial z} = -ig_B \bar{A}_s \bar{B}, \tag{3}$$

$$\frac{\partial \bar{A}_s}{\partial t} - v_o \frac{\partial \bar{A}_s}{\partial z} = -ig_B^* \bar{B}^* \bar{A}_p. \tag{4}$$

Here, $\bar{A}_p(z,t)$, $\bar{A}_s(z,t)$ and $\bar{B}(z,t)$ are the slowly varying envelopes for the optical and acoustic fields. $\Gamma_B/2$ is the phonon dissipation rate and $v_o(v_a)$ is the phase velocity of light (sound) in the bulk medium. $g_B$ characterizes the strength of the electrostrictive coupling.

At room temperature, elastic waves decay rapidly due to intrinsic dissipation. In this limit, propagation of phonons in space can be neglected i.e. $\partial \bar{B}/\partial z \to 0$. With this assumption, the steady-state phonon envelope due to the optical driving is

$$\bar{B}(z) = \frac{g_B^* \bar{A}_s^*(z) \bar{A}_p(z)}{\left(\Omega - \Omega_B + i\frac{\Gamma_B}{2}\right)}. \tag{5}$$

This equation shows that the mechanical response to the optical driving is local because the phonon envelope at $z$ depends only on the optical driving term $\bar{A}_s^* \bar{A}_p$ also at $z$. It is important to note that, $B^\dagger(z)B(z)$ ($A_\gamma^\dagger(z)A_\gamma$) here corresponds to the phonon (photon) number per unit length. The power in the acoustic (optical) field along $z$ in terms of the envelope functions is given by $P^{\text{ph}} = \hbar\Omega_0 v_a B^\dagger(z)B(z)$ ($P^{\text{opt}} = \hbar\omega_\gamma v_o A_\gamma^\dagger(z)A_\gamma(z)$) [3]. Using this definition for power and substituting Eq. (5) into Eq. (4), we obtain

$$\frac{\partial P_s(z)}{\partial z} = -G_{\text{SBS}}(\Omega) P_p(z) P_s(z). \tag{6}$$

Here, $P_s(z)$ ($P_p(z)$) denotes the power in the Stokes (pump) wave at position $z$. The Brillouin gain coefficient, $G_{\text{SBS}}$, that determines the energy transfer is [1, 2]

$$G_{\text{SBS}}(\Omega) = \frac{\omega^2 n^7 p_{13}^2}{c^3 v_a \rho \Gamma_B A^{ao}} \frac{(\Gamma_B/2)^2}{(\Omega - \Omega_B)^2 + (\Gamma_B/2)^2}. \tag{7}$$

Here, $n$ is the index of refraction of the material, $p_{13}$ is the relevant photo-elastic constant, $\rho$ is the density, and $A^{ao}$ is the effective acousto-optic area which is determined by the mode-overlap of the acoustic and optical fields. In the undepleted pump regime ($P_p(z) \approx$ constant) with weak-signal gain ($\Delta P_s \ll P_s$), the change in Stokes power over the interaction length $L$ becomes

$$\Delta P_s(L) = P_s(0) - P_s(L) \simeq G_{\text{SBS}}(\Omega) P_p(0) P_s(L) L. \tag{8}$$

Therefore, as the frequency detuning between the pump and the Stokes (i.e. $\Omega = \omega_p - \omega_s$) is varied, back-scattered Stokes power has a Lorentzian shape centered at the Brillouin frequency (see Fig. 1 c)). The linewidth of this gain spectrum can be used to measure the phonon dissipation rate ($\Gamma_B$). The height of the gain spectrum along with the phonon dissipation rate can then be used to infer the strength of the electrostrictive coupling.

At low temperatures the coherence length of phonons increases dramatically. If we consider a waveguide of infinitely long length we obtain these equations of motion for the pump, Stokes, and the acoustic field

$$\frac{\partial \bar{B}}{\partial t} + v_a \frac{\partial \bar{B}}{\partial z} = -i(\Omega_s - \Omega)\bar{B} - ig_B^* \bar{A}_s^* \bar{A}_p e^{-i\Delta q z} - \frac{\Gamma_B}{2} B, \tag{9}$$

$$\frac{\partial \bar{A}_p}{\partial t} + v_o \frac{\partial \bar{A}_p}{\partial z} = -ig_B \bar{A}_s \bar{B} e^{i\Delta q z}, \tag{10}$$

$$\frac{\partial \bar{A}_s}{\partial t} - v_o \frac{\partial \bar{A}_s}{\partial z} = -ig_B^* \bar{B}^* \bar{A}_p e^{-i\Delta q z}. \tag{11}$$

Here, $\Delta q = q(\Omega) - (k(\omega_p) - (-k(\omega_s)))$ is the phase mismatch between the optical driving and the phonon field. In this intermediate regime, propagation of phonons can no longer be neglected ($\partial \bar{B}/\partial z \neq 0$). The response of the phonon envelope at a given position now depends on the optical force throughout the system, (i.e. the phonon response is non-local in space).

Real experimental systems are finite in length. When coherence length of phonons at low temperatures become comparable to the crystal length, the traveling wave treatment with left and right moving phonon fields (i.e. $\bar{B}_L(z)$ and $\bar{B}_R(z)$) with self-consistent boundary conditions can be used to describe electrostriction mediated acousto-optic interactions. However, when the coherence length of phonons far exceeds the crystal length, the left and right moving phonons form standing waves or macroscopic discrete phonon modes. In the next section, we discuss this limit when the phonon coherence far exceeds the length of the crystal.

## 1.2 Bulk Crystalline Optomechanics: 'coherent-phonon limit'

At low temperatures, the coherence length of phonons in a pristine crystalline solid can be many times longer (order of meters) than the length $L$ of the solid (order of cm). As discussed above, the conventional assumption that the mechanical response of phonons is local is no longer valid. The acoustic waves with long coherence length reflect off the edges of the medium to form standing wave phonon modes much like the optical modes in a Fabry-Pérot resonator. In this section, we will show

how the character of energy transfer in this new regime of high-coherence for phonons differs from that in the Brillouin limit, where phonons have a relatively short coherence length.

The coupled mode equations for the interaction of the standing wave bulk phonon modes with elastic displacement $\mathbf{u}_m(\mathbf{r},t) = \mathbf{U}_m(\mathbf{r})b_m(t)(e^{iq_mt} + e^{-iq_mt})e^{-i\Omega_mt} + c.c.$, to the (forward propagating) pump wave with electric displacement field $\mathbf{D}_p(\mathbf{r},t) = \mathbf{D}_p(\mathbf{r}_\perp)\bar{A}_p(z,t)e^{i(k_pz-\omega_pt)} + c.c.$, and (backward propagating) Stokes wave with electric displacement field $\mathbf{D}_s(\mathbf{r},t) = \mathbf{D}_s(\mathbf{r}_\perp)\bar{A}_s(z,t)e^{i(-k_sz-\omega_st)} + c.c.$ in the limit where $l_{ph} \gg L$ is described by (see Section 2.6 for details)

$$\frac{\partial \bar{b}_m}{\partial t} = -i(\Omega_m - \Omega)\bar{b}_m - \frac{\Gamma_m}{2}\bar{b}_m - i\int_0^L dz\, (g_0^m)^* e^{-i\Delta qz}\bar{A}_s^\dagger \bar{A}_p, \tag{12}$$

$$\frac{\partial \bar{A}_p}{\partial t} + v_o\frac{\partial \bar{A}_p}{\partial z} = -i\sum_m g_0^m e^{i\Delta qz}\bar{A}_s\bar{b}_m, \tag{13}$$

$$\frac{\partial \bar{A}_s}{\partial t} - v_o\frac{\partial \bar{A}_s}{\partial z} = -i\sum_m g_{0m}^* e^{-i\Delta qz}\bar{b}_m^\dagger\bar{A}_p. \tag{14}$$

Here, $\bar{b}_m(t)$ and $\bar{A}_\gamma(z,t)$ ($\gamma$ = p, s) represent the slowly varying envelopes for the acoustic and optical fields respectively. While we have treated optical fields as traveling waves with mode profile $\mathbf{D}_\gamma(\mathbf{r})$, the acoustic field is treated as a discrete mode with mode profile $\mathbf{U}_m(\mathbf{r})$. Therefore, $\bar{b}(t)$ has no spatial dependence. $g_0^m$ is the electrostrictive coupling rate of the optical fields to a single $m^{\text{th}}$ phonon mode. $\Delta q = q(\Omega_m) - (k(\omega_p) - (-k(\omega_s))) = q_m - k_p - k_s$ is the phase-mismatch between the optical driving and the right moving component of the standing wave phonon mode. We assume that $\Gamma_m$ for each phonon mode encompasses all forms of losses such as intrinsic losses, losses at the surface, diffraction losses and so on. The integral over $z$ in Eq. (12) suggests that the phonon mode is driven by the optical beat tone throughout the crystal. Therefore, the susceptibility for the phonon field is non-local as the optical driving all along the crystal affects the phonon mode amplitude $\bar{b}_m$. The uncertainty in the phonon wavevector resulting from the finite length of a crystal results in a non-zero phase matching bandwidth. A small set of acoustic modes within the phase-matching bandwidth can interact with the pump and the Stokes light field. The integral term in Eq. (12) incorporates this salient feature of bulk-crystalline optomechanics.

The actual set of standing wave phonon modes within the phase-matching bandwidth for the bulk crystalline optomechanical system not only depends on the geometry (i.e. how you shape the boundaries of the crystalline media) but also on the anisotropy of the crystal. We will discuss this problem in greater detail in Section 5. Nevertheless, frequencies of these standing wave acoustic modes in 1D can be approximately calculated by fitting acoustic half-wavelengths ($\lambda_a$) inside the crystal length ($L$) (i.e. $m\lambda_a/2 = L$). So the frequency of the $m^{\text{th}}$ acoustic mode assuming a linear acoustic dispersion ($q(\Omega) = \Omega/v_a$) is $\Omega_m = 2\pi \times (mv_a)/(2L)$, where $v_a$ is the longitudinal sound velocity.

In the limit of weak signal gain and undepleted pump (i.e. $\Delta P_s(0) \ll P_s(L)$ and $A_p(z) \approx A_p(0) \approx$ constant), the equation for the acoustic mode amplitude and the Stokes power change for a crystalline medium of length $L$ becomes

$$\bar{b}_m \simeq (g_0^m)^* L \frac{\bar{A}_p(0)\bar{A}_s^*(L)}{(\Omega - \Omega_m + i\frac{\Gamma_m}{2})} e^{-i\frac{\Delta qL}{2}} \text{sinc}\left(\frac{\Delta qL}{2}\right), \tag{15}$$

$$\Delta P_s(L) = P_s(0) - P_s(L) \approx P_p(0)P_s(L)\sum_m \frac{4|g_0^m|^2 L^2}{\hbar\omega_p v_o^2 \Gamma_m} \frac{(\Gamma_m/2)^2}{(\Omega - \Omega_m)^2 + (\Gamma_m/2)^2}\text{sinc}^2\left(\frac{\Delta qL}{2}\right), \tag{16}$$

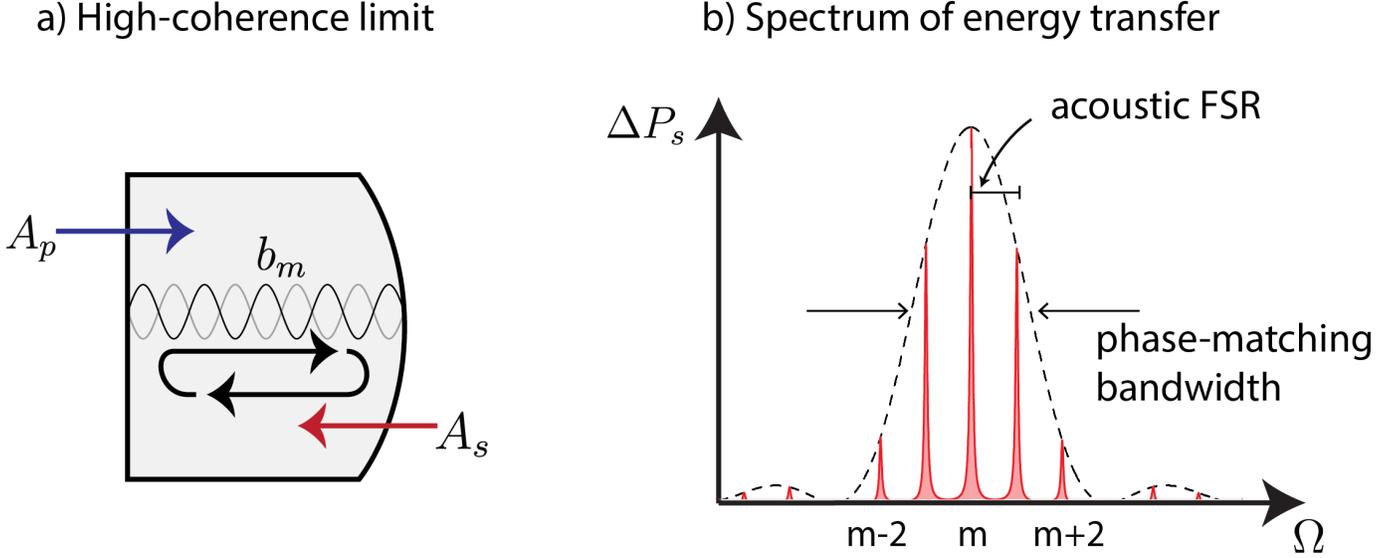

Figure 3: coherent-phonon limit. a) Acousto-optic interaction between traveling wave pump ($A_p$) and Stokes light ($A_s$) with discrete standing wave phonon modes ($b_m$). b) Several standing wave acoustic modes within the phase matching bandwidth resonantly interact with the light fields resulting in a multi-peaked spectrum. FSR: Free Spectral Range.

where the phase mismatch $\Delta q_m = q_m - k_p - k_s \simeq \Omega_m/v_a - 2\omega_p/v_o$ to an excellent approximation. In deriving Eqs. (15) and (16) we have assumed that the coupling rate $g_0^m$ is invariant in $z$. Eq. (16) shows that the spectrum of the power transfer has multiple Lorentzians under a sinc squared envelope (See Fig. 3). This spectrum is markedly different than just the Lorentzian response obtained in the low coherence limit (See Eq. (8)). The Lorentzian response for each discrete phonon mode within the phase-matching bandwidth still contains information about the phonon dissipation rate, $\Gamma_m$. For each phonon mode, $g_0^m$ is determined by the overlap integrals between the optical and acoustic mode profiles (see Section 2.4 for more details).

## 2  Hamiltonian Treatment

In this section, we derive the coupled-mode equations for the light and sound in the coherent-phonon limit, identify coupling rate, nonlinear susceptibility and co-operativity using a generalizable quantum Hamiltonian treatment. Using this treatment, we also derive well-know coupled-mode equations in the Brillouin limit (or classical Brillouin scattering) to point out the differences between the dynamics of the light-sound interaction in the coherent-phonon limit and the dynamics of light-sound interaction in the Brillouin limit.

The Hamiltonian in a system with photoelasticity/electrostriction mediated opto-acoustic interaction can be written as [3]

$$H = H^{\text{ph}} + H^{\text{opt}} + H^{\text{int}}, \tag{17}$$

where $H^{\text{ph}}$, $H^{\text{opt}}$, and $H^{\text{int}}$ characterize the dynamics of acoustic fields, the dynamics of optical fields and the acousto-optic interaction respectively.

$$H^{\text{ph}} = \int \frac{\pi^i(\mathbf{r})\pi^i(\mathbf{r})}{2\rho(\mathbf{r})} d\mathbf{r} + \int \frac{1}{2} S^{ij}(\mathbf{r}) c^{ijkl}(\mathbf{r}) S^{ij}(\mathbf{r}) d\mathbf{r}, \tag{18}$$

where $\pi(\mathbf{r})$ is the conjugate momentum of the acoustic displacement field operator $\mathbf{u}(\mathbf{r})$, $\rho(\mathbf{r})$ is the density, $c^{ijkl}(\mathbf{r})$ is the elastic constant tensor and $S^{ij}(\mathbf{r}) = 1/2(\partial u^i(\mathbf{r})/\partial r^j + \partial u^j(\mathbf{r})/\partial r^i)$ is the strain operator.

$$H^{\text{opt}} = \frac{1}{2\mu_0} \int B^i(\mathbf{r})B^i(\mathbf{r}) d\mathbf{r} + \frac{1}{2\epsilon_0} \int D^i(\mathbf{r})\beta_r^{ij}(\mathbf{r})D^j(\mathbf{r}) d\mathbf{r}, \quad (19)$$

where $\mathbf{D}(\mathbf{r})$ is the electric displacement field operator, $\mathbf{B}(\mathbf{r})$ is the magnetic field operator and $\epsilon_r^{ij}(\mathbf{r}) = 1/\beta^{ij}(\mathbf{r})$ is the relative dielectric constant tensor. The photoelastic/electrostrictive interaction for a bulk medium is

$$H^{\text{int}} = \int \frac{1}{2\epsilon_0} D^i(\mathbf{r})D^j(\mathbf{r})p^{ijkl}(\mathbf{r})S^{kl}(\mathbf{r}) d\mathbf{r}. \quad (20)$$

For a system of finite length, we use normal mode expansion of fields to write

$$\mathbf{D}(\mathbf{r}) = \sum_\gamma \sqrt{\frac{\hbar\omega_\gamma}{2}} a_\gamma(t) \tilde{\mathbf{D}}_\gamma(\mathbf{r}) + \text{H.c}, \quad (21)$$

$$\mathbf{u}(\mathbf{r}) = \sum_\Lambda \sqrt{\frac{\hbar\Omega_\lambda}{2}} b_\Lambda(t) \tilde{\mathbf{u}}_\Lambda(\mathbf{r}) + \text{H.c}, \quad (22)$$

where $\tilde{D}_\gamma(\mathbf{r})$ is the eigenmode obtained by solving the Maxwell's equations, $a_\gamma$ is the optical mode amplitude operator and $\omega_\gamma$ is the optical mode frequency. $\tilde{u}_m(\mathbf{r})$ is the acoustic mode obtained by solving the elastic-displacement equation, $b_m$ is the acoustic mode amplitude operator, $\Omega_m$ is the phonon frequency, and $\gamma$ and $\Lambda$ are collective mode indices. The optical and acoustic modes are normalized such that

$$\frac{1}{\epsilon_0} \int d^3 r \; \beta_r(\mathbf{r}) \tilde{\mathbf{D}}_\gamma^*(\mathbf{r}) \cdot \mathbf{D}_{\gamma'}(\mathbf{r}) = \delta_{\gamma,\gamma'} \quad (23)$$

$$\int d^3 r \; \rho(\mathbf{r})\Omega_\Lambda \Omega_{\Lambda'} \tilde{\mathbf{u}}_\Lambda^*(\mathbf{r}) \cdot \tilde{\mathbf{u}}_{\Lambda'}(\mathbf{r}) = \delta_{\Lambda,\Lambda'} \quad (24)$$

The mode amplitude operators satisfy these commutation relations

$$[a_\gamma, a_{\gamma'}] = 0, \; [a_\gamma, a_{\gamma'}^\dagger] = \delta_{\gamma,\gamma'}, \quad (25)$$

$$[b_\Lambda, b_{\Lambda'}] = 0, \; [b_\Lambda, b_{\Lambda'}^\dagger] = \delta_{\Lambda,\Lambda'}. \quad (26)$$

Using the normal mode expansion, the normalizations, and the commutation relations we can express the optical, acoustic, and interaction Hamiltonian in terms of the mode amplitude operators as

$$H^{\text{ph}} = \sum_\Lambda \hbar\Omega_\Lambda (b_\Lambda^\dagger b_\Lambda + 1/2), \quad (27)$$

$$H^{\text{opt}} = \sum_\gamma \hbar\omega_\gamma (a_\gamma^\dagger a_\gamma + 1/2), \quad (28)$$

$$H^{\text{int}} = \sum_{\gamma,\gamma',m} \int d\mathbf{r} \; \frac{1}{2\epsilon_0} \sqrt{\frac{\hbar\omega_\gamma \hbar\omega_{\gamma'} \hbar\Omega_m}{8}} p^{ijkl} \left(a_\gamma \tilde{D}_\gamma^i(\mathbf{r}) + \text{H.c.}\right) \left(a_{\gamma'} \tilde{D}_{\gamma'}^j(\mathbf{r}) + \text{H.c.}\right) \left(b_m \tilde{s}_m^{kl}(\mathbf{r}) + \text{H.c}\right). \quad (29)$$

8

The above treatment is valid for any structure with finite size having discrete optical and acoustic modes. However, we are interested in looking at the interaction between traveling wave pump photons with mode profile, wave-vector, and frequency given by $(\mathbf{D}_p(\mathbf{r}), k_p, \omega_p)$ with the Stokes wave photons with mode profile, wave-vector, and frequency $(\mathbf{D}_s(\mathbf{r}), -k_s, \omega_s)$ mediated by phonons. Therefore, the optical eigenmodes of interest, ignoring small reflections at the crystal end facets, are traveling-waves of infinite extent in $z$ (equivalent to a system with $l \to \infty$)

$$\tilde{\mathbf{D}}_\gamma(\mathbf{r}) \mapsto \mathbf{D}_\gamma(\mathbf{r}) e^{ikz}. \tag{30}$$

We also assume that the traveling optical waves are sharply peaked about their carrier wavevectors: $k_p$ and $-k_s$ for pump and Stokes respectively. Therefore, we represent the field using the following normal mode expansion

$$\mathbf{D}(\mathbf{r}) = \int \frac{dk}{\sqrt{2\pi}} \sqrt{\frac{\hbar \omega_p(k)}{2}} a_{pk}(t) \mathbf{D}_p(\mathbf{r}) e^{ikz} + \int \frac{dk}{\sqrt{2\pi}} \sqrt{\frac{\hbar \omega_s(k)}{2}} a_{sk}(t) \mathbf{D}_s(\mathbf{r}) e^{ikz} + \text{H.c.} \tag{31}$$

to write the optical Hamiltonian as

$$H^{\text{opt}} = \int dk \ \hbar \omega_p(k) a_{pk}^\dagger a_{pk} + \int dk \ \hbar \omega_s(k) a_{sk}^\dagger a_{sk}. \tag{32}$$

It is important to note that, in this continuum picture, the mode amplitude operator $a_k$'s have units of $\sqrt{L}$. In writing the above equation we assumed that the eigenmode profiles $\mathbf{D}_p(\mathbf{r})$ and $\mathbf{D}_s(\mathbf{r})$ for the pump and Stokes traveling waves in a narrow band about their respective carrier wave-vector remains unchanged. The optical mode profiles are normalized at each point in $z$ such that

$$\frac{1}{\epsilon_o} \int d\mathbf{r}_\perp \ \beta_r(\mathbf{r}) \mathbf{D}_p^*(\mathbf{r}) \cdot \mathbf{D}_p(\mathbf{r}) = 1, \tag{33}$$

$$\frac{1}{\epsilon_o} \int d\mathbf{r}_\perp \ \beta_r(\mathbf{r}) \mathbf{D}_s^*(\mathbf{r}) \cdot \mathbf{D}_s(\mathbf{r}) = 1. \tag{34}$$

Extended coherence length of the phonons and the formation of discrete standing wave cavity modes for phonons results in a markedly different nonlinear optical susceptibility in the coherent-phonon limit. We will explore this in the next two subsections.

## 2.1 Coherent-Phonon Limit

In the coherent-phonon limit, where the coherence length of phonons is much longer than crystal length ($l_{ph} \gg L$), phonons can form standing wave cavity modes that extend throughout the crystal. The modes of interest (i.e. symmetric longitudinal modes) are of the form

$$\tilde{\mathbf{u}}_\Lambda(\mathbf{r}) \mapsto \mathbf{U}_m(\mathbf{r})(e^{iq_m z} + e^{-iq_m z})/\sqrt{2}, \tag{35}$$

where the standing-wave acoustic modes have discrete wavevector $q_m = 2\pi m/L$, for any integer m, and $\mathbf{U}_m(\mathbf{r})$ is the mode profile that varies slowly in $z$ compared to the carrier wavevector. Since we have discrete acoustic modes in the coherent-phonon limit, the acoustic Hamiltonian is

$$H^{\text{ph}} = \sum_m \hbar \Omega_m b_m^\dagger b_m. \tag{36}$$

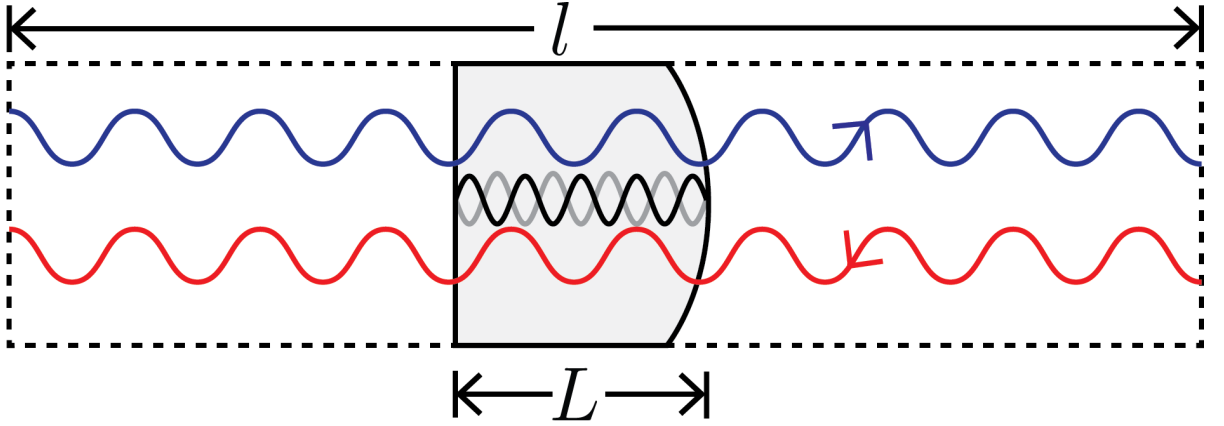

Figure 4: A discrete picture for the optical modes. We take the optical quantization length to be much larger than the crystal length ($l \gg L$).

We used the following normal mode expansion for the acoustic displacement field

$$\mathbf{u}(\mathbf{r}) = \sum_m \sqrt{\frac{\hbar \Omega_m}{2}} b_m \frac{\mathbf{U}_m(\mathbf{r})}{\sqrt{2}} (e^{iq_m z} + e^{-iq_m z}) + \text{H.c.}, \tag{37}$$

and acoustic mode profiles are normalized so that

$$\int \mathrm{d}^3 r \ \Omega_m^2 \rho(\mathbf{r}) \mathbf{U}_m(\mathbf{r})^* \cdot \mathbf{U}_m(\mathbf{r}) \simeq 1. \tag{38}$$

The $k$-space interaction Hamiltonian in the coherent-phonon limit using the normal mode expansion in Eq. (37) and Eq. (31) is then given by

$$H^{\text{int}} = \sum_m \int \frac{\mathrm{d}k \mathrm{d}k'}{2\pi} \int \mathrm{d}z \ \hbar g_0^m(z) e^{i(k'-k+q_m)z} a_{pk}^\dagger a_{sk'} b_m + \text{H.c.} \tag{39}$$

$g_0^m(z)$, that characterizes the electrostriction-mediated coupling rate between traveling optics fields and a discrete phonon mode, is

$$g_0^m(z) \simeq \frac{\sqrt{\hbar \omega_p \omega_s \Omega_m}}{4\epsilon_0} \int \mathrm{d}\mathbf{r}_\perp (D_p^i(\mathbf{r}))^* D_s^j(\mathbf{r}) p^{ijkl}(\mathbf{r}) \left( \frac{\partial U_m^k(\mathbf{r})}{\partial r^l} + iq_m \delta_{lz} U_m^k(\mathbf{r}) \right). \tag{40}$$

As before, we have assumed that the coupling rate in the narrow band of wavevectors around the carrier wavevectors for pump and Stokes remains unchanged. This coupling rate between traveling wave photons and a discrete phonon mode has dimensions of Hz. Note that the wavevector $k$ and $k'$ for photons are continuous variables whereas the wavevector $q_m$ for a phonon mode is discrete.

## 2.2 Nonlinear optical susceptibility

To derive the nonlinear susceptibility, we examine interaction between monochromatic pump and Stokes fields as they impinge on the crystal. We assume that $P_p \gg P_s$. Throughout, we also assume

10

a weak interaction between the coupling fields such that their coupling produces negligible depletion of the pump wave; we also require that $\Delta P_s/P_s \ll 1$, where $\Delta P_s$ increase in Stokes-wave power as it traversed the crystal.

As an accessible starting point, we begin by using $k$-space representation for the modes, and we apply first-order time-dependent perturbation theory to derive the optical susceptibility. Alternatively, through the analyses presented Sections 2.6, we will see that the real-space description provides a more natural way of formulating the coupling dynamics in non-perturbative regimes of interaction.

For simplicity, we assume that the transverse spatial mode profiles are uniform along $z$ such that $g_0^m$ is a constant. This simplification allows us to express Eqs. (32), (36) and (39) as

$$H^{\text{opt}} = \int \mathrm{d}k \; \hbar\omega_p(k) a_{pk}^\dagger a_{pk} + \int \mathrm{d}k \; \hbar\omega_s(k) a_{sk}^\dagger a_{sk}, \tag{41}$$

$$H^{\text{ph}} = \sum_m \hbar\Omega_m b_m^\dagger b_m, \tag{42}$$

$$H^{\text{int}} = \sum_m \int \frac{\mathrm{d}k\mathrm{d}k'}{2\pi} \hbar g_m(k,k') a_{pk}^\dagger a_{sk'} b_m + \text{H.c.} \tag{43}$$

Here, $g_m(k,k') = g_0^m L \operatorname{sinc}\left((k'-k+q_m)L/2\right) e^{i\phi}$, where $\phi = ((k'-k+q_m)L/2) + \pi/2$. From this Hamiltonian, we derive equations of motion and the nonlinear optical susceptibility associated with the coherent-phonon limit.

We use the Heisenberg equations of motion in conjunction with equal-time commutator relations, $[a_k, a_{k'}^\dagger] = \delta(k-k')$ and $[b_m, b_{m'}^\dagger] = \delta_{m,m'}$ one obtains these equations of motion [3,4]

$$\dot{a}_{sk} = -\frac{i}{\hbar}[a_{sk}, H] = -i\omega_s(k)a_{sk} - i\sum_m \int \frac{\mathrm{d}k'}{2\pi} g_m^*(k',k) a_{pk'} b_m^\dagger, \tag{44}$$

$$\dot{a}_{pk} = -\frac{i}{\hbar}[a_{pk}, H] = -i\omega_p(k)a_{pk} - i\sum_m \int \frac{\mathrm{d}k'}{2\pi} g_m(k,k') a_{sk'} b_m, \tag{45}$$

$$\dot{b}_m = -\frac{i}{\hbar}[b_m, H] = -i\Omega_m b_m - i\int \frac{\mathrm{d}k\mathrm{d}k'}{2\pi} g_m^*(k,k') a_{sk'}^\dagger a_{pk}. \tag{46}$$

To simplify our derivation, we will find it convenient to begin by treating our optical fields as a discrete set of traveling waves within a system of length, $l$, where $l \gg L$. (See Fig.4 for a sketch of the optical modes relative to the crystal.) At the conclusion of this derivation, we will see that the continuum limit is recovered by taking $l \to \infty$.

To switch from continuous to discrete representations of our modes, we observe that $\int \hbar\omega_p(k) a_{pk}^\dagger a_{pk} \, \mathrm{d}k$ is equivalent to $\sum_m \omega_m a_{k_m}^\dagger a_{k_m} \Delta k$ in the limit as $\Delta k \equiv 2\pi/l \to 0$. Hence, the mode amplitude operators in the continuum picture, $a_{pk}$, can be translated to the dimensionless mode amplitude operator in the discrete picture, $\tilde{a}_k$, by making the replacements

$$a_k \to \sqrt{\frac{l}{2\pi}} \tilde{a}_k \tag{47}$$

and

$$\int \mathrm{d}k \to \frac{2\pi}{l} \sum_k. \tag{48}$$

Using Eq. (47) and Eq. (48) to express Eq.(44)-(46) in discrete form, we obtain

$$\dot{\tilde{a}}_{sk} = -i\omega_s(k)\tilde{a}_{sk} - \frac{i}{l}\sum_m\sum_{k'} g_m^*(k',k)\tilde{a}_{pk'}b_m^\dagger, \tag{49}$$

$$\dot{\tilde{a}}_{pk} = -i\omega_p(k)\tilde{a}_{pk} - \frac{i}{l}\sum_m\sum_{k'} g_m(k',k)\tilde{a}_{sk'}b_m, \tag{50}$$

$$\dot{b}_m = (-i\Omega_m - \Gamma_m/2)b_m - \frac{i}{l}\sum_k\sum_{k'} g_m^*(k,k')\tilde{a}_{sk'}^\dagger \tilde{a}_k. \tag{51}$$

Note that we have phenomenologically added the phonon dissipation rate $\Gamma_m/2$ in the equations of motion. In the spirit of perturbation theory, we make substitutions $\tilde{a}_{jk} \to \tilde{a}_{jk}^{(0)} + \lambda\tilde{a}_{jk}^{(1)} + \lambda^2\tilde{a}_{jk}^{(2)} + (...)$, $b_m \to b_m^{(0)} + \lambda b_m^{(1)} + \lambda^2 b_m^{(2)} + (...)$, and $g_m(k',k) \to \lambda g_m(k',k)$, where $\lambda$ is a unitless order-parameter. Our initial conditions require that only a single pump-mode ($\tilde{a}_{pk_p}$) and a single Stokes-mode ($\tilde{a}_{sk_s}$) are occupied while the phonon modes ($b_m$) are unoccupied. Hence, $\tilde{a}_{pk_p}^{(0)}(t)$, $\tilde{a}_{sk_s}^{(0)}(t)$, and $b_m^{(0)}(t)$ undergo free oscillation at frequencies $\omega_p(k_p)$, $\omega_s(k_s)$, and $\Omega_m$. However, since the phonon mode is unoccupied, (i.e., $b_m^{(0)}(0) = 0$) the phonon mode amplitude vanishes. To find the third-order effective optical susceptibility, we seek $\tilde{a}_{sk_s}^{(2)}$ for $t > 0$ in the undepleted pump approximation. The first order equations of motion become

$$\dot{\tilde{a}}_{k_s}^{(1)} = -i\omega_s(k)\tilde{a}_{k_s}^{(1)} - \frac{i}{l}\sum_m g_m^*(k_p,k_s)\tilde{a}_{k_p}^{(0)}(b_m^{(0)})^\dagger, \tag{52}$$

$$\dot{\tilde{a}}_{k_p}^{(1)} = -i\omega_s(k)\tilde{a}_{k_s}^{(1)} - \frac{i}{l}\sum_m g_m(k_p,k_s)\tilde{a}_{k_s}^{(0)}b_m^{(0)}, \tag{53}$$

$$\dot{b}_m^{(1)} = (-i\Omega_m - \Gamma_m/2)b_m^{(1)} - \frac{i}{l}g_m^*(k_p,k_s)(\tilde{a}_{k_s}^{(0)})^\dagger \tilde{a}_{k_p}^{(0)}. \tag{54}$$

Notice that, since $b_m^{(0)}(t) = 0$, the sourcing terms in Eq. (52) and Eq. (53) vanish; only the sourcing term on the right hand side of Eq. (54) is nonzero. The steady state phonon mode amplitude of $b_m^{(1)}$ after factoring out the fast oscillation is given by

$$b_m^{(1)} = \frac{1}{(\Omega - \Omega_m + i\Gamma_m/2)} \frac{g_m^*(k_p,k_s)}{l} (\tilde{a}_{k_s}^{(0)})^\dagger \tilde{a}_{k_p}^{(0)}. \tag{55}$$

Here, $\Omega \equiv \omega_p(k_p) - \omega_s(k_s)$. Substituting this result into the Equation of motion for $\tilde{a}_{s,k_s}^{(2)}$, we have

$$\dot{\tilde{a}}_{s,k_s}^{(2)}(t) = -\frac{i}{l^2}\sum_m \frac{|g_m(k_p,k_s)|^2}{(\Omega - \Omega_m - i\Gamma_m/2)} |\tilde{a}_{p,k_p}^{(0)}|^2 \tilde{a}_{s,k_s}^{(0)}. \tag{56}$$

We now use the dynamics of $\tilde{a}_{s,k_s}$ to find single-pass change in Stokes power. The Stokes photon generation rate using Eq. (56) is

$$\dot{\tilde{n}}_s \simeq \frac{4}{l^2}\sum_m \frac{|g_m(k_p,k_s)|^2}{\Gamma_m} \frac{(\Gamma_m/2)^2}{(\Omega - \Omega_m)^2 + (\Gamma_m/2)^2} \tilde{n}_p \tilde{n}_s. \tag{57}$$

The change in Stokes power for a single-pass is then given by

$$\Delta P_s = \frac{4}{l^2}\Delta t \sum_m \frac{|g_m(k_p,k_s)|^2}{\Gamma_m} \frac{(\Gamma_m/2)^2}{(\Omega - \Omega_m)^2 + (\Gamma_m/2)^2} \tilde{n}_p P_s, \tag{58}$$

where $\Delta t$ is the single pass transit time of the Stokes across the optical system and is given by $\Delta t = l/v_{gs}$. Since $P_p = \hbar\omega_p v_{gp} \tilde{n}_p/l$, we find

$$\Delta P_s = \sum_m \frac{4|g_m(k_p, k_s)|^2}{\hbar\omega_p v_{gp} v_{gs} \Gamma_m} \frac{(\Gamma_m/2)^2}{(\Omega - \Omega_m)^2 + (\Gamma_m/2)^2} P_p P_s. \tag{59}$$

Therefore, the nonlinear optical susceptibility in the weak signal limit is given by

$$\zeta = \sum_m \frac{4L^2|g_0^m|^2}{\hbar\omega_p v_{gp} v_{gs} \Gamma_m} \text{sinc}^2\left(\frac{(k_s - k_p + q_m)L}{2}\right) \frac{(\Gamma_m/2)^2}{(\Omega - \Omega_m)^2 + (\Gamma_m/2)^2} P_p. \tag{60}$$

Notice that, since we have expressed this result in terms of the incident pump- and Stokes-wave powers this result is independent of size ($l$) of the optical system. In other words, we can readily take the limit as $l \to \infty$ to return to the continuum limit. We will see that the expression for the susceptibility is obtained using the real-space treatment of this interaction, as described in Section 2.6.

## 2.3 Co-operativity

In this section, we derive co-operativity, which is a figure of merit that compares the ratio of optical contribution to the mechanical damping and the intrinsic mechanical damping. We follow a derivation of co-operativity similar to that in cavity optomechanics [5–7] by finding an effective mechanical susceptibility in presence of the optomechanical interaction.

We look at the co-operativity for a phonon mode labeled by index $m$. We start by considering Eqs. (44) - (46) with optical dissipation rate $\kappa/2$ and acoustic dissipation rate $\Gamma_m/2$ added phenomenologically as

$$\dot{a}_{sk} = \left(-i\omega_s(k) - \frac{\kappa}{2}\right) a_{sk} - i\int \frac{dk'}{2\pi} g_m^*(k', k) a_{pk'} b_m^\dagger, \tag{61}$$

$$\dot{a}_{pk} = \left(-i\omega_p(k) - \frac{\kappa}{2}\right) a_{pk} - i\int \frac{dk'}{2\pi} g_m(k, k') a_{sk'} b_m, \tag{62}$$

$$\dot{b}_m = \left(-i\Omega_m - \frac{\Gamma_m}{2}\right) b_m - i\int \frac{dk dk'}{2\pi} g_m^*(k, k') a_{sk'}^\dagger a_{pk} + f_\text{ext}(t). \tag{63}$$

We have phenomenologically added external force $f_\text{ext}(t)$ in the equation of motion for the phonon field to derive the effective mechanical susceptibility under optomechanical coupling. In our system, we assume that the crystalline medium is essentially transparent to the travelling wave optical beams. The optical beams experience negligible loss inside the crystal. We will eventually take the limit of zero optical loss to derive our co-operativity.

We derive co-operativity in the limit of weak coupling and assuming undepleted pump. It is convenient to work in an interaction frame that is rotating at $\omega_p$ (i.e. $a_{sk}(t) \to \bar{a}_{sk}(t)e^{-i\omega_p t}$ and $a_{pk}(t) \to \bar{\alpha}_{pk}e^{-i\omega_p t}$). Note that the pump field is freely evolving and not changing in time. Hence, $\bar{\alpha}_{pk}$ is just a constant. So, we arrive at these equations of motion for Stokes and the phonon field

$$\dot{\bar{a}}_{sk}(t) = \left(-i\omega_s(k) + i\omega_p - \frac{\kappa}{2}\right) \bar{a}_{sk}(t) - i\int \frac{dk'}{2\pi} g_m^*(k', k) \bar{\alpha}_{pk'} b_m^\dagger(t), \tag{64}$$

$$\dot{b}_m(t) = \left(-i\Omega_m - \frac{\Gamma_m}{2}\right) b_m - i\int \frac{dk dk'}{2\pi} g_m^*(k, k') \bar{a}_{sk'}^\dagger(t) \bar{\alpha}_{pk} + f_\text{ext}(t). \tag{65}$$

Now we make the undepleted pump approximation, so that $\bar{\alpha}_{pk} = \bar{\alpha}_{k_p}\delta(k - k_p)$ is a delta function that is peaked at the wavevector $k_p$. It is important to note that because $\delta(k - k_p)$ has units of $L$ and our mode amplitude operators $\bar{\alpha}_{pk}$ has units of $\sqrt{L}$, $\bar{\alpha}_{k_p}$ has units of $1/\sqrt{L}$. Now, we perform the integrals in $k$-space in Eqs. (64) and (65) to obtain

$$\dot{\bar{a}}_{sk}(t) = \left(-i\Delta(k) - \frac{\kappa}{2}\right)\bar{a}_{sk}(t) - \frac{i}{2\pi}\bar{\alpha}_{k_p}\, g_m^*(k_p, k)b_m^\dagger(t), \tag{66}$$

$$\dot{b}_m(t) = \left(-i\Omega_m - \frac{\Gamma_m}{2}\right)b_m - i\bar{\alpha}_{k_p}\int\frac{\mathrm{d}k'}{2\pi}g_m^*(k_p, k')\bar{a}_{sk'}^\dagger(t) + f_{\mathrm{ext}}(t), \tag{67}$$

where $\Delta(k) = -\omega_p + \omega_s(k)$ is the detuning between the pump and Stokes wave. $\bar{\alpha}_{k_p}$ is related to the input pump power incident on the crystal as [3]

$$P_p = \int\frac{\mathrm{d}k\mathrm{d}k'}{2\pi}\,\hbar\omega_p v_{gp}a_{pk'}^\dagger(t)a_{pk}(t)e^{i(k-k')z} = \frac{\hbar\omega_{pk}v_{gp}|\bar{\alpha}_{k_p}|^2}{2\pi}. \tag{68}$$

In the Fourier domain, using the the transformation $\bar{b}_m[\Omega] = \int_{-\infty}^{+\infty}\mathrm{d}t\, e^{i\Omega t}\bar{b}_m[t]$ and $\bar{a}_{sk}[\Omega] = \int_{-\infty}^{+\infty}\mathrm{d}t\, e^{i\Omega t}\bar{a}_{sk}(t)$, we obtain

$$\bar{a}_{sk}[\Omega] = \frac{g_m^*(k_p, k)\bar{\alpha}_{k_p}}{2\pi}\frac{1}{\Omega - \Delta(k) + i\kappa/2}b_m^\dagger[\Omega]. \tag{69}$$

$$b_m[\Omega] = \frac{\bar{\alpha}_{k_p}}{\Omega - \Omega_m + i\Gamma_m/2}\int\frac{\mathrm{d}k'}{2\pi}g_m^*(k_p, k')\bar{a}_{sk'}^\dagger[\Omega] + \frac{1}{\Omega - \Omega_m + i\Gamma_m/2}f_{\mathrm{ext}}[\Omega] \tag{70}$$

Substituting Eq. (69) in Eq. (70) gives

$$b_m[\Omega] = \chi_{m,\mathrm{eff}}[\Omega]f_{\mathrm{ext}}[\Omega] \tag{71}$$

$$= \frac{1}{\Omega - \Omega_m + i\Gamma_m/2 + \sum(\Omega)}f_{\mathrm{ext}}[\Omega] \tag{72}$$

$$= \frac{1}{\Omega - (\Omega_m + \delta\omega[\Omega]) + i(\Gamma_m + \Gamma_{\mathrm{opt}}(\Omega))/2}f_{\mathrm{ext}}[\Omega], \tag{73}$$

where

$$\sum(\Omega) = -|\bar{\alpha}_{k_p}|^2\int\frac{\mathrm{d}k'}{4\pi^2}\,|g_m(k_p, k')|^2\left[\frac{\Omega - \Delta(k')}{(\Omega - \Delta(k'))^2 + (\kappa/2)^2} + i\frac{\kappa/2}{(\Omega - \Delta(k'))^2 + (\kappa/2)^2}\right]. \tag{74}$$

and the frequency dependent mechanical frequency $\delta\omega[\Omega] = -\mathrm{Re}\sum[\Omega]$ and the optical contribution to the damping is $\Gamma_{\mathrm{opt}}[\Omega] = 2\mathrm{Im}\sum[\Omega]$. We define the co-operativity in our system as the ratio of optical contribution to the phonon damping and the instrinsic phonon damping as $\mathcal{C}^{\mathrm{fs}} = \Gamma_{\mathrm{opt}}/\Gamma_m$. The optical contribution to the phonon damping is

$$\Gamma_{\mathrm{opt}}[\Omega] = 2\mathrm{Im}[\sum(\Omega)] = -2|\bar{\alpha}_{k_p}|^2\int\frac{\mathrm{d}k'}{4\pi^2}\,\frac{\kappa/2}{(\Omega - \Delta(k'))^2 + (\kappa/2)^2}|g_m(k_p, k')|^2. \tag{75}$$

$|g_m(k_p, k')|^2$, which is a sinc squared function, is peaked at the Stokes wavevector $k_s = -(\omega_p - \Omega)/v_s$. If we now take the limit of negligible optical loss (i.e. the limit where the sharply peaked Lorentzian

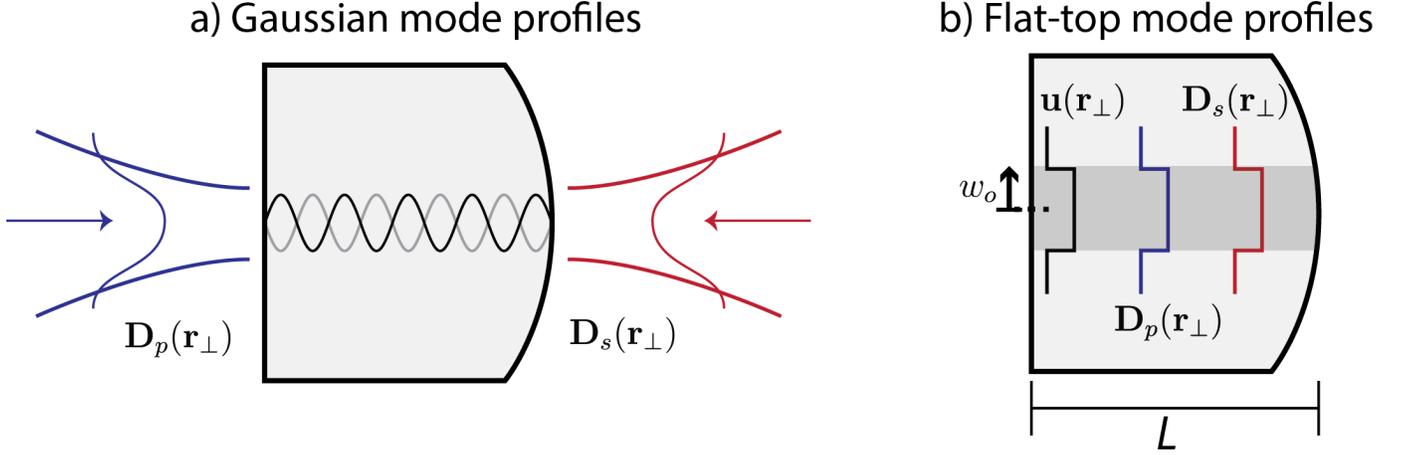

Figure 5: a) Transverse profiles of both the optical and acoustic fields vary with $z$ because of the optical and acoustic diffraction inside the crystal. b) We consider a simpler case (say when the diffraction is weak or the Rayleigh range for both photons and phonons is larger than the crystal length) when the mode profiles are essentially unchanging along $z$.

in Eq.(75) samples the sinc squared function), we obtain

$$\Gamma_{\text{opt}}[\Omega] \approx -2|\bar{\alpha}_{k_p}|^2 |g_m(k_p,k_s)|^2 \int \frac{\mathrm{d}k'}{4\pi^2} \frac{\kappa/2}{(\Omega - v_s k' - v_p k_p)^2 + (\kappa/2)^2} \tag{76}$$

$$= -\frac{|\bar{\alpha}_{k_p}|^2}{2\pi v_{gs}} |g_m(k_p,k_s)|^2 \tag{77}$$

$$= -\frac{P_p}{\hbar \omega_p v_{gp} v_{gs}} |g_m(k_p,k_s)|^2. \tag{78}$$

When the frequency detuning between the pump and Stokes light matches the phonon mode frequency the co-operativity is given by

$$\mathcal{C}^{\text{fs}} = \frac{\Gamma_{\text{opt}}(\Omega_m)}{\Gamma_m} = \frac{P_p L^2 |g_0^m|^2}{\hbar \omega_p v_{gs} v_{gp} \Gamma_m}. \tag{79}$$

It is interesting to note that if we consider an optical cavity of length $L$ with light only making a single pass, standard cavity-optomechanical definition of cooperativity $\mathcal{C}^{\text{om}} = 4n_p |g_0^m|^2 / (\Gamma_m \kappa)$, where $\kappa = 4v_g/L$ and $n_p = P_p L / (\hbar \omega_p v_{gp})$ give us the same co-operativity derived in Eq. (79).

In the next section we find a simple expression for the coupling rate $g_0^m$ and relate it to the coupling rate when we place our bulk crystalline system in an optical cavity in Section 2.5.

## 2.4 Coupling rate: coherent-phonon limit

The coupling rate for electrostriction-mediated coupling between traveling-wave photons and a standing wave phonon is given by the acousto-optic overlap in Eq. (40). However, it is instructive to derive a simple expression for the coupling rate to extract dependencies between coupling rate and material parameters. To this end, we consider the case when mode profiles are essentially invariant along $z$ (say when the diffraction for photons and phonons is weak) (See Fig. 5). We also assume that the pump field, $\mathbf{D}_p$, and Stokes fields, $\mathbf{D}_s$, are co-polarized along the $\hat{x}$-direction, and the phonon modes, $\mathbf{U}_m$ are longitudinally polarized along $\hat{z}$. In this case Eq. (40) reduces to

$$g_0^m \simeq \frac{iq_m}{4\epsilon_o}\sqrt{\hbar\Omega_m\omega_p\omega_s}\int d\mathbf{r}_\perp \left(D_p^x(\mathbf{r})\right)^* D_s^x(\mathbf{r}) p^{1133} U_m^z(\mathbf{r}). \tag{80}$$

In deriving the above expression we assumed that the $\partial_x u_m^z, \partial_y u_m^z \ll q_m u_m^z$. The transverse derivatives are small if the transverse acoustic mode profiles (typically $\sim$ tens of microns for our experimental system) are much larger than the wavelength of phonons ($\sim$sub micron). The relevant photoelastic constant depends on the anisotropy of the crystal as well as the crystal axis along which the longitudinal phonons propagate. For both $z$-cut $\alpha$-quartz and $z$-cut TeO$_2$, with optical fields polarized along the $\hat{x}$ direction, $p^{1133} = p^{13}$ is the relevant photoelastic constant.

Before we use the actual mode profiles of the optical and acoustic fields, we consider a simpler problem by considering modes with a flat-top profile (plane-wave like) with waist $w_o$ (See Fig. 5 b)).

$$D_p^x(x,y) = D_s^x(x,y) = \epsilon_o\epsilon_r E_o \text{Rect}\left(\frac{r}{w_o}\right), \tag{81}$$

$$u_m^z = U_o \text{Rect}\left(\frac{r}{w_o}\right). \tag{82}$$

From the normalization conditions in Eqs. (33), (34) and (38) we arrive at

$$|U_o| = \frac{1}{\Omega_m\sqrt{\rho A L}}, \tag{83}$$

$$|E_o| = \frac{1}{\sqrt{\epsilon_o\epsilon_r A}}. \tag{84}$$

Here, $A = \pi w_o^2$ is the transverse mode area. Since $\omega_p \simeq \omega_s$, $q_m \simeq 2\omega_p/v_o$ and $\epsilon_r = n^2$ we find the approximate (order of magnitude) expression for the coupling rate given by

$$g_0^m \approx i\frac{\omega^2 n^3 p_{13}}{2c}\sqrt{\frac{\hbar}{\Omega_m \rho A L}}. \tag{85}$$

As an example consider a $z$-cut TeO$_2$ crystal with $L = 1$ mm, $A = \pi \times (20\ \mu\text{m})^2$, $n = 2.33$, $p_{13} = 0.34$, $\rho = 6040$ kgm$^{-3}$, $\lambda = 1549$ nm, and $\Omega = 2\pi \times 12$GHz. The coupling rate for this system $|g_0^m| = 2\pi \times 725$ Hz.

For a $z$-cut quartz crystal with $L = 1$ mm, $A = \pi \times (20\ \mu\text{m})^2$, $n = 1.54$, $p_{13} = 0.27$, $\rho = 2648$ kgm$^{-3}$, $\lambda = 1549$ nm, and $\Omega = 2\pi \times 12$ GHz, we find a coupling rate $|g_0^m| = 2\pi \times 250$ Hz. In deriving the coupling rate above we made a simplifying assumption that the mode profiles were invariant in the direction of propagation. In Section 6, we compute $g_0^m$ more rigorously using the full 3-dimensional acoustic and optical mode overlaps.

Note that we are coupling to macroscopic phonon modes with large effective masses [8]. Assuming an effective transverse mode area of $\pi \times (20\ \mu\text{m})^2$ and a length of 1 mm, we find 3.3 $\mu$g as effective mass for a crystalline quartz optomechanical system. In fact, for the quartz crystal in our experiment the effective phonon mode volume is $V_{\text{eff}} = \pi \times (38.3\ \mu\text{m})^2 \times 5$ mm resulting in an effective mass of 61 $\mu$g (See Section 5 for the acoustic mode calculations).

## 2.5 Bulk Crystalline optomechanical system in an optical cavity

Next we look at the case when you place the bulk-crystalline optomechanical system in a Fabry-Pérot optical cavity of length $l$ with pump mode ($\omega_p, k_p$) and Stokes mode ($\omega_s, k_s$) being the resonant optical

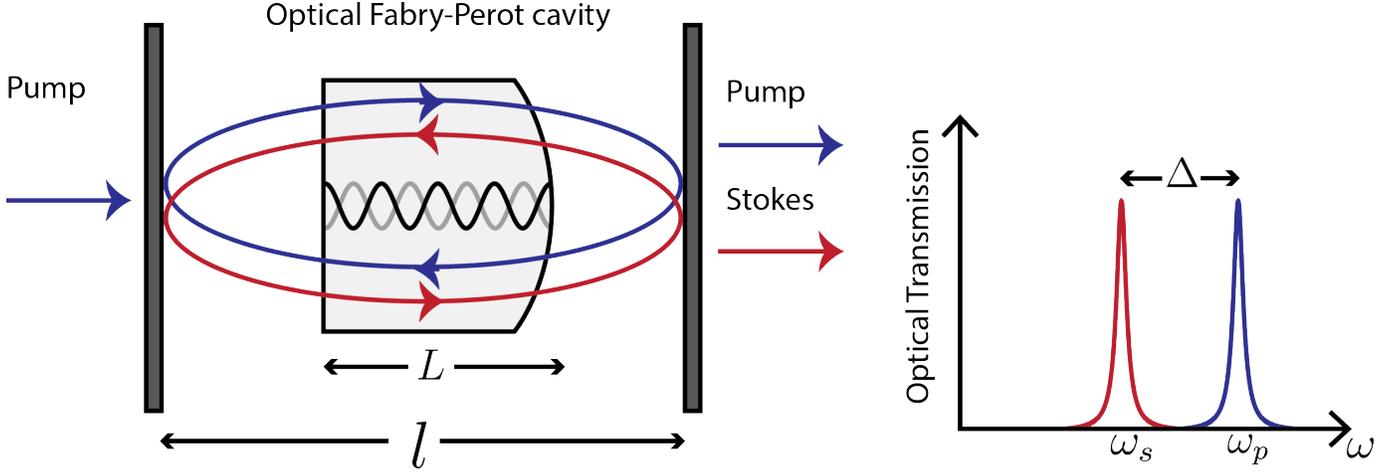

Figure 6: A simple schematic where the bulk-crystalline system is placed in an optical Fabry-Perot cavity with both the pump and the Stokes light as resonant modes.

cavity modes. We look at the interaction of these optical modes with a single longitudinal phonon mode $(\Omega_m, q_m)$ on resonance, meaning both the energy conservation $\Omega_m = \omega_p - \omega_s$ and phase-matching $q_m = k_p - k_s$ are satisfied. The Hamiltonian for this system in the rotating wave approximation from Eqs. (27), (28), and (29) are

$$H^{\text{ph}} = \hbar\Omega_m b_m^\dagger b_m, \tag{86}$$

$$H^{\text{opt}} = \hbar\omega_p a_p^\dagger a_p + \hbar\omega_s a_s^\dagger a_s, \tag{87}$$

$$H^{\text{int}} = \hbar g_{12} a_p^\dagger a_s b_m + \text{H.c.}. \tag{88}$$

If we assume that both the optical and acoustic beams are paraxial (i.e. plane-wave like) having the same cross-sectional area $A$, one finds the following coupling rate for our system

$$g_{12} \simeq \frac{i\omega^2 n^3 p_{13}}{2c}\sqrt{\frac{\hbar}{2\Omega_m \rho AL}}\left(\frac{L}{l}\right). \tag{89}$$

Note that, since the optical cavity length $l$ is different than the acoustic cavity length $L$, the coupling rate acquires an extra factor of $(L/l)$ when we put an optical cavity in our bulk-crystalline system. When the optical cavity is the same length as the crystalline system ($l = L$) we see that the $g_{12}$ derived considering a cavity [5] is similar to that we derived for the case without a cavity up to a constant factor.

Now we derive co-operativity for this optomechanical system consisting of a bulk-crystalline system inside an optical cavity. The Heisenberg equations of motion yield

$$\dot{a}_s(t) = \left(-i\omega_s - \frac{\kappa}{2}\right)a_s(t) - ig_{12}^* a_p(t)b_m^\dagger(t), \tag{90}$$

$$\dot{a}_p(t) = \left(-i\omega_s - \frac{\kappa}{2}\right)a_s(t) - ig_{12} a_s(t)b_m, \tag{91}$$

$$\dot{b}_m(t) = \left(-i\Omega_m - \frac{\Gamma_m}{2}\right)b_m(t) - ig_{12}^* a_s^\dagger(t)a_p(t). \tag{92}$$

17

We have added the optical loss rate $\kappa/2$ and the phonon loss rate $\Gamma_m/2$ phenomenologically. We derive the co-operativity in the limit of weak coupling when the pump is undepleted and the laser driving happens at the pump frequency. It is easier to work in a frame that is rotating at the pump frequency (i.e. $a_p(t) = \bar{\alpha} e^{-i\omega_p t}$ and $a_s(t) = \bar{a}_s(t) e^{-i\omega_s t}$). Note that since the pump field is driven, we assume a constant average field amplitude $\bar{\alpha}_p$. We now look at the dynamics of the Stokes and the phonon field. So, Eqs. (90) and (92) can be written as

$$\dot{\bar{a}}_s(t) = \left(-i\Delta - \frac{\kappa}{2}\right) \bar{a}_s(t) - i g_{12}^* \bar{\alpha}_p b_m^\dagger(t), \tag{93}$$

$$\dot{b}_m(t) = \left(-i\Omega_m - \frac{\Gamma_m}{2}\right) b_m(t) - i g_{12}^* \bar{a}_s^\dagger(t) \bar{\alpha}_p. \tag{94}$$

where $\Delta = \omega_s - \omega_p$. To solve these equations we take the Fourier transforms (i.e. $b_m[\Omega] = \int_{-\infty}^{\infty} e^{i\Omega t} b_m(t)$ and $\bar{a}_s[\Omega] = \int_{-\infty}^{\infty} e^{i\Omega t} \bar{a}_s(t)$) which results in

$$\bar{a}_s[\Omega] = \frac{g_{12}^* \bar{\alpha}_p}{\Omega - \Delta + i\kappa/2} b_m[\Omega], \tag{95}$$

$$b_m[\Omega] = \frac{g_{12}^* \bar{\alpha}_p}{\Omega - \Omega_m + i\Gamma_m/2} a_s^\dagger \, \bar{a}_s^\dagger[\Omega]. \tag{96}$$

Using Eqs. (95) and (96), we can write the effective mechanical susceptibility as $\chi_m^{\text{eff}}(\Omega) = \Omega - \Omega_m + i\Gamma_m/2 + \sum(\Omega)$, where

$$\sum(\Omega) = -|g_{12}|^2 |\bar{\alpha}_p|^2 \left[ \frac{(\Omega - \Delta)}{(\Omega - \Delta)^2 + (\kappa/2)^2} + i \frac{\kappa/2}{(\Omega - \Delta)^2 + (\kappa/2)^2} \right]. \tag{97}$$

So the optical contribution to the mechanical damping is given by

$$\Gamma_{\text{opt}}(\Omega) = 2\text{Im}[\sum(\Omega)] = -2|g_{12}|^2 |\bar{\alpha}_p|^2 \frac{\kappa/2}{(\Omega - \Delta)^2 + (\kappa/2)^2}. \tag{98}$$

Since we assumed that the pump and Stokes waves are detuned to match the phonon resonance (i.e. $\Omega_m = \omega_p - \omega_s$), we obtain the following maximum co-operativity for our cavity opto-mechanical system

$$\mathcal{C}^{\text{om}} = \frac{\Gamma_{\text{opt}}}{\Gamma_m} = 4 \frac{|g_{12}|^2 |\bar{\alpha}_p|^2}{\kappa \Gamma_m}. \tag{99}$$

So far we have derived the equations of motion for our optical fields in $k$-space. However, the optical fields are traveling waves with spatial dependence. So, a more natural treatment of such a traveling wave acousto-optic interaction is in real space using mode envelope operators. In the next two sections we consider a description of the acousto-optic interaction in real space for the case of the coherent-phonon limit as well as the Brillouin limit.

## 2.6 Coherent-phonon limit: real-space description

Traveling wave acousto-optic interaction results in the spatial evolution of optical fields in $z$. In $k$-space this amounts to mode amplitude operators around respective carrier wavevectors being populated. Therefore, the $k$-space optical Hamiltonian in Eq. (32) written in terms of the sum over a continuous set of wavevector captures the spatial evolution of fields due to acousto-optic interaction. However, we

know that the optical mode amplitude operators are sharply peaked about their carrier wave-vectors. So, we define the following mode envelope operators

$$A_p(z) = \int \frac{\mathrm{d}k}{\sqrt{2\pi}} \, a_{pk} e^{i(k-k_p)z}, \tag{100}$$

$$A_s(z) = \int \frac{\mathrm{d}k}{\sqrt{2\pi}} \, a_{sk} e^{i(k+k_s)z}. \tag{101}$$

Taking the Fourier transform to write $a_{pk}$ and $a_{sk}$ in terms of $A_p(z)$ and $A_s(z)$ we can write Eq. (32) in terms of the mode envelope operators as [3, 4, 9]

$$\begin{aligned} H^{\mathrm{opt}} &= \int \mathrm{d}z \, \hbar\omega_p A_p^\dagger(z) A_p(z) - i \int \mathrm{d}z \, \hbar v_o A_p^\dagger(z) \partial_z A_p(z) \\ &+ \int \mathrm{d}z \, \hbar\omega_s A_s^\dagger(z) A_s(z) + i \int \mathrm{d}z \, \hbar v_o A_s^\dagger(z) \partial_z A_s(z), \end{aligned} \tag{102}$$

where we have assumed linear dispersion for optical waves in a bulk medium (i.e $\omega_p(k) = \omega_p(k_p) + v_o(k - k_p)$ and $\omega_p(k) = \omega_p(k_p) - v_o(k - (-k_s))$). Finally, the real-space optical Hamiltonian in Eq. (102) can be written succinctly in terms of the mode envelope operators as

$$H^{\mathrm{opt}} = \int \mathrm{d}z \, \hbar A_p^\dagger(z) \hat{\omega}_{p,z} A_p(z) + \int \mathrm{d}z \, \hbar A_s^\dagger(z) \hat{\omega}_{s,z} A_s(z), \tag{103}$$

where the spatial operators $\hat{\omega}_{p,z} = \omega_p - iv_o\partial_z$ and $\hat{\omega}_{s,z} = \omega_s + iv_o\partial_z$. Note that the electric displacement field in Eq. (31) can be expressed in terms of the mode envelope operators as

$$\mathbf{D}(\mathbf{r}) \simeq \sqrt{\frac{\hbar\omega_p}{2}} \mathbf{D}_p(\mathbf{r}) A_p(z) e^{ik_p z} + \sqrt{\frac{\hbar\omega_s}{2}} \mathbf{D}_s(\mathbf{r}) A_s(z) e^{-ik_s z} + \mathrm{H.c.}. \tag{104}$$

Finally, the interaction Hamiltonian in Eq. (39) can be written in real space in terms of the mode envelope operators as

$$H^{\mathrm{int}} = \sum_m \int \mathrm{d}z \, \hbar g_0^m(z) e^{i(q_m - k_p - k_s)z} A_p^\dagger(z) A_s(z) b_m + \mathrm{H.c.}. \tag{105}$$

In Table 1 we have summarized two equivalent Hamiltonian descriptions for coherent-phonon-limit in the $k$-space as well as in the real space.

The interaction between traveling wave pump photon ($\mathbf{D}_p(\mathbf{r}), k_p, \omega_p$) with counter-propagating Stokes photon ($\mathbf{D}_s(\mathbf{r}), -k_s, \omega_s$) mediated by a standing wave acoustic wave ($\mathbf{U}_m(\mathbf{r}), q_m, \Omega_m$) results in the spatio-temporal evolution of the optical fields. This is easily captured in the real-space description of Hamiltonian, which we will discuss next.

### 2.6.1 Equations of motion: coherent-phonon limit

Using the Heisenberg equations of motion ($\dot{b}_m(t) = -(i/\hbar)[b_m, H]$, and $\dot{A}_\gamma(z,t) = -(i/\hbar)[A_\gamma, H]$) and commutation relations ($[b_m, b_n^\dagger] = \delta_{mn}$, and $[A_\gamma(z,t), A_{\gamma'}^\dagger(z',t)] = \delta_{\gamma\gamma'}\delta(z-z')$), we can write down

| operators | $k$-space | real-space |
|---|---|---|
| $H^{\text{opt}}$ | $\int \mathrm{d}k \; \hbar\omega_p(k) a_{pk}^\dagger a_{pk} + \int \mathrm{d}k \; \hbar\omega_s(k) a_{sk}^\dagger a_{sk}$ | $\int \mathrm{d}z \; \hbar A_p^\dagger(z) \hat{\omega}_{p,z} A_p(z) + \int \mathrm{d}z \; \hbar A_s^\dagger(z) \hat{\omega}_{s,z} A_s(z)$ |
| $H^{\text{ph}}$ | $\sum_m \hbar \Omega_m b_m^\dagger b_m$ | $\sum_m \hbar \Omega_m b_m^\dagger b_m$ |
| $H^{\text{int}}$ | $\sum_m \int \frac{\mathrm{d}k \mathrm{d}k'}{2\pi} \int \mathrm{d}z \; \hbar g_0^m(z) e^{i(k'-k+q_m)z} a_{pk}^\dagger a_{sk'} b_m + \text{H.c.}$ | $\sum_m \int \mathrm{d}z \; \hbar g_0^m(z) e^{i(q_m-k_p-k_s)z} A_p^\dagger(z) A_s(z) b_m + \text{H.c.}$ |

Table 1: Two equivalent representations of the Hamiltonian for the coherent-phonon limit in $k$-space and in real space.

the equations of motion for the phonon mode amplitude operator and the optical mode envelope operators as

$$\partial_t b_m = -i\Omega_m b_m - i \int_0^L \mathrm{d}z \; (g_0^m)^* e^{-i\Delta q z} A_s^\dagger A_p - \Gamma_m/2, \tag{106}$$

$$\partial_t A_p = -i\hat{\omega}_{p,z} A_p - i \sum_m g_0^m e^{i\Delta q z} A_s b_m, \tag{107}$$

$$\partial_t A_s = -i\hat{\omega}_{s,z} A_s - i \sum_m (g_0^m)^* e^{-i\Delta q z} b_m^\dagger A_p. \tag{108}$$

Here, $\Delta q = q_m - k_p - k_s$. Recall, that the spatial operators $\hat{\omega}_{p,z} = \omega_p - iv_o \partial_z$ and $\hat{\omega}_{s,z} = \omega_s + iv_o \partial_z$. In Eq. (106), we phenomenologically added a phonon dissipation rate $\Gamma_m/2$ so that it models all forms of phonon losses such as intrinsic losses in the bulk medium, losses at the surface due to roughness, losses due to diffraction and so on. Finally, after factoring out the fast-oscillating component by letting $\bar{b}_m(t) = b_m(t) e^{i\Omega t}$, where $\Omega = \omega_p - \omega_s$ is the frequency detuning between the pump and the Stokes fields, and $\bar{A}_\gamma(z,t) = A_\gamma(z,t) e^{i\omega_\gamma t}$, we obtain the following spatio-temporal evolution for the phonon mode amplitude operators and the envelope fields

$$\partial_t \bar{b}_m = -i(\Omega_m - \Omega) \bar{b}_m - \frac{\Gamma_m}{2} \bar{b}_m - i \int_0^L \mathrm{d}z \; (g_0^m)^* e^{-i\Delta q z} \bar{A}_s^\dagger \bar{A}_p, \tag{109}$$

$$\partial_t \bar{A}_p + v_o \partial_z \bar{A}_p = -i \sum_m g_0^m e^{i\Delta q z} \bar{A}_s \bar{b}_m, \tag{110}$$

$$\partial_t \bar{A}_s - v_o \partial_z \bar{A}_s = -i \sum_m (g_0^m)^* e^{-i\Delta q z} \bar{b}_m^\dagger \bar{A}_p. \tag{111}$$

We have assumed negligible optical absorption inside the crystalline medium because the optical crystals are transparent. As before, we assume that the mode profiles are uniform along $z$ such that $g_0^m$ is a constant. If we now make the undepleted pump approximation ($A_p(z,t) \approx A_p$) and small signal gain approximation ($\Delta P_s(0) \ll P_s(L)$), Eqs. (109) and (111) result in the following steady-state equations for the phonon mode amplitude and the Stokes field envelope

$$\bar{b}_m \simeq (g_0^m)^* L \frac{\bar{A}_p(0) \bar{A}_s^*(L)}{\left(\Omega - \Omega_m + i\frac{\Gamma_m}{2}\right)} e^{-i\frac{\Delta q L}{2}} \operatorname{sinc}\left(\frac{\Delta q L}{2}\right), \tag{112}$$

$$\bar{A}_s(z) = \bar{A}_s(L) - \sum_m \frac{|g_o^m|^2 L}{v_o} \frac{|\bar{A}_p^*(0)|^2 \bar{A}_s(L)}{\left(\Omega - \Omega_m - i\frac{\Gamma_m}{2}\right)} e^{i\frac{\Delta q L}{2}} \operatorname{sinc}\left(\frac{\Delta q L}{2}\right) \frac{\left(e^{-i\Delta q z} - e^{-i\Delta q L}\right)}{\Delta q}. \tag{113}$$

Eq. (113) can be further simplified to calculate Stokes field envelope at $z = 0$ as

$$\bar{A}_s(0) \approx \bar{A}_s(L) - i \sum_m \frac{|g_0^m|^2 L^2}{v_o} \frac{|\bar{A}_p^*(0)|^2 \bar{A}_s(L)}{\left(\Omega - \Omega_m - i\frac{\Gamma_m}{2}\right)} \operatorname{sinc}^2\left(\frac{\Delta q L}{2}\right). \tag{114}$$

Using Eq. (114) and the expression for the optical power $P_\gamma^{\text{opt}}(z) = \hbar\omega_\gamma v_o A_\gamma^\dagger(z) A_\gamma(z)$, the Stokes power exiting the system at $z = 0$ is given by

$$P_s(0) \approx P_s(L) + P_p(0)P_s(L)\sum_m \frac{4|g_0^m|^2 L^2}{\hbar\omega_p v_o^2 \Gamma_m}\frac{(\Gamma_m/2)^2}{(\Omega - \Omega_m)^2 + (\Gamma_m/2)^2}\text{sinc}^2\left(\frac{\Delta q L}{2}\right) \quad (115)$$

In deriving the above expression, we have ignored higher order terms proportional to $|\bar{g}_o^m|^4 P_p^2 L^4/(\hbar^2\omega^2 v_o^4)$ because we are considering the case of weak signal gain. Note that the optical susceptibility in the weak signal gain regime derived here is the same as that in Eq. (60) using the $k$-space formulation.

In the next section we look at the difference in the acoustic Hamiltonian when the phonon coherence is much shorter than the crystal length. We will derive the familiar dynamics in the Brillouin limit to compare it to the dynamics in the coherent-phonon limit derived above.

## 2.7 Brillouin limit: real-space description

In the low coherence limit for phonons (or the Brillouin limit), where the system length is much longer than coherence length of phonons, the acoustic waves are traveling waves. The acoustic modes of interest are of the form

$$\tilde{\mathbf{u}}_\Lambda(\mathbf{r}) \mapsto \mathbf{U}_\Lambda(\mathbf{r})e^{iqz}. \quad (116)$$

Similar to the Hamiltonian for the traveling optical fields in Eq. (32), the acoustic Hamiltonian for this system can be expressed as

$$H^{\text{ph}} = \int dq\ \hbar\Omega(q) b_q^\dagger b_q, \quad (117)$$

where we have used the following normal mode expansion for the acoustic displacement field

$$\mathbf{u}(\mathbf{r}) = \int \frac{dq}{\sqrt{2\pi}}\sqrt{\frac{\hbar\Omega(q)}{2}} b_q(t) \mathbf{U}(\mathbf{r}) e^{iqz} + \text{H.c.}. \quad (118)$$

Note that although the full Hamiltonian includes sum over many transverse phonon modes, we focus on the dynamics of single acoustic field with dispersion $\Omega(q)$. As before we have assumed that the eigenmode profile $\mathbf{U}(\mathbf{r})$ for the acoustic field in a narrow band about its respective carrier wavevector remains unchanged. The acoustic mode profiles are normalized at each point in $z$ such that

$$\Omega^2 \int d\mathbf{r}_\perp\ \rho(\mathbf{r}) \mathbf{U}^*(\mathbf{r}) \cdot \mathbf{U}(\mathbf{r}) = 1 \quad (119)$$

Defining an acoustic envelope operator $B(z) = 1/\sqrt{2\pi}\int dq\ b_q e^{i(q-q_s)z}$ peaked around the carrier wavevector $q_s = k_p + k_s$, the acoustic Hamiltonian in Eq. (117) in real space can be written similar to Eq. (102) as

$$H^{\text{ph}} = \int dz\ \hbar B^\dagger(z)\hat{\Omega}_z B(z). \quad (120)$$

where the spatial operator $\hat{\Omega}_z = \Omega_s - iv_a\partial_z$ is obtained by assuming linear dispersion (i.e. $\Omega(q) = \Omega(q_s) + (q - q_s)v_a$) of the acoustic field of interest around the carrier wavevector. The acoustic displacement field in terms of the mode envelope operator can be written as

$$\mathbf{u}(\mathbf{r}) \simeq \sqrt{\frac{\hbar\Omega_s}{2}}\mathbf{U}(\mathbf{r})B(z)e^{iq_s z} + \text{H.c.}. \quad (121)$$

| operators | $k$-space | real-space |
|---|---|---|
| $H^{\text{opt}}$ | $\int \mathrm{d}k\ \hbar\omega_p(k)a_{pk}^\dagger a_{pk} + \int \mathrm{d}k\ \hbar\omega_s(k)a_{sk}^\dagger a_{sk}$ | $\int \mathrm{d}z\ \hbar A_p^\dagger(z)\hat{\omega}_{p,z}A_p(z) + \int \mathrm{d}z\ \hbar A_s^\dagger(z)\hat{\omega}_{s,z}A_s(z)$ |
| $H^{\text{ph}}$ | $\int \mathrm{d}q\ \hbar\Omega(q)b_q^\dagger b_q$ | $\int \mathrm{d}z\ \hbar B^\dagger(z)\hat{\Omega}_z B(z)$ |
| $H^{\text{int}}$ | $\int \frac{\mathrm{d}k\mathrm{d}k'\mathrm{d}q}{(2\pi)^{3/2}} \int \mathrm{d}z\ \hbar g_B(z)e^{i(k'-k+q)z}a_{pk}^\dagger a_{sk'}b_q + \text{H.c.}$ | $\int \mathrm{d}z\ \hbar g_B(z)A_p^\dagger(z)A_s(z)B(z) + \text{H.c.}$ |

Table 2: Two equivalent representations of the Hamiltonian for the Brillouin-limit in $k$-space and in real space.

In the Brillouin limit, the interaction Hamiltonian using the normal mode expansion in Eq. (31) and (118) and after making the rotating wave approximation is

$$H^{\text{int}} = \int \frac{\mathrm{d}k\mathrm{d}k'\mathrm{d}q}{(2\pi)^{3/2}} \int \mathrm{d}z\ \hbar g_B(z)e^{i(k'-k+q)z}a_{pk}^\dagger a_{sk'}b_q + \text{H.c.}. \tag{122}$$

where the coupling strength $g_B(z)$ is

$$g_B(z) \simeq \frac{1}{\epsilon_0}\sqrt{\frac{\hbar\omega_p\omega_s\Omega_s}{8}} \int \mathrm{d}\mathbf{r}_\perp (D_p^i(\mathbf{r}))^* D_s^j(\mathbf{r}) p^{ijkl}(\mathbf{r}) \left(\frac{\partial U^k(\mathbf{r})}{\partial r^l} + iq\delta_{lz}U^k(\mathbf{r})\right). \tag{123}$$

We assumed that the coupling strength is constant in the narrow band of wavevectors around the carrier wavevectors. Note that the coupling strength which characterizes the coupling between continuum of modes has dimensions of L$^{1/2}$Hz [9].

Finally, the interaction Hamiltonian for the low coherence limit (Eq. 122) can be written in terms of the mode amplitude operators as

$$H^{\text{int}} = \int \mathrm{d}z\ \hbar g_B(z)A_p^\dagger(z)A_s(z)B(z) + \text{H.c.}. \tag{124}$$

In Table (2) we have summarized the two equivalent Hamiltonian descriptions of traveling-wave photon-phonon interaction in $k$-space and in real space for the Brillouin limit.

The interaction between traveling wave pump photon $(\mathbf{D}_p(\mathbf{r}), k_p, \omega_p)$ with counter-propagating Stokes photon $(\mathbf{D}_s(\mathbf{r}), -k_s, \omega_s)$ mediated by traveling acoustic wave $(\mathbf{U}(\mathbf{r}), q_s, \Omega_s)$ results in a spatio-temporal evolution of these optical and acoustic fields. This is easily captured in the real-space description of Hamiltonian. We will discuss this next.

### 2.7.1 Brillouin limit: Equations of motion

The equations of evolution for the acoustic and optical fields can be calculated using the Heisenberg equations of motion ($\dot{B}(z,t) = -(i/\hbar)[B,H]$, and $\dot{A}_\gamma(z,t) = -(i/\hbar)[A_\gamma,H]$) along with the equal-time commutation relations for the mode envelope operators ($[B(z,t), B^\dagger(z',t)] = \delta(z-z')$ and $[A_\gamma(z,t), A_{\gamma'}^\dagger(z',t)] = \delta_{\gamma\gamma'}\delta(z-z')$) [3]. The spatio-temporal evolution of the envelope fields is [4]

$$\partial_t B(z,t) = -i\hat{\Omega}_z B - ig_B^* A_s^\dagger A_p, \tag{125}$$
$$\partial_t A_p(z,t) = -i\hat{\omega}_{p,z}A_p - ig_B A_s B, \tag{126}$$
$$\partial_t A_s(z,t) = -i\hat{\omega}_{s,z}A_s - ig_B^* B^\dagger A_p. \tag{127}$$

Finally, after factoring out the fast-oscillating component of the envelope field operators by letting $\bar{B}(z,t) = B(z,t)e^{i\Omega t}$, where $\Omega = \omega_p - \omega_s$ is the detuning between the pump and the Stokes light, and $\bar{A}_\gamma(z,t) = A_\gamma(z,t)e^{i\omega_\gamma t}$, we obtain the following spatio-temporal evolution for the mode envelopes

$$\partial_t \bar{B} + v_a \partial_z \bar{B} = -i(\Omega_s - \Omega)\bar{B} - ig_B^* \bar{A}_s^\dagger \bar{A}_p, \tag{128}$$

$$\partial_t \bar{A}_p + v_o \partial_z \bar{A}_p = -ig_B \bar{A}_s \bar{B}, \tag{129}$$

$$\partial_t \bar{A}_s - v_o \partial_z \bar{A}_p = -ig_B^* \bar{B}^\dagger \bar{A}_p. \tag{130}$$

The dissipation rate for the acoustic field, $\Gamma_B/2$, is added phenomenologically in Eq. (128) which results

$$\frac{\partial \bar{B}}{\partial t} + v_a \frac{\partial \bar{B}}{\partial z} = -i(\Omega_s - \Omega)\bar{B} - ig_B^* \bar{A}_s^\dagger \bar{A}_p - \frac{\Gamma_B}{2}\bar{B}, \tag{131}$$

$$\frac{\partial \bar{A}_p}{\partial t} + v_o \frac{\partial \bar{A}_p}{\partial z} = -ig_B \bar{A}_s \bar{B}, \tag{132}$$

$$\frac{\partial \bar{A}_s}{\partial t} - v_o \frac{\partial \bar{A}_s}{\partial z} = -ig_B^* \bar{B}^\dagger \bar{A}_p. \tag{133}$$

These equations are the same as the ones derived classically using nonlinear polarization and density variations due to electrostriction [1, 2]. This type of traveling wave treatment for *both* acoustic and optical fields is appropriate in the classical Brillouin limit. It is important to note that $B^\dagger(z)B(z)$ ($A_\gamma^\dagger(z)A_\gamma(z)$) here corresponds to the phonon (photon) number per unit length. The power in the acoustic (optical) field along $z$ in terms of the envelope operators is given by $P^\mathrm{ph} = \hbar\Omega|v_a|B^\dagger(z)B(z)$ ($P^\mathrm{opt} = \hbar\omega_\gamma|v_o|A_\gamma^\dagger(z)A_\gamma(z)$) [3].

As discussed before in Section 1.1, these equations can be solved in the regime of undepleted pump and weak-signal gain to derive an optical susceptiblity that has a Lorentzian lineshape.

## 3 Experiment

### 3.1 Experimental apparatus

The optomechanical coupling is characterized with the pumb-probe experimental apperatus depicted in Fig. 7. The optical setup consists of fiber and free-space sections, with the fiber displayed in green. The pump and probe beams are delivered from fiber to free-space with fiber-pigtailed collimators before being focused and counter-propagated through the bulk-crystalline sample under test. Half-wave plates are used to match the pump and probe polarization states, which is achieved with a polarizing beam splitter (the splitter is removed during measurement to avoid interferring with counter-propagating light). The sample is mounted on the cold finger of a continuous-flow liquid helium cryostat with a 4 K base temperature and is accessed through two anti-reflection coated fused quartz windows.

The pump and probe beams are both derived from the same amplified narrow-linewidth laser source centered around 1550 nm. The pump has a frequency, $\omega_p$, which is identical to the source laser and provides ∼ 200 mW of power at the crystalline sample. The probe frequency, $\omega_{pr}$, is created by first phase modulating the laser source to create new frequencies, and then by applying a fiber-Bragg grating filter to select a single desired frequency. Since the phase modulator can be driven with an arbitrary frequency up to 20 GHz, this technique allows for the probe frequency to be swept through the acousto-optical resonance under investigation. The probe power is designed to provide

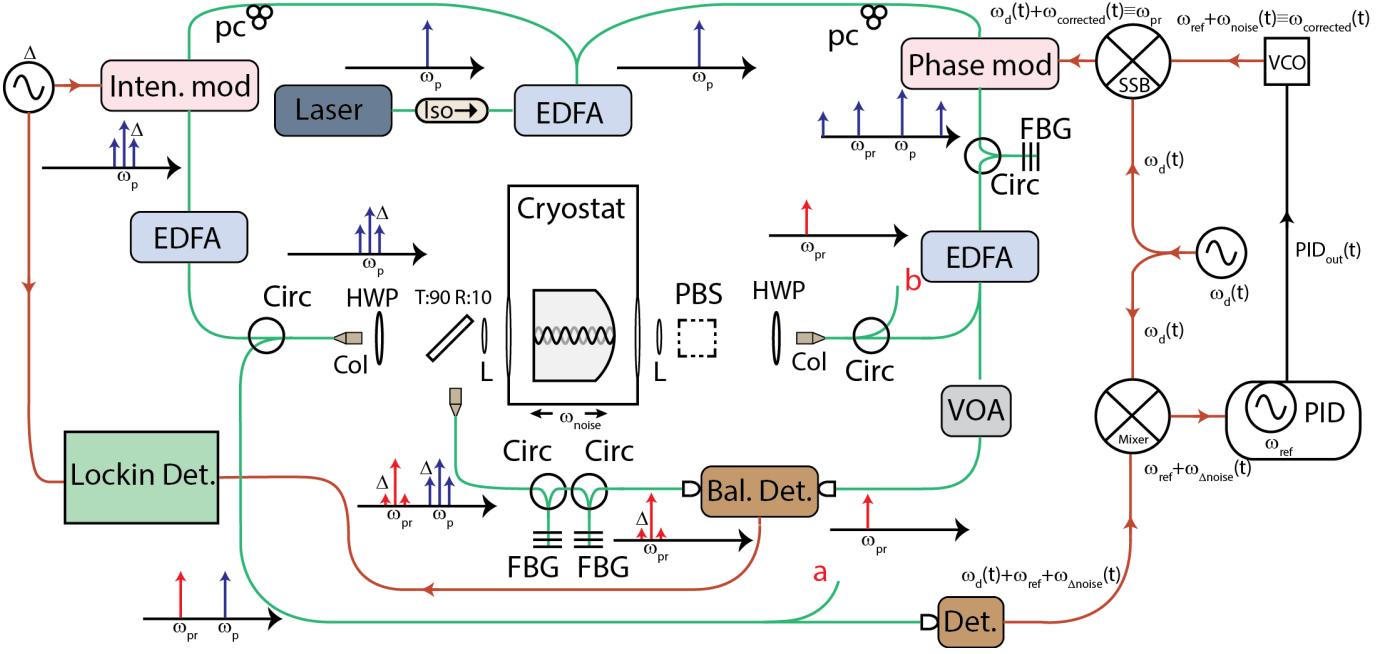

Figure 7: Experimental apparatus for sensitive detection of the bulk crystalline optomechanical susceptibility. Green lines represent optical fiber and red lines indicate RF cables. pc: polarization controller; Xol: collimator; iso: isolator; EDFA: erbium-doped fiber amplifier; HWP: half-wave plate; PBS: polarizing beam splitter; L: lens; det: detector; FBG: fiber bragg grating; bal det: balanced detector; Inten. mod: intensity modulator; phase mod: phase modulator.

the maximum power that can be measured by the detector without damage, which results in $\sim 40$ mW at the crystalline sample.

The optomechanically-amplified probe signal is reflected from a 90:10 dichroic mirror to an output collimator. Since this output collimator also contains light from pump back-reflected from the crystal, two additional fiber-bragg grating filters are used to remove this unwanted light. The resultant signal is input to one port of a balanced detector, where the other is derived from the probe before amplification. This probe reference arm is power matched to the signal arm through the use of a variable optical attenuator. The output of the balanced detector is measured with a lock-in detector.

To enhance the sensitivity of the apparatus, lock-in detection is used. For this purpose, the pump light is modulated with an optical intensity modulator. This modulation frequency is driven by the lock-in detector's RF frequency generator. This is the same frequency that is measured on the output signal through lock-in detection. For a complete description of how the pump modulation frequency becomes imprinted on the measured amplified probe for the two limits of operation, see Section 4.

Neglecting the phase noise cancelling circuit (see details regarding this in the subsection below), the probe frequency is given by $\omega_{pr} = \omega_p - \omega_d$. $\omega_d$ is repeatedly swept through the optomechanical frequency, $\Omega_B$, at a period, $T_{\text{sweep}} \sim 1$ second. Finally, the lock-in detector output is measured with an oscilloscope triggered with period $T_{\text{sweep}}$ and the accumulating swept optomechanical signal is averaged for $\sim 1$ minute. This signal corresponds to the optomechanical susceptibility of the system. Typical measurement results are displayed in Fig. 5 of the main manuscript.

## 3.2 Alignment technique

For maximum optomechanical interaction, the pump, probe, and acoustic mode all must overlap. For this to occur, the optical beams must be aligned to each other, as well as to the preferred axis of the shaped crystalline sample. To this end, a circulator is place before both the pump and the probe collimators, and the ports labeled **a** and **b** are observed during alignment. Several procedures are followed prior to each measurement.

First, the pump beam is aligned to the plano-convex crystal. The beam is aligned to the flat surface normal by maximizing the reflection observed at point **a**. The beam is then centered on the crystal by maximizing the reflection from the second curved surface as observed through point **a**. To confirm that the pump beam is well aligned to the crystal, a tunable laser is swept through the pump collimator in order to confirm the existence of an optical Fabrey-Pérot cavity from the crystal.

Second, the probe beam is aligned to the pump beam, and therefore also to the crytstal. This is done by maximizing the power observed at point **b**. Directly, this indicates that the pump light is optimally coupled into the probe collimator. Indirectly, this also ensures that the probe light overlaps with the pump light.

Finally, after aligning the polarization state of each beam, the output collimator is aligned by maximizing the received power from the probe collimator. These procedures are repeated as necessary before measurement. For example, the setup is realigned if the temperature changes or slow alignment drifts accumulate. To accurately predict the experimental results, despite the nominally perfect alignment, the actual beam profiles are nonetheless measured as detailed in Section [6].

## 3.3 Phase-noise cancelling

If the crystal were stationary, the relative frequency between the pump and the probe would be precise. As a result, the linewidth of the optomechanical response could be measured with excellent precision. In practice, however, the crystal is attached to the long arm of a cryostat and is highly sensitive to the vibrations in the environment. Therefore, as these vibrations translate to phase noise on the optical tones, they place a lower limit to the measurable linewidths of the optomechanical response. Experimentally, in this system, this limit corresponds to $\sim 1$ kHz. In order to surpass this limit, we developed a phase-noise canceling technique.

Phase noise is accounted for through the use of a feedback loop and a proportional integral derivative (PID) controller. The ability to control the noise lies in the fact that the pump light reflecting from the vibrating crystal acquires the same phase noise as the optomechanical response. Therefore, we use the back-reflecting pump light in order to provide an error signal to our PID circuit. The complete circuit is depicted in Fig. 7.

The new probe frequency consists of a swept signal, $\omega_d$, in addition to a varying frequency, $\omega_{\text{corrected}}$, which stems from the vibrations of the crystal. Specifically, a single sideband mixer mixes $\omega_d$ with the output of a voltage controlled oscillator (VCO) that is controlled by the PID output. The error frequency for the PID is given by the beat frequency of the back-reflected pump and the probe, after mixing with $\omega_d$. This error frequency is compared with a stable reference frequency, $\omega_{\text{ref}}$, to generate the phase difference as the error signal for the VCO drive. When the PID loop is activated the new probe has an identical frequency response to that of the back-reflecting pump. The resulting linewidth resolution that we have measured is $\sim 300$ Hz, corresponding to our narrowest linewidth. However, from our detection limited beat frequency between the back-reflecting pump and the probe, we expect the measurement scheme to have a linewidth resolution limit of sub Hz.

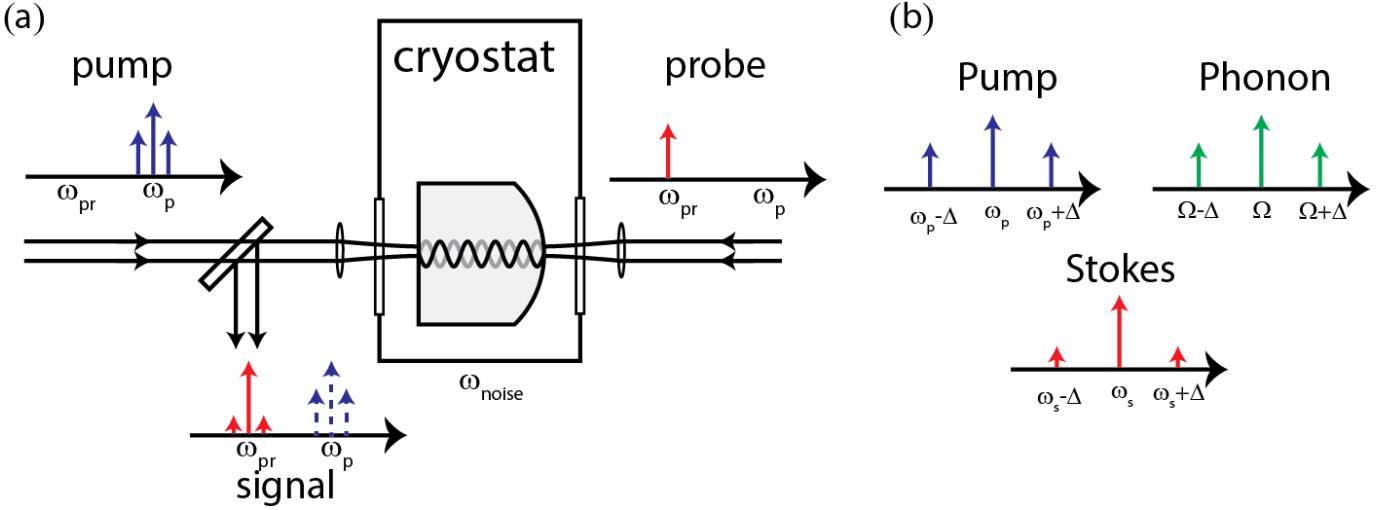

Figure 8: (a) Overview of relevant optical tones. (b) Participating optical and acoustic tones.

## 4 Measurement Theory

As discussed in Section 1 and Section 2, the spectrum of the Stokes signal gain has both information about the optomechanical coupling strength as well as the phonon dissipation rate. The fractional change in Stokes signal as it passes through most crystals is small. To perform zero-background measurement of the small-signal gain (as low as $\Delta P_s/P_s \sim 10^{-6}$), we modulate the pump light and use lock-in detection of the modulated Stokes gain signal. In this section we describe the measurement theory in both the classical Brillouin limit (high temperature) and the coherent-phonon limit (low temperature).

### 4.1 Measurement theory in the low-coherence limit

In this section, we develop measurement theory that describes stimulated Brillouin scattering in the low-coherence limit when the pump wave consists of three modes (i.e. pump wave is modulated). While we can generalize the Hamiltonian treatment in Section 2.7 to include multiple pump fields, for simplicity, we use a classical treatment using coupled nonlinear electromagnetic wave equation and the elastic wave equation. Our treatment is an extension to the classical Brillouin treatment with the pump wave consisting of two modes [10]. We consider the case when the pump modulation frequency, $\Delta$, is much smaller than the Brillouin frequency ($\Delta \ll \Omega_B$). We assume here that the forward propagating pump waves and backward propagating Stokes wave propagate in the $z$-axis and are both co-polarized along the $x$-axis of the medium. We assume that the acoustic wave, which propagates along positive $z$-axis, is longitudinal and polarized along the direction of propagation. For simplicity we treat the modes as plane-waves with slowly varying envelopes depending only on the longitudinal coordinate. Optical fields in the experiment have transverse mode profiles. This results in an effective acousto-optic area, $A^{ao}$.

Let us look at the interaction of the incident pump waves and scattered Stokes waves in this medium which are given by

$$\mathbf{E}_p(z,t) = [E_{p-1}(z)e^{i\Delta(t-z/v_o)} + E_{p0}(z) + E_{p+1}(z)e^{-i\Delta(t-z/v_o)}]e^{i(k_p z - \omega_p t)}\hat{x} + \text{c.c.}, \quad (134)$$

$$\mathbf{E}_s(z,t) = [E_{s-1}(z)e^{i\Delta'(t+z/v_o)} + E_{s0}(z) + E_{s+1}(z)e^{-i\Delta'(t+z/v_o)}]e^{i(-k_s z - \omega_s t)}\hat{x} + \text{c.c.} \quad (135)$$

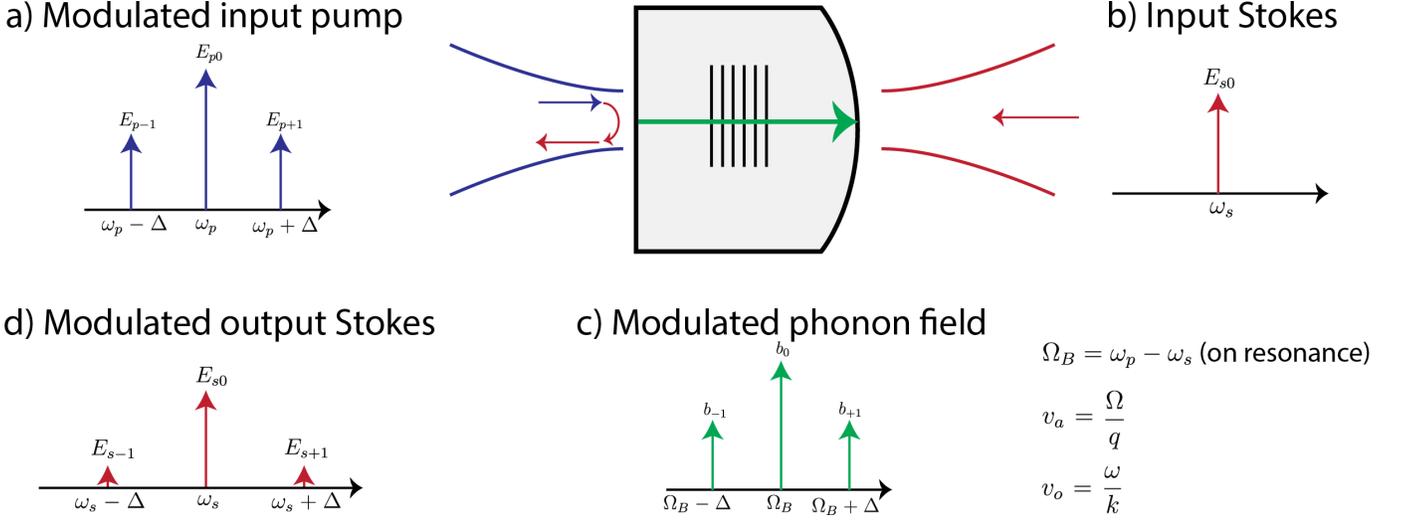

Figure 9: A schematic of different waves involved in the measurement of the Brillouin gain signal at room temperature. The input pump field is modulated whereas the Stokes field is not modulated. The beating of input pump- and Stokes- fields results in a modulated phonon field. The back-scattered Stokes light from this modulated phonon field has side-bands even though the input Stokes light is single-toned.

where $v_o$ is the velocity of the light in the medium, and $k_p$ and $-k_s$ are the wave-vectors for the pump and the Stokes light (See Fig. 9). We have assumed a linear dispersion, $\omega(k) = v_o k$, for light in the bulk medium. The frequency separation for the modulated Stokes tones for the strongest resonant interaction is $\Delta' = \Delta(1 - 2(v_o/v_a))$ [10]. Since, $v_o/v_a \sim 10^{-5}$, $\Delta' \simeq \Delta$ to an excellent approximation. Since the input Stokes light in the experiment is not modulated (i.e. $E_{s-1}(L) = E_{s-1}(L) = 0$) the acoustic wave generated by the beating of the three pump waves and a single Stokes wave to leading order in frequency is given by

$$\mathbf{u}(z,t) = [b_{-1}(z)e^{i\Delta(t-z/v_a)} + b_0(z) + b_{+1}(z)e^{-i\Delta(t-z/v_a)}]e^{i(q_0 z - \Omega_B t)}\hat{z} + c.c. \tag{136}$$

Here, $v_a$ is the velocity of sound in a bulk medium, $\Omega_B$ is the resonant Brillouin frequency, and $q_0 = k_p - (-k_s)$ is the acoustic wavevector. We assumed a linear dispersion, $\Omega(q) = v_a q$, for sound in the bulk medium. The interaction between $E_{p0}$ and $E_{s0}$ generates the acoustic wave $b_0$, whereas the cross interactions of $E_{p-1}$ and $E_{s0}$ and $E_{p+1}$ and $E_{s0}$ generates $b_{-1}$ and $b_{+1}$ respectively. After making the slowly varying envelope approximation and taking steady-state limit, we can derive the following coupled mode equations for the acoustic fields and the Stokes fields using the nonlinear electromagnetic wave equation and the elastic wave equation [2]

**Equations of motion for the acoustic fields:**

$$\frac{\partial b_{-1}(z)}{\partial z} + \frac{\Gamma_B}{2v_a}b_{-1}(z) = -\frac{\chi_1}{v_a}E_{p-1}(z)E_{s0}^*(z)e^{i\Delta z/v_a}, \tag{137}$$

$$\frac{\partial b_0(z)}{\partial z} + \frac{\Gamma_B}{2v_a}b_0(z) = -\frac{\chi_1}{v_a}E_{p0}(z)E_{s0}^*(z), \tag{138}$$

$$\frac{\partial b_{+1}(z)}{\partial z} + \frac{\Gamma_B}{2v_a}b_{+1}(z) = -\frac{\chi_1}{v_a}E_{p+1}(z)E_{s0}^*(z)e^{-i\Delta z/v_a}. \tag{139}$$

27

Here, $\Gamma_B$ is the dissipation rate of the acoustic waves, and $\chi_1 = \epsilon_o n^4 p_{13} q_0/(2\Omega_0 \rho)$, where $n$ is the refractive index, and $\rho$ is the density of the medium.

**Equations of motion for the Stokes fields:**

$$\frac{\partial E_{s-1}(z)}{\partial z} = \chi_2 \left( b_0^*(z) E_{p-1}(z) e^{-\frac{2i\Delta z}{v_o}} + b_{+1}^*(z) E_{p0}(z) e^{-\frac{i\Delta z}{v_a}} \right), \tag{140}$$

$$\frac{\partial E_{s0}(z)}{\partial z} = \chi_2 \left( b_{-1}^*(z) E_{p-1}(z) e^{\frac{i\Delta z}{v_a}} + b_0^*(z) E_{p0}(z) + b_{+1}^*(z) E_{p+1}(z) e^{-\frac{i\Delta z}{v_a}} \right), \tag{141}$$

$$\frac{\partial E_{s+1}(z)}{\partial z} = \chi_2 \left( b_0^*(z) E_{p+1}(z) e^{\frac{2i\Delta z}{v_o}} + b_{-1}^*(z) E_{p0}(z) e^{+\frac{i\Delta z}{v_a}} \right). \tag{142}$$

Here, $\chi_2 = q_0 n^2 p_{13} \omega_s/(2v_o)$.

In the low coherence limit (i.e at room temperatures) the coherence length ($\propto v/\Gamma_B$) of the phonons is much shorter than both the crystal length (i.e. $(\Gamma_B/v)L \gg 1$) and the spatial variation of the envelope fields. Therefore, Eq. (137-139) when further simplified become

$$b_{-1}(z) = -\frac{\chi_1}{\left(\frac{\Gamma_B}{2} + i\Delta\right)} E_{p-1}(z) E_{s0}^* e^{\frac{i\Delta z}{v_a}}, \tag{143}$$

$$b_0(z) = -\frac{2\chi_1}{\Gamma_B} E_{p0}(z) E_{s0}^*, \tag{144}$$

$$b_{+1}(z) = -\frac{\chi_1}{\left(\frac{\Gamma_B}{2} - i\Delta\right)} E_{p+1}(z) E_{s0}^* e^{\frac{-i\Delta z}{v_a}}. \tag{145}$$

It is important to note that acoustic side-tones $b_{-1}$ and $b_{+1}$ have Lorentzian responses as a function of the pump modulation frequency $\Delta$. Now substituting Eqs. (143-145) in Eqs. (140-142), we obtain the following equations for the amplitudes of the Stokes field

$$\frac{\partial}{\partial z} \begin{bmatrix} E_{s-1} \\ E_{s0} \\ E_{s+1} \end{bmatrix} = -\chi_1 \chi_2 \begin{bmatrix} \frac{2}{\Gamma_B} E_{p0}^* E_{s0} E_{p-1} e^{-\frac{2i\Delta z}{v_o}} + \frac{1}{\left(\frac{\Gamma_B}{2}+i\Delta\right)} E_{p+1}^* E_{s0} E_{p0} \\ \frac{1}{\left(\frac{\Gamma_B}{2}-i\Delta\right)} |E_{p-1}|^2 E_{s0} + \frac{2}{\Gamma_B} |E_{p0}|^2 E_{s0} + \frac{1}{\left(\frac{\Gamma_B}{2}+i\Delta\right)} |E_{p+1}|^2 E_{s0} \\ \frac{2}{\Gamma_B} E_{p0}^* E_{s0} E_{p+1} e^{+\frac{2i\Delta z}{v_o}} + \frac{1}{\left(\frac{\Gamma_B}{2}-i\Delta\right)} E_{p-1}^* E_{s0} E_{p0} \end{bmatrix}. \tag{146}$$

In the limit $\Delta \gg \Gamma_B$, Eq. (146) reduces to

$$\frac{\partial}{\partial z} \begin{bmatrix} E_{s-1} \\ E_{s0} \\ E_{s+1} \end{bmatrix} \approx -\chi_1 \chi_2 \begin{bmatrix} \frac{2}{\Gamma_B} E_{p0}^* E_{s0} E_{p-1} e^{-\frac{2i\Delta z}{v_o}} \\ \frac{2}{\Gamma_B} |E_{p0}|^2 E_{s0} \\ \frac{2}{\Gamma_B} E_{p0}^* E_{s0} E_{p+1} e^{+\frac{2i\Delta z}{v_o}} \end{bmatrix}. \tag{147}$$

Therefore, even when the modulation frequency is larger than the linewidth of the Brillouin resonace, there is a non-zero gain for the Stokes side-bands $E_{s-1}$ and $E_{s+1}$. Since the Stokes light in the experiment is not modulated (i.e. $E_{s-1}(L) = E_{s+1}(L) = 0$), the change in Stokes field $E_{s-1}$ (or $E_{s+1}$) allows zero-background measurement of the coupling strength. Note that $\chi_1 \chi_2$ can be related to the Brillouin gain coefficient derived in Eq. (7) as

$$G_{\text{SBS}} = \frac{2\chi_1 \chi_2}{A \epsilon_o nc \Gamma_B}. \tag{148}$$

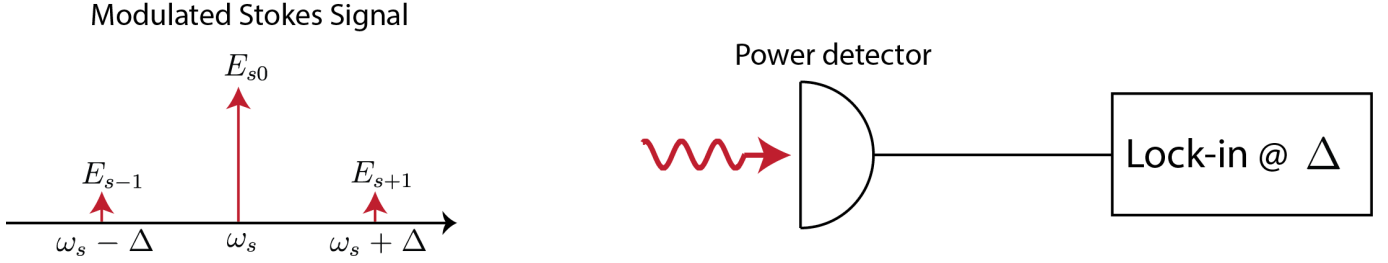

Figure 10: Lock-in detection of the Stokes gain.

Let us assume that $|E_{p+1}(z)| = |E_{p-1}(z)|$ (i.e. pump side-bands have same field strength) and $E_{s-1}(L) = 0$ (i.e. input Stokes wave is not modulated) then Eq. (146) can be solved for $E_{s-1}$ in the weak gain limit. Thus,

$$E_{s-1}(0) = \chi_1 \chi_2 L \left[ \frac{2}{\Gamma_B} E_{p0}^*(0) E_{s0}(L) E_{p-1}(0) + \frac{1}{\frac{\Gamma_B}{2} + i\Delta} E_{p+1}^*(0) E_{s0}(L) E_{p0}(0) \right]. \tag{149}$$

Since our modulation frequency $\Delta$ is of the order of 10 MHz we assume that $v_o/\Delta \gg L$ (or $e^{-\frac{2i\Delta L}{v_o}} \approx 1$) in deriving the equation above. Eq. (149) written in terms of the optical intensity, $I(z) = 2n\epsilon_0 c E^*(z) E(z)$ [1], can be used to calculate Stokes side-band intensity

$$I_{s-1}(0) = \frac{\chi_1^2 \chi_2^2 L^2}{(2n\epsilon_0 c)^2} \left| \frac{2}{\Gamma_B} + \frac{1}{\frac{\Gamma_B}{2} + i\Delta} \right|^2 I_{p0}(0) I_{p-1}(0) I_{s0}(L) \tag{150}$$

$$= G_{\text{SBS}}^2 A^2 L^2 \left( \frac{\Gamma_B^2 + \Delta^2}{\Gamma_B^2 + 4\Delta^2} \right) I_{p0}(0) I_{p-1}(0) I_{s0}(L). \tag{151}$$

The side-band Stokes gain is a second order process as it depends on $G_{\text{SBS}}^2$. In our measurement we use lock-in detection to measure the heterodyne beat-tone between $E_s(0)$ and $E_{s-1}(0)$ at pump modulation frequency $\Delta$. The measured heterodyne signal then becomes a direct measure of $G_{\text{SBS}}$. We explain this in the next subsection.

The optical intensity is given by $I = P/A$, where $P$ is the optical power for a mode with area $A$. So, we can rewrite Eq. (151) as

$$P_{s-1}(0) = G_{\text{SBS}}^2 L^2 \left( \frac{\Gamma_B^2 + \Delta^2}{\Gamma_B^2 + 4\Delta^2} \right) P_{p0}(0) P_{p-1}(0) P_{s0}(L). \tag{152}$$

From Eq. (152) we see that $P_{s-1}$ is largest for $\Delta \to 0$. To maximize signal-to-noise ratio at room-temperature we modulate the pump such that $\Delta \ll \Gamma_B$ (See Fig. 11) In this limit, the side-band Stokes power is given by

$$P_{s-1}(0) \simeq G_{\text{SBS}}^2 L^2 P_{p0}(0) P_{p-1}(0) P_{s0}(L). \tag{153}$$

### 4.1.1 Lock-in detection

The power measured by an optical detector is proportional to the square of the optical fields (i.e. $P \propto \langle E^* E \rangle$). Therefore, when a Stokes field with two side-bands separated by $\Delta$ is measured (See Fig. 10) using a power detector we observe

$$P_s(\text{detector}) = P_{s0} + P_{s-1} + P_{s+1} + 2\sqrt{P_{s0} P_{s-1}} \cos(\Delta t) + 2\sqrt{P_{s0} P_{s+1}} \cos(\Delta t) + 2\sqrt{P_{s+1} P_{s-1}} \cos(2\Delta t). \tag{154}$$

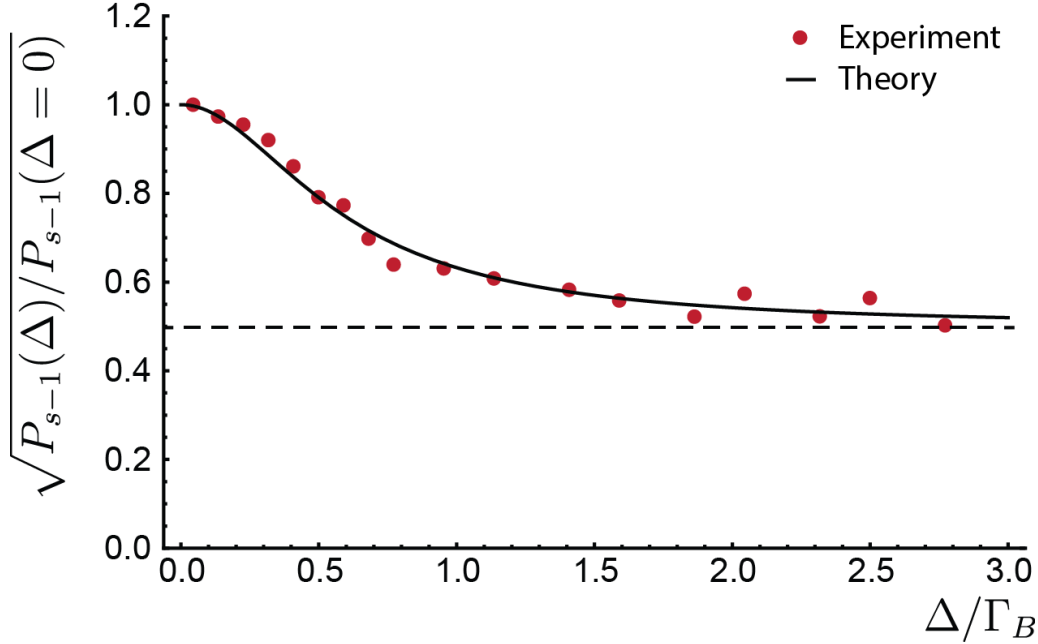

Figure 11: Peak gain for $TeO_2$ at room-temperature measured using heterodyne lock-in detection as a function of $\Delta/\Gamma_B$. $\Gamma_B \simeq 11$ MHz obtained by fitting Eq. (152) agrees with the dissipation rate for phonons in $TeO_2$ at room-temperature.

In our experiments, $P_{s-1} = P_{s+1} \ll P_{s0}$. Furthermore, in the small-gain limit $\Delta P_{s0}(0) \ll P_{s0}(L)$ is also an excellent approximation. By demodulating the detected power at frequency $\Delta$ we obtain the following expression for the power detected at the lock-in amplifier

$$P_s(\text{heterodyne}) = 4\sqrt{P_{s0} P_{s-1}} \tag{155}$$

Now substituting Eq. (152) in Eq. (155) we arrive at the following expression for the power detected in the lock-in measurement

$$P_s(\text{heterodyne}) = 4 G_{\text{SBS}} L P_{s0} \sqrt{P_{p0} P_{p-1}} \tag{156}$$

Therefore, the heterodyne measurement of the zero-background side-band Stokes tone using the lock-in detection gives a direct measure of the coupling strength. Next we discuss, measurement theory in the high-coherence limit.

## 4.2 Measurement theory in the high-coherence limit

In this section, we discuss measurement theory in the long phonon coherence limit for phonons when the pump wave consists of three modes (i.e. the pump wave is modulated.) We derive a theory for the case when the modulation frequency, $\Delta$, is much larger than the phase matching bandwidth. We do this because modulation within the phase-matching bandwidth results in a more complicated spectrum of Stokes scattering that will have overlapping contributions from the three pump tones. For our analysis, we consider a simple case where forward propagating pump wave, $D_p(\mathbf{r},t)$, and a backward propagating Stokes wave, $D_s(\mathbf{r},t)$, is in resonance with *one* standing wave phonon mode $u_0(\mathbf{r},t)$ (i.e. $\omega_p - \omega_s = \Omega_m$) at the center of the phase-matching bandwidth (See Fig. 12) . The pump wave is modulated at frequency $\Delta$ so it consists of side-bands $D_{p-1}(\mathbf{r},t)$ $(D_{p+1}(\mathbf{r},t))$ at $\omega_p - \Delta$

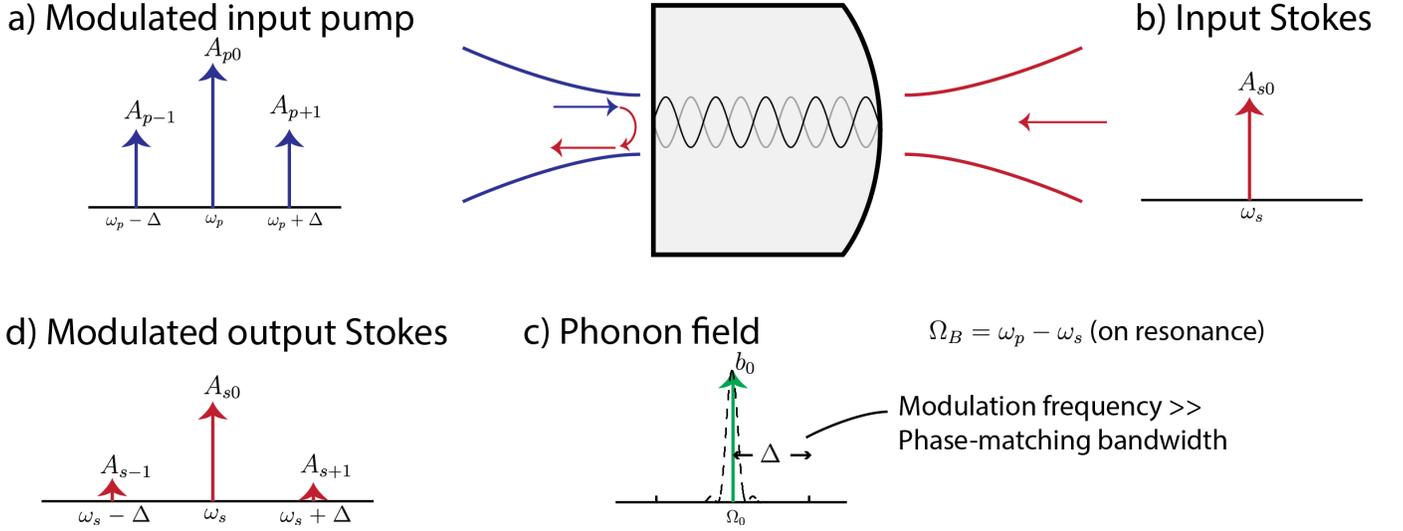

Figure 12: A schematic of different waves involved in the measurement of Brillouin gain signal at low temperature. When the pump modulation is much larger than the phase matching bandwidth the side-bands on the phonon field are vanishingly small.

$(\omega_p + \Delta)$. The phonon mediated resonant energy transfer between the modulated pump wave and the Stokes wave results in two side-bands for the Stokes waves at $D_{s-1}(\mathbf{r}, t)$ ($D_{s+1}(\mathbf{r}, t)$) at $\omega_s - \Delta$ ($\omega_s + \Delta$). The Hamiltonian for this system is given by

$$H^{\text{ph}} = \hbar\Omega_0 b_0^\dagger b_0, \tag{157}$$

$$H^{\text{opt}} = \sum_\gamma \int_0^L dz\, \hbar A_\gamma^\dagger \hat{\omega}_{\gamma,z} A_\gamma, \text{ where } \gamma = p-1, p, p+1, s-1, s, s+1, \tag{158}$$

$$H^{\text{int}} = \int_0^L dz g_0 e^{i(q_0 - \Delta k)z} A_{p0}^\dagger A_{s0} b_0 + \int_0^L dz g_0 e^{i(q_0 - \Delta k_{-1})z} A_{p-1}^\dagger A_{s-1} b_0 + \int_0^L dz g_0 e^{i(q_0 - \Delta k_{+1})z} A_{p+1}^\dagger A_{s+1} b_0. \tag{159}$$

Here, $b_0(t)$ is the mode amplitude operator for the standing wave phonon mode, and $A_\gamma(z, t)$ is the mode envelope operator for the optical field, and $\Delta k_{-1} = k_{p-1} - (-k_{s-1})$ and $\Delta k_{+1} = k_{p+1} - (-k_{s+1})$. We assumed that the center pump and Stokes tones are in resonance (i.e. $q_0 - (k_p + k_s) = 0$) so $q_0 - \Delta k_{-1} = 2\Delta/v_o$ and $q_0 - \Delta k_{+1} = -2\Delta/v_o$.

Following the approach outlined in Section [2.6] and using the Hamiltonian in Eqs.(157)-(159) one can derive the following equations of motion for the phonon field and the Stokes fields

$$\partial_t b_0 = -i\Omega_0 b_0 - \frac{\Gamma_0}{2} b_0 - i\int_0^L dz g_0 \left(A_{s0}^\dagger A_{p0} + e^{-\frac{2i\Delta}{v_o}z} A_{s-1}^\dagger A_{p-1} + e^{\frac{2i\Delta}{v_o}z} A_{s+1}^\dagger A_{p+1}\right), \tag{160}$$

$$\partial_t A_{s0} = -i\hat{\omega}_{s,z} A_{s0} - ig_0 b_0^\dagger A_{p0}, \tag{161}$$

$$\partial_t A_{s-1} = -i\hat{\omega}_{s-1,z} A_{s-1} - ig_0 e^{-\frac{2i\Delta}{v_o}z} b_0^\dagger A_{p-1}, \tag{162}$$

$$\partial_t A_{s+1} = -i\hat{\omega}_{s+1,z} A_{s+1} - ig_0 e^{\frac{2i\Delta}{v_o}z} b_0^\dagger A_{p+1}. \tag{163}$$

As before we phenomenologically added the phonon dissipation rate $\Gamma_0/2$ and assumed the optical losses inside the crystal are negligible. Note that for our experiments in the high-coherence limit we

modulate the pump such that the intensity beat length is much longer than the crystal length (i.e. $v_o/(\Delta) \gg L$). This means that $e^{\frac{2i\Delta}{v_o}z} \approx 1$ to an excellent approximation. As before, after factoring out the fast-oscillating frequency components, we obtain the following steady state equations for the phonon field and the Stokes side-bands

$$\bar{b}_0 \simeq -\frac{2i}{\Gamma_0} \int_0^L dz\, g_0^* \bar{A}_{s0}^\dagger \bar{A}_{p0}, \tag{164}$$

$$\frac{\partial \bar{A}_{s0}}{\partial z} \simeq g_0^* \bar{b}_o^\dagger \bar{A}_{p0}, \tag{165}$$

$$\frac{\partial \bar{A}_{s-1}}{\partial z} \simeq g_0^* \bar{b}_o^\dagger \bar{A}_{p-1}, \tag{166}$$

$$\frac{\partial \bar{A}_{s+1}}{\partial z} \simeq g_0^* \bar{b}_o^\dagger \bar{A}_{p+1}. \tag{167}$$

In writing Eq. (164), we assumed that in the undepleted pump regime with weak Stokes gain the phonon mode, $b_0$ is primarily driven by the beating of the center pump tone, $E_{p0}$ and the center Stokes tone, $E_{s0}$. In this limit, the phonon amplitude in the crystal and Stokes fields at $z = 0$ is given by

$$b_0 = -\frac{2i}{\Gamma_0} A_{p0}(0) A_s(L)^* L, \tag{168}$$

$$A_{s0}(0) = -\frac{2i|g_0|^2 L^2}{\Gamma_0 v_o} A_{p0}(0)^* A_{p-1}(0) A_s(L), \tag{169}$$

$$A_{s-1}(0) = -\frac{2i|g_0|^2 L^2}{\Gamma_0 v_o} A_{p0}(0)^* A_{p-1}(0) A_s(L). \tag{170}$$

The power in the optical field in terms of the envelope fields is

$$P_\gamma^{\text{opt}} = \hbar \omega v_o A_\gamma^\dagger(z)_\gamma A(z). \tag{171}$$

Finally, using Eq. (155), the zero-background heterodyne power for our experiments in the high coherence limit is

$$P_s(\text{heterodyne}) = \frac{8}{\hbar \omega v_o^2 \Gamma_0} |g_0|^2 L^2 \sqrt{P_p(0) P_{p-1}(0)} P_s(L). \tag{172}$$

# 5   Acoustic Modes in an anisotropic medium

For a crystalline medium with flat faces at low temperatures, phonons reflect of the edges to form standing acoustic modes much like the optical modes in a Fabry-Pérot resonator. Not surprisingly, these acoustic modes suffer from diffraction losses (See Fig. 13). However, the boundaries of crystalline medium can be shaped to "trap" these longitudinal acoustic modes in a crystalline medium much like using curved mirrors to confine light in a laser cavity.

A rich spectrum of standing-wave acoustic modes exists when the boundaries of a crystalline medium are shaped to confine longitudinal acoustic phonons. For such systems, the actual set of standing wave phonon modes for a crystalline medium depends on the geometry (i.e. the shape of boundaries) and also on the cut and anisotropy of the crystal. In this section, we will discuss an approach to finding the bulk-acoustic modes in a shaped crystalline medium.

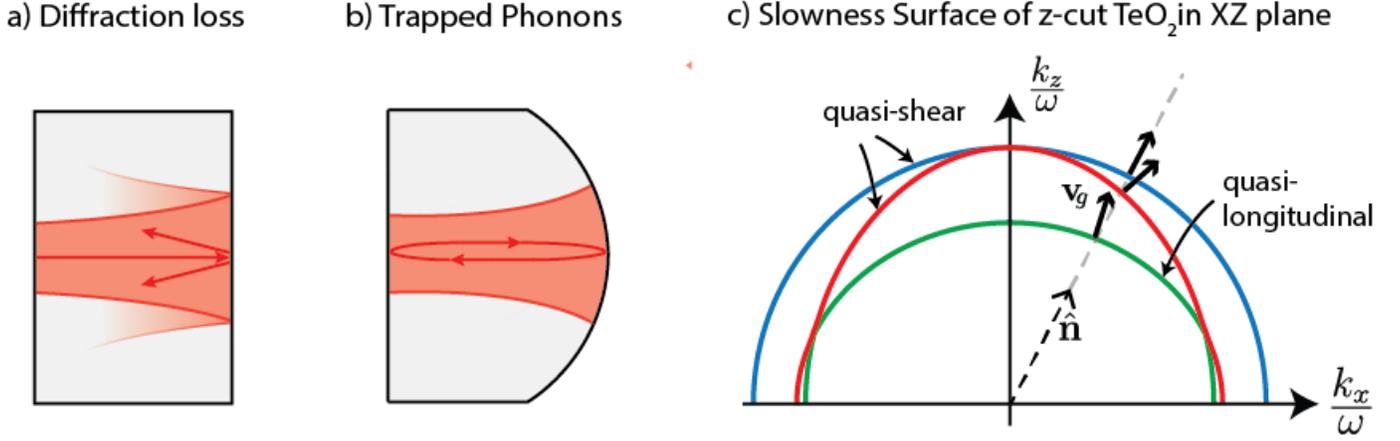

Figure 13: a) Acoustic beam of finite transverse area diffracts inside an acoustic medium. b) Crystal geometry can be shaped to formed standing acoustic modes without diffraction losses. c) Slowness surfaces obtained by solving the Christoeffel's equation for plane waves can be used to understand acoustic beam propagation in anisotropic solids.

## 5.1 Elastic-wave theory

The propagation of elastic waves in solids is described by the elastic wave equation [11]

$$\rho \frac{\partial u_i}{\partial t^2} = c_{ijlm} \frac{\partial u_m}{\partial x_j \partial x_l}, \tag{173}$$

where $u_i$ is the $i^{\text{th}}$ component of the acoustic displacement vector, $c_{ijlm}$ is the isothermal elastic constant tensor and $\rho$ is the density of the medium. For a plane wave $\mathbf{u} = \mathbf{u}_0 e^{i(\mathbf{k}\cdot\mathbf{r}-\omega t)}$ propagating in the direction $\hat{\mathbf{n}}$, where $\mathbf{k} = k\hat{\mathbf{n}}$, Eq. (173) reduces to the following eigenvalue equation

$$\left(\frac{c_{ijlm}}{\rho} n_j n_l - v^2 \delta_{im}\right) u_{0m} = 0, \tag{174}$$

where $v = \omega/k$ is the phase velocity of the acoustic wave. The nontrivial solutions to the equation above determine the dispersion relation for the acoustic wave in the medium. For each direction $\hat{\mathbf{n}}$, the solution to the eigenvalue equation yields three orthogonal eigenvalues and eigenvectors corresponding to the three speeds and polarizations of the acoustic waves. The plot of $1/v$ (or $k/\omega$) as a function of $\hat{\mathbf{n}}$, yields the slowness surface which describes the magnitudes and directions of the phase as well as group velocities (see Fig. 13 (c)). The group velocity direction points in the direction normal to the slowness surface. When $\hat{\mathbf{n}}$ lies in certain symmetry directions, the group velocity direction is in the same direction as the phase-velocity direction (or direction of $\hat{\mathbf{n}}$). Only in this case do we have one longitudinally polarized and two shear polarized acoustic waves. Generally however the waves are quasi-longitudinal ("longitudinal like") or quasi-shear ("shear like").

In piezoelectric crystals, the elastic wave equation and electromagnetic equations are intimately coupled. This type of piezoelectric coupling affects acoustic propagation in solids because slowness surfaces are modified upon such coupling [11]. We have ignored this effect because the modification to slowness surfaces (corresponding to longitudinal acoustic waves) due to piezoelectric coupling is negligible in the crystals we measured experimentally.

We use plane-wave decomposition and beam propagation to solve the Christoffel equation in anisotropic crystals with arbitrary boundary conditions. We use these quasi-numerical techniques

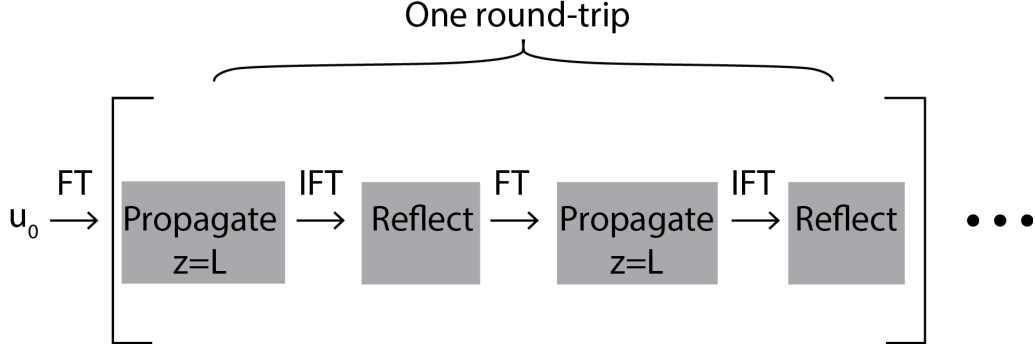

Figure 14: Schematics showing the steps in the beam propagation algorithm. FT = Fourier Transform and IFT = Inverse Fourier Transform.

because they not only allows us to find acoustic modes of our system with great speed but also lets us include non-linear interactions.

## 5.2 Beam Propagation: Finding stable acoustic modes

An acoustic field $\mathbf{u}(x, y, z, t)$ propagating in the $z$ direction, can be written in the plane wave basis as [12]

$$\mathbf{u}(x,y,z,t) = \left(\frac{1}{2\pi}\right)^3 \int_{-\infty}^{\infty} d\omega \int_{-\infty}^{\infty} dk_x \int_{-\infty}^{\infty} dk_y \sum_{i=1}^{3} \phi^n(k_x, k_y, \omega) \hat{d}^n\left(\frac{k_x}{\omega}, \frac{k_y}{\omega}\right) e^{i(\omega t - k_x x - k_y y - k_z^n z)}. \tag{175}$$

where $n$ indexes the three slowness surfaces, $\hat{d}^n(k_x/\omega, k_y/\omega)$ is the polarization vector for the plane wave propagating along $(k_x, k_y, k_z^n)$, where $k_z^n$ is determined by the corresponding slowness surfaces, and $\phi^n(k_x, k_y, \omega)$ corresponds to the amplitude for each polarization. For propagation along any other direction than $z$ direction, we rotate the coordinate system (and the elastic tensor) such that the $z$-axis always corresponds to the direction of beam propagation. To find resonant acoustic modes of a system, we look for solutions of the form $\mathbf{u}(x,y,z,t) = \mathbf{u}(x,y,z)e^{i\Omega t}$, where $\Omega$ is the frequency of the acoustic mode. Once the slowness surfaces, $k_z^n$, and the corresponding polarization vectors, $d^n$, are calculated using Eq. (174), we calculate the amplitudes, $\phi^n$, for the plane wave decomposition of the input field $\mathbf{u}^0(x, y, z = 0)$ at $z=0$ using the following equation [12]

$$\phi^{n=1}(k_x, k_y, \Omega) = \frac{\left\{(\mathbf{H} \cdot \hat{d}^1)[1 - (\hat{d}^2 \cdot \hat{d}^3)^2] + (\mathbf{H} \cdot \hat{d}^2)[(\hat{d}^1 \cdot \hat{d}^3)(\hat{d}^2 \cdot \hat{d}^3) - (\hat{d}^1 \cdot \hat{d}^2)] + (\mathbf{H} \cdot \hat{d}^3)[(\hat{d}^1 \cdot \hat{d}^2)(\hat{d}^3 \cdot \hat{d}^2) - (\hat{d}^1 \cdot \hat{d}^3)]\right\}}{\left\{1 + 2(\hat{d}^1 \cdot \hat{d}^2)(\hat{d}^2 \cdot \hat{d}^3)(\hat{d}^3 \cdot \hat{d}^1) - (\hat{d}^1 \cdot \hat{d}^2)^2 - (\hat{d}^1 \cdot \hat{d}^3)^2 - (\hat{d}^2 \cdot \hat{d}^3)^2\right\}}, \tag{176}$$

where $\mathbf{H}(k_x, k_y, \Omega)$ is the Fourier transform of the initial displacement field at plane $z = 0$ and is given by

$$\mathbf{H}(k_x, k_y, \Omega) = \int_{-\infty}^{\infty} dt \int_{-\infty}^{\infty} dx \int_{-\infty}^{\infty} dy\, \mathbf{u}^0(x, y, z=0) e^{i\Omega t} e^{i(-\omega t + k_x x + k_y y)}. \tag{177}$$

$\phi^2$ and $\phi^3$ are obtained similarly by permutation of indices in Eq. (176). Once the amplitude coefficients $\phi^n$ of the input field at $z = 0$ is calculated, the field at $z = L$ is calculated using Eq. (175). Then a phase-profile corresponding to spherical geometry is applied to each displacement component

$$\begin{bmatrix} u'_x(x,y,L) \\ u'_y(x,y,L) \\ u'_z(x,y,L) \end{bmatrix} = \begin{bmatrix} u_x(x,y,L)e^{ik_{t1}(x^2+y^2)/R} \\ u_y(x,y,L)e^{ik_{t2}(x^2+y^2)/R} \\ u_z(x,y,L)e^{ik_l(x^2+y^2)/R} \end{bmatrix}. \tag{178}$$

where $k_{t1} = \Omega/v_{t1}$, $k_{t2} = \Omega/v_{t2}$, and $k_l = \Omega/v_l$ are the wave-vectors of the two shear polarizations and one longitudinal polarization for a plane wave propagating along the $z$ direction. The spherical surface has a radius of curvature given by $R$. After computing the phase-shift due to the boundary, we re-compute the projection amplitudes and propagate the beam further by a distance of $L$. As before, we apply a phase profile due to the flat surface (i.e. a constant phase). This results in a displacement profile $\mathbf{u}^{m=1}(x, y, z = 0)$ after one round-trip inside the crystal. For a standing-wave acoustic mode, both the phase and the amplitude profile of the displacement field should remain unchanged after a single round-trip inside the crystal (i.e. $\mathbf{u}^1(x, y, z = 0) = \mathbf{u}^0(x, y, z = 0)$). Propagation over multiple round trips can be used to figure out both the resonant frequencies and mode profiles. This technique is analogous to a "Fox and Li" calculation used to find transverse mode patterns in an optical cavity [13]. We discuss this in detail next.

To find the resonant mode frequencies, we take an input field $\mathbf{u}^0(x, y, z = 0)e^{i\Omega t}$. $k_z^n$ is then computed for this particular value of $\Omega$. Using the procedure outlined above, we perform multiple round trips of the input field $\mathbf{u}^0$. We then calculate an interferometric sum of displacement fields after each round-trip at $z = 0$ as

$$\mathbf{u}_{\text{sum}} = \sum_m \mathbf{u}^m(x, y, z = 0). \tag{179}$$

The total intensity in the interfered field is then given by $I = \int dx\, dy |\mathbf{u}_{\text{sum}}|^2$. We then sweep the acoustic frequency $\Omega$ and calculate intensity as a function of $\Omega$, $I(\Omega)$. A resonant build-up of intensity occurs at frequencies corresponding to the acoustic modes that remain unchanged after each round-trip (See Fig. 15). Once a resonant frequency $\Omega_m$ is found, the input field $\mathbf{u}^0(x, y, z = 0)e^{i\Omega_m t}$ is propagated over multiple round-trips and the interferometric sum of displacement fields $\mathbf{u}_{\text{sum}}(x, y, z = 0)$ is computed. This displacement profile, $\mathbf{u}_{\text{sum}}$, is then used as an input to another beam-propagation over multiple round-trips. After several iterations, the interference profile does not change and converges to the mode profiles of the standing wave acoustic modes. We choose a Gaussian profile for the input acoustic field to find resonant acoustic modes because the electrostrictive force produced by the beating of the pump and Stokes light with Gaussian profiles also has a Gaussian shape. The input acoustic field is also off-centered relative to the crystalline axis to simulate the imperfect alignment in our experiment.

We used this general approach to find the acoustic modes accounting for all three slowness surfaces and considering all three polarization directions in our plano-convex system. For the longitudinal modes of our interest we find that only a small amount ($\sim 1\%$) of energy resides on the other polarizations (i.e. $u_x$ and $u_y$). Assuming all the energy in the polarization other than the z-direction is lost, we find a linewidth of $\sim$ 10Hz from our simulations, corresponding to a Q-factor of $10^9$. For the acoustic modes of interest, we are well within the paraxial limit as the beam profiles varies only in a small range of $k_x$ and $k_y$. Therefore, we neglect the variation of polarization with propagation direction (i.e. $\hat{d}^n(k_x/\omega, k_y/\omega) = \hat{d}^i(0,0)$) [12]. To calculate longitudinal acoustic modes we can then use the slowness surface corresponding to just the longitudinal polarization (i.e. $u_z(x, y, z)$). This

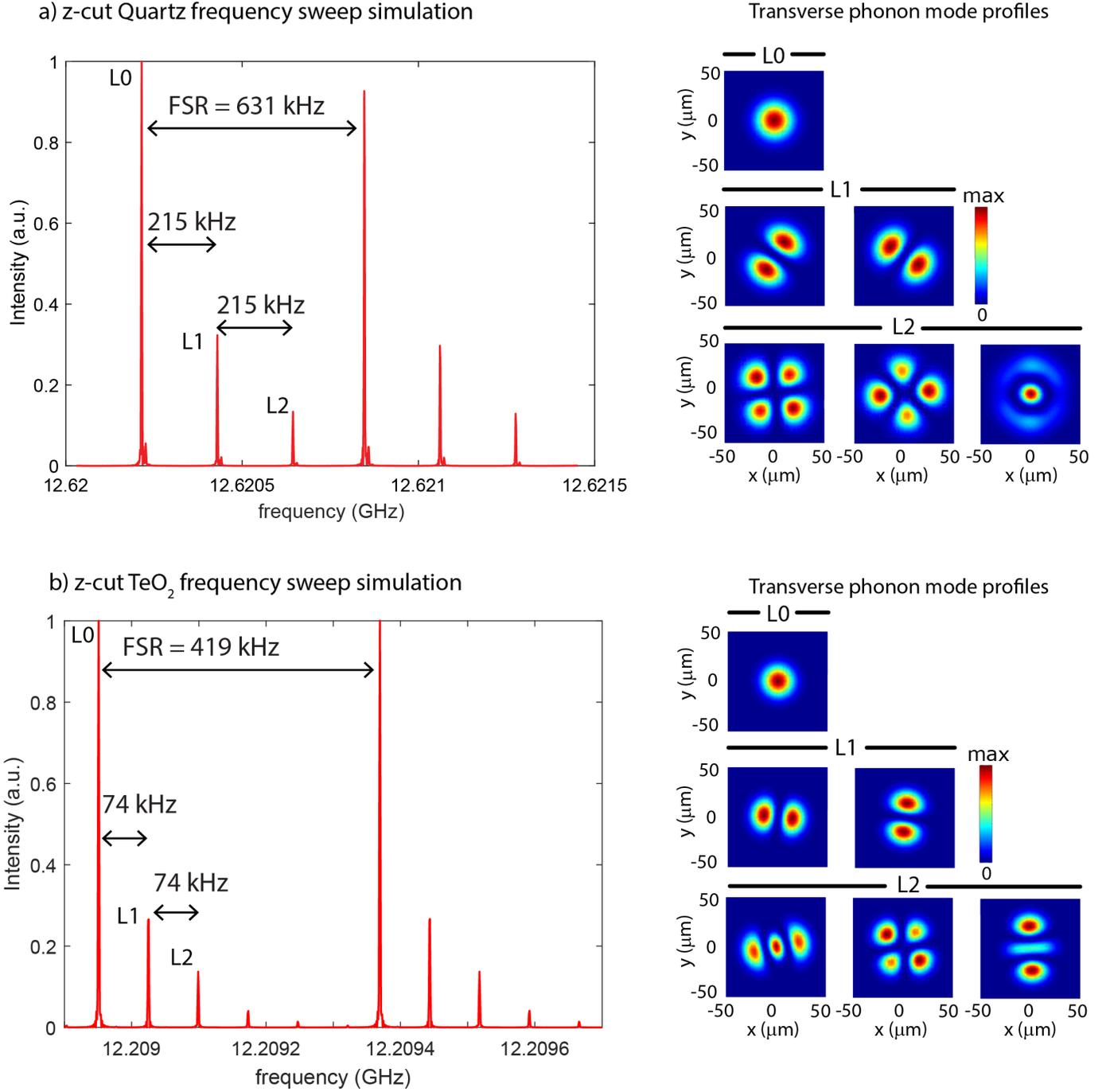

Figure 15: Simulated frequency sweep to identify resonant modes and mode profiles for a plano-convex crystal of 5 mm length and 10 mm radius of curvature. a) Simulation of $z$-cut Quartz with $\rho = 2648$ kgm$^{-3}, c_{11} = 86.8$ GPa, $c_{12} = 7.04$ GPa, $c_{13} = 11.91$ GPa, $c_{33} = 105.75$ GPa, $c_{44} = 58.20$ GPa, $c_{14} = -18.04$ GPa, and $c_{66} = 39.88$ GPa [14]. b) Simulation of $z$-cut $\alpha$-TeO$_2$ with $\rho = 6040$ kgm$^{-3}, c_{11} = 55.7$ GPa, $c_{12} = 51.2$ GPa, $c_{13} = 21.8$ GPa, $c_{33} = 105.8$ GPa, $c_{44} = 26.5$ GPa, and $c_{66} = 65.9$ GPa [15]. Here, $c_{ij}$ are elastic constants written in the reduced notation.

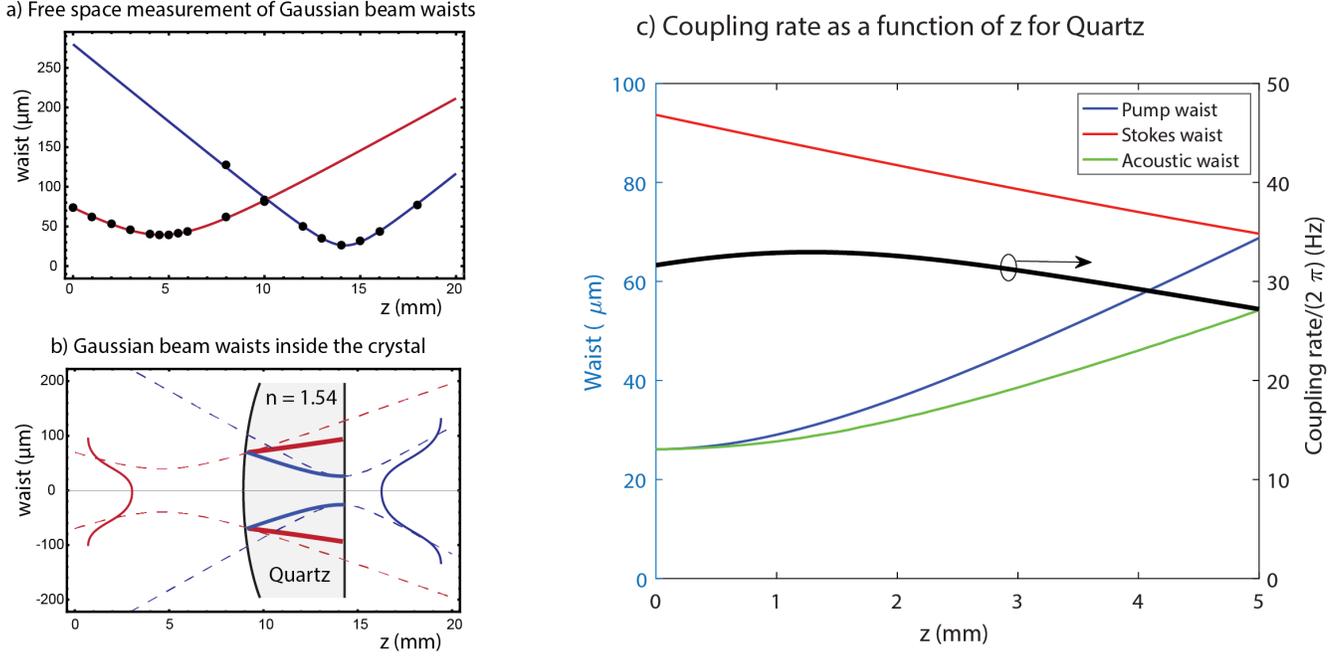

Figure 16: Acousto-optic overlap for quartz. a) The optical beams are Gaussian beams and their beam waists are measured as a function of $z$ in free-space. b) Using Gaussian beam optics we calculate the propagation of the optical beams inside the quartz-crystal. The optical beams are well within the paraxial approximation. c) The acoustic waist as a function of $z$ for the fundamental longitudinal mode for crystalline z-cut quartz is obtained from the numerical simulations and the coupling rate $g_0^m$ at each point on $z$ is calculated numerically.

process is computationally much faster and allows us to calculate transverse mode profiles for the longitudinal modes shown in Figure. 15. We compared the mode found using this approximation to the ones found using the general approach and find an excellent agreement.

# 6 Theory-Experiment Comparison

## 6.1 Low Coherence limit (Room Temperature)

Our experiment consists of co-polarized forward moving pump light and backward moving Stokes light. The pump light is modulated at 487 kHz, which is well within the linewidth of the Brillouin gain spectrum at room temperature. In this limit, the measured heterodyne Stokes power is given by Eq (156)

$$P_s(\text{lock-in}) = 4G_B L P_{s0} \sqrt{P_{p0} P_{p-1}}, \tag{180}$$

where $G_B$ in Eq. (7) can be re-written in terms of the effective acousto-optic area as

$$G_{\text{SBS}} = \frac{2\omega^3 n^8 p_{13}^2}{c^4 \rho \Omega_B \Gamma_B A_{\text{eff}}^{ao}}. \tag{181}$$

37

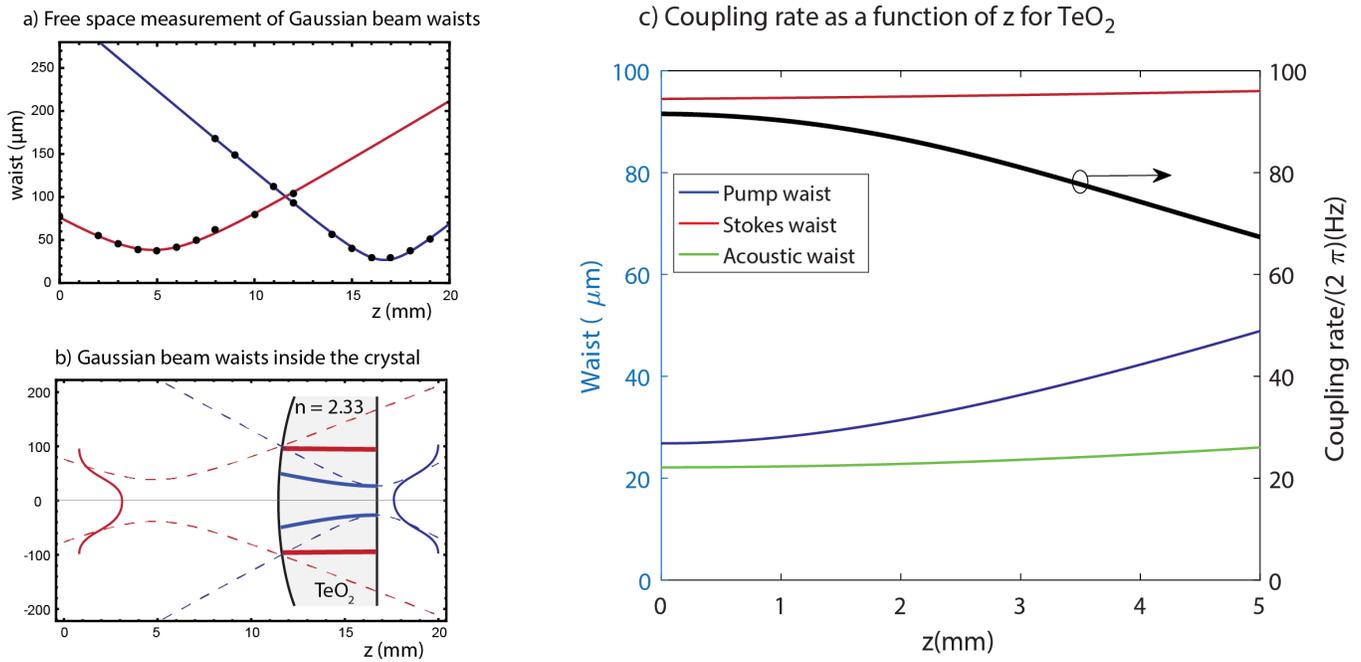

Figure 17: Acousto-optic overlap for $TeO_2$. a) The optical beams are Gaussian beams and their beam waists are measured as a function of $z$ in free-space. b) Using Gaussian beam optics we calculate the propagation of the optical beams inside the z-cut $TeO_2$ crystal. The optical beams are well within the paraxial approximation. c) The acoustic waist as a function of $z$ for the fundamental longitudinal mode for crystalline z-cut quartz is obtained from the numerical simulations and the coupling rate $g_0^m$ at each point on $z$ is calculated numerically.

### 6.1.1 Quartz

We used a pump light of free-space wavelength, $\lambda_p = 1548.99$ nm. The total pump power (including the side tones) before and after the crystal was 180 mW and 165 mW respectively. The Stokes power before and after the crystal was 41.3 mW and 38.6 mW respectively. The optical reflectivity at each surface due to refractive index mismatch is 4.52%. We take into account the cavity enhancement due to this optical relection to calculate the input optical powers for pump and Stokes light inside the crystal [13]. This calculation gives 184 mW of total pump light and 42.4 mW of Stokes light inside the crystal. Each side-band on the pump has 4.33% of the total pump power. So, $P_{s0} = 42.4$ mW, $P_{p0} = 0.9134 \times 184$ mW $= 168$ mW, $P_{p-1} = 0.0433 \times 184$ mW $= 7.97$ mW. Using $n = 1.54$, $p_{13} = 0.27$, $\rho = 2648$ kg/m$^3$, $\Omega_B = 2\pi \times 12.43$ GHz, $\Gamma_B = 2\pi \times 6.1$ MHz, $A_{\text{eff}}^{ao} = \pi \times (65.2 \ \mu\text{m})^2$, $L = 5$ mm we find that $G_{\text{SBS}} = 9.7 \times 10^{-3}$ (Wm)$^{-1}$ and $P_s(\text{lock-in}) = 0.31$ $\mu$W. This is in very good agreement with $P_s(\text{lock-in}) = 0.38$ $\mu$W measured in the experiment on resonance at room-temperature for $z$-cut Quartz.

### 6.1.2 TeO$_2$

We used a pump light of free-space wavelength, $\lambda_p = 1548.98$ nm. The total pump power (including the side tones) before and after the crystal was 186 mW and 135 mW respectively. The Stokes power before and after the crystal was 40.1 mW and 29.2 mW respectively. The optical reflectivity at each surface due to refractive index mismatch is 16%. We take into account the cavity enhancement due to this optical relection to calculate the input optical powers for pump and Stokes light inside the crystal [13]. This calculation gives 200 mW of total pump light and 43.2 mW of Stokes light inside the crystal. Each side-band on the pump has 3.69% of the total pump power. So, $P_{s0} = 43.2$ mW, $P_{p0} = 0.962 \times 200$ mW $= 185$mW, $P_{p-1} = 0.0369 \times 200$ mW $= 7.38$ mW. Using $n = 2.33$, $p_{13} = 0.34$, $\rho = 6040$ kg/m$^3$, $\Omega_B = 2\pi \times 11.86$ GHz, $\Gamma_B = 2\pi \times 10$ MHz, $A_{\text{eff}}^{ao} = \pi \times (71.1 \ \mu\text{m})^2$, $L = 5$ mm we find that $G_{\text{SBS}} = 0.10$ (Wm)$^{-1}$ and $P_s(\text{lock-in}) = 3.2$ $\mu$W. This is in good agreement with $P_s(\text{lockin}) = 3.3$ $\mu$W measured in the experiment on resonance at room-temperature for $z$-cut TeO$_2$.

## 6.2 High Coherence limit (Low Temperature)

Our experiment consists of co-polarized forward moving pump light and backward moving Stokes light. The pump light is modulated at 20.041 MHz, which is well outside the phase matching bandwidth (approximately 1 MHz for 5 mm long crystals). In this limit, the heterodyne Stokes power is given by Eq. (172).

### 6.2.1 Quartz

We used a pump light of free-space wavelength, $\lambda_p = 1548.98$ nm. The total pump power (including the side tones) before and after the crystal was 11.6 mW and 10.8 mW respectively. The Stokes power before and after the crystal was 0.83 mW and 0.75 mW respectively. The optical reflectivity at each surface due to index mismatch is 4.52%. We take into account the cavity enhancement due to this optical relection to calculate the input optical powers for pump and Stokes light inside the crystal [13]. This calculation gives 11.9 mW of total pump light and 0.84 mW of Stokes light inside the crystal. Each side-band on the pump has 4.33% of the total pump power. So, $P_{s0} = 0.84$ mW, $P_{p0} = 0.9134 \times 11.9$ mW $= 10.9$ mW, $P_{p-1} = 0.0433 \times 11.9$ mW $= 0.52$ mW. Using $n = 1.54$, $p_{13} = 0.27$, $\rho = 2648$ kg/m$^3$, $\Omega_B = 2\pi \times 12.62$ GHz, $\Gamma_B = 2\pi \times 300$ Hz, $|g_0| = 2\pi \times$

31.4 Hz (or $A_{\text{eff}}^{ao} = \pi \times (69.7 \ \mu\text{m})^2$), $L = 5$ mm we find that $P_s(\text{heterodyne}) = 1.7 \ \mu\text{W}$. This is in good agreement with $P_s(\text{lockin}) = 2.2 \ \mu\text{W}$ measured in the experiment on resonance at low-temperature for $z$-cut quartz.

### 6.2.2  TeO$_2$

We used a pump light of free-space wavelength, $\lambda_p = 1548.98$ nm. The total pump power (including the side tones) before and after the crystal was 205 mW and 165 mW respectively. The Stokes power before and after the crystal was 40.9 mW and 30.6 mW respectively. The optical reflectivity at each surface due to index mismatch is 16%. We take into account the cavity enhancement due to this optical relection to calculate the input optical powers for pump and Stokes light inside the crystal [13]. This calculation gives 229 mW of total pump light and 44.5 mW of Stokes light inside the crystal. Each sideband on the pump has 4.33% of the total pump power. So, $P_{s0} = 44.5$ mW, $P_{p0} = 0.9134 \times 229$ mW $= 209$ mW, $P_{p-1} = 0.0369 \times 229$ mW $= 9.92$ mW. Using $n = 2.33$, $p_{13} = 0.34$, $\rho = 6040$ kg/m$^3$, $\Omega_B = 2\pi \times 12.2$ GHz, $\Gamma_B = 2\pi \times 23$ kHz, $|g_0| = 2\pi \times 83.5$ Hz (or $A_{\text{eff}}^{ao} = \pi \times (77.0 \ \mu\text{m})^2$), $L = 5$ mm we find that $P_s(\text{heterodyne}) = 335 \ \mu\text{W}$. This is close to the experimental value of 240 $\mu$W measured in the experiment on resonance at low-temperature for $z$-cut TeO$_2$.

## 7  Dispersive Symmetry Breaking

In backward Brillouin scattering, incoming pump light can be either red-shifted (Stokes light) or blue shifted (anti-Stokes light) (see Fig. (18)). However, such processes are mediated by phonons moving in opposite directions (or phonons having opposite wavevectors). Therefore, it is clear from just the wavevector mismatch that the dynamics of Stokes and Anti-Stokes fields are uncoupled. However, since ultra-long lived standing wave phonons have both $+q$ and $-q$ wavevector components, it is natural to wonder if the Stokes and the anti-Stokes process are no longer decoupled in the high coherence limit. Careful re-consideration of the energy conservation and phase matching for the Stokes and anti-Stokes processes reveals that the phonon mode $(\Omega_s, q)$ involved in Stokes scattering is different than the phonon mode $(\Omega_{as}, -q)$ involved in the anti-Stokes scattering. Solving $k_p = q - (-k_s)$, $\omega_p = \omega_s + \Omega_s$, $k_{as} = q - k_p$, and $\omega_{as} = \omega_p + \Omega_{as}$, where $(\omega_p, k_p)$, $(\omega_{as}, -k_{as})$, $(\omega_s, -k_s)$ corresponds to frequency and wave-vector of pump, anti-Stokes and Stokes light, and $(\Omega_s, q)$ and $(\Omega_{as}, -q)$ and using linear acoustic and optical dispersion for light we find that

$$\Omega_s = \frac{2v_a \omega_p}{v_o} \frac{1}{1 + \frac{v_a}{v_o}}, \tag{182}$$

$$\Omega_{as} = \frac{2v_a \omega_p}{v_o} \frac{1}{1 - \frac{v_a}{v_o}}. \tag{183}$$

The Stokes/anti-Stokes splitting, $\Omega_{as} - \Omega_s$, is then given by

$$\Omega_{as} - \Omega_s = \left(\frac{2v_a}{v_o}\right)^2 \frac{\omega_p}{1 - (\frac{v_a}{v_o})^2}. \tag{184}$$

For a $z$-cut quartz with refractive index of $n = 1.54$ and $v_a = 6319$ m/s, the Stokes-anti-Stokes splitting for a pump light with a free-space wavelength of $\lambda_p = 1548.98$ nm is given by $\Omega_{as} - \Omega_s = 2\pi \times 815$ kHz. This splitting is not resolvable at room temperatures because the dissipation rate of phonons is large (i.e. order of $2\pi \times 10$ MHz). However, at cryogenic temperatures phonon dissipation rates can be

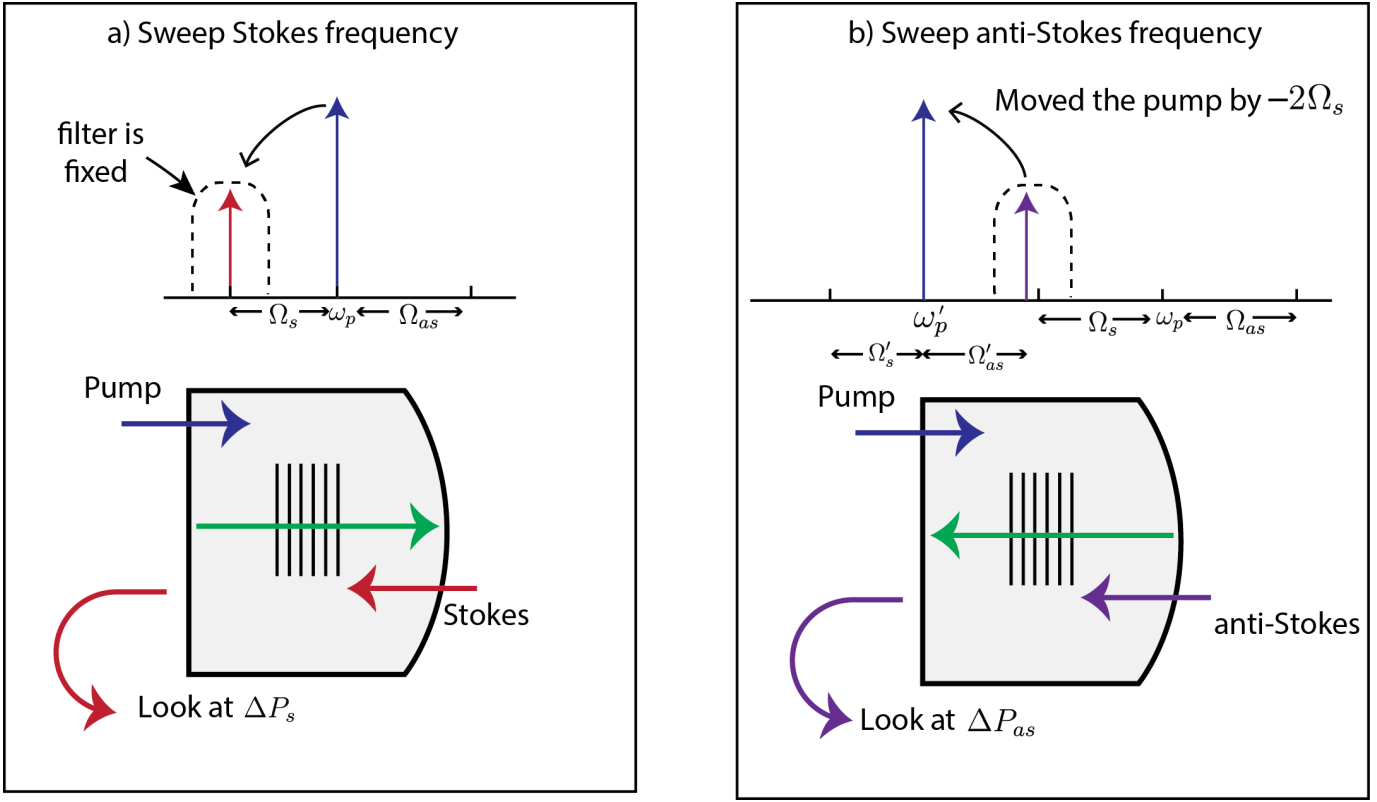

Figure 18: Measurement of the Stokes-anti-Stokes splitting. a) We measure the change in power of the Stokes light at $\omega_s$ as we sweep the frequency around $\omega_p - \Omega_s$. b) The pump is now moved to $\omega_p' = \omega_p - 2\Omega_s$ since the fiber Bragg grating filter remains fixed. Now we sweep the tone at a frequency higher than new pump frequency at $\omega_p'$, which gives us a Brillouin resonance around $\Omega_{as}'$. $\Omega_s \neq \Omega_{as}'$ and is related to the Stokes/anti-Stokes splitting in the crystal.

be much smaller than the Stokes-anti-Stokes splitting and phonons involved in the Stokes and the anti-Stokes processes can be resolved.

To determine $\Omega_{as}$ we could measure the energy-transfer from the anti-Stokes field to the pump field. However, this approach is not possible with our apparatus because the fiber Bragg grating filter used to select the Stokes frequency tone is fixed in frequency. Alternatively, we move the laser frequency to a lower value, $\Omega'_p = \omega_p - 2\Omega_s$ (see Fig. 18). We then sweep the anti-Stokes frequency around $\omega'_p + \Omega$ to obtain resonant energy-transfer between the anti-Stokes which is mediated by the phonon at frequency $\Omega'_{as}$. We will now show that $|\Omega'_{as} - \Omega_s| = |\Omega_{as} - \Omega_s|$,

$$\Omega'_{as} - \Omega_s = \left(\frac{2v_a}{v_o}\right) \frac{(\omega_p - 2\Omega_s)}{1 - \frac{v_a}{v_o}} - \left(\frac{2v_a}{v_o}\right) \frac{\omega_p}{1 + \frac{v_a}{v_o}}. \tag{185}$$

Substituting $\Omega_s$ in Eq. (182) in Eq. 185, we find

$$\Omega'_{as} - \Omega_s = -\left(\frac{2v_a}{v_o}\right)^2 \frac{\omega_p}{1 - \left(\frac{v_a}{v_o}\right)^2} \tag{186}$$

$$= -(\Omega_{as} - \Omega_s). \tag{187}$$

Therefore, the Stokes/anti-Stokes frequency splitting, $|\Omega_{as} - \Omega_s|$ can also be measured by measuring $|\Omega'_{as} - \Omega_s|$. The experimental measurement of $2\pi \times 830$ kHz for the Stokes/anti-Stokes splitting matches well with the predicted theoretical value of $2\pi \times 815$ kHz.



%\newpage

\end{document}